\documentclass[a4paper,11pt]{article}
\pdfoutput=1 % if your are submitting a pdflatex (i.e.~if you have
\usepackage{jheppub,braket} % for details on the use of the package, please
\usepackage[T1]{fontenc} % if needed
\usepackage{longtable}
\usepackage{amsfonts,amsthm,amsmath,amssymb,graphicx,color}
\usepackage{xcolor,colortbl}
\usepackage{float}
\usepackage{graphicx}  % needed for figures
\usepackage{xcolor}
\usepackage{braket}
\usepackage{amssymb}   % for math
\usepackage{amsmath,bm}
\usepackage{amsfonts}
\usepackage{bbm}

\usepackage{setspace}
\usepackage{mathrsfs}

\definecolor{Gray}{gray}{0.95}
\definecolor{LightCyan}{rgb}{0.88,1,1}

\newcommand{\be}{\begin{equation}}
\newcommand{\ee}{\end{equation}}
\newcommand{\bea}{\begin{eqnarray}}
\newcommand{\eea}{\end{eqnarray}}

\newcommand{\Tr}{\text{Tr}}

\newcommand{\ba}{\begin{eqnarray}}
\newcommand{\ea}{\end{eqnarray}}

\newcommand{\f}{\frac}

\newcommand{\prl}{Physical Review Letters}

\newcommand{\prd}{Physical Review D}

 \def\f {\frac}

\newcolumntype{a}{>{\columncolor{Gray}}c}
\newcolumntype{b}{>{\columncolor{white}}c}

\title{\boldmath Quantum vs. classical information: operator negativity as a probe of scrambling}

\author[a]{Jonah Kudler-Flam,}
\author[b,c]{Masahiro Nozaki}
\author[a]{Shinsei Ryu}
\author[a]{Mao Tian Tan}

\affiliation[a]{Kadanoff Center for Theoretical Physics, University of Chicago, IL~60637, USA}
\affiliation[b]{Berkeley Center for Theoretical Physics, Berkeley, CA 94720, USA}
\affiliation[c]{iTHEMS Program, RIKEN, Wako, Saitama 351-0198, Japan}

\emailAdd{jkudlerflam@uchicago.edu}
\emailAdd{mnozaki@yukawa.kyoto-u.ac.jp}
\emailAdd{ryuu@uchicago.edu}
\emailAdd{mtan1@uchicago.edu}

\abstract{We consider the logarithmic negativity and related quantities of time evolution operators. We study free fermion, compact boson, and holographic conformal field theories (CFTs) as well as numerical simulations of random unitary circuits and integrable and chaotic spin chains. The holographic behavior strongly deviates from known non-holographic CFT results and displays clear signatures of maximal scrambling. Intriguingly, the random unitary circuits display nearly identical behavior to the holographic channels. Generically, we find the ``line-tension picture'' to effectively capture the entanglement dynamics for chaotic systems and the ``quasi-particle picture'' for integrable systems. With this motivation, we propose an effective ``line-tension'' that captures the dynamics of the logarithmic negativity in chaotic systems in the spacetime scaling limit. We compare the negativity and mutual information leading us to find distinct dynamics of quantum and classical information. The ``spurious entanglement'' we observe may have implications on the ``simulatability'' of quantum systems on classical computers. Finally, we elucidate the connection between the operation of partially transposing a density matrix in conformal field theory and the entanglement wedge cross section in Anti-de Sitter space using geodesic Witten diagrams.
}

\begin{document} 
\maketitle
\flushbottom

%%%%%%%%%%%%%%%%%%%%%%%%%%%%%%%%%%%%%%%%%%%%%%%%%%%%%
%%%%%%%%%%%%%%%%%%%%%%%%%%%%%%%%%%%%%%%%%%%%%%%%%%%%%
\section{Introduction}
\label{intro}
The Hilbert space of quantum many-body systems is infamously enormous. Fortunately, when asking certain questions, one can restrict focus to small corners of the Hilbert space. For example, the ground states of gapped many-body systems are generically described by states with area-law entanglement. This is useful because we can discard the vast majority of Hilbert space that has volume law entanglement, allowing for accurate variational ansatzes. Furthermore, the smaller entanglement translates to efficient numerical algorithms on classical computers. On the other side of the spectrum, highly excited states in a many-body system can be well approximated by (generalized) Gibbs ensembles while fluctuations about stationary states are understandable using hydrodynamics. While these two extremes are well understood, our understanding of the chasm of far-from equilibrium excited states that lie between is limited and computationally intractable.

An important 
``order parameter''
that captures the complexity of simulating a quantum system classically is the entanglement. Frequently, entanglement is quantified by the von Neumann entropy. For example, if one restricts to a subregion of space after a global quantum quench, the von Neumann entropy generically undergoes linear growth at an ``entanglement velocity,'' $v_E$, and saturates to the classical thermal entropy that is extensive with subsystem size \cite{2005JSMTE..04..010C}. The early-time behavior corresponds to the area law regime while the late-time behavior lies within the hydrodynamical regime. It is at intermediate times that we are concerned with in this paper. 

An important caveat of the von Neumann entropy is that it captures both quantum and classical (thermal) correlations. While quantum correlations pose threats to simulatability, classical correlations are innocuous. It is thus important to isolate the quantum contribution to the entropy. We therefore study the logarithmic negativity, a proper measure of entanglement in mixed states. The logarithmic negativity is an information-theoretic quantity based on the \textit{Positive Partial Transpose} (PPT) criterion for the separability of mixed states \cite{PhysRevLett.77.1413,1996PhLA..223....1H,1999JMOp...46..145E,2000PhRvL..84.2726S,2002PhRvA..65c2314V,2005PhRvL..95i0503P} and has been used in many contexts. For example, it has been used to characterize 1+1d conformal field theories, 2+1d topological field theories, many-body localized spin chains, variational states, and holographic conformal field theories \cite{2014JHEP...10..060R,2019PhRvD..99j6014K, 2012PhRvL.109m0502C, 2013JSMTE..02..008C, 2013JSMTE..05..002C, 2013JSMTE..05..013A, 2014PhRvB..90f4401C, 2016PhRvB..94s5121R, 2016PhRvB..94c5152R, 2016JPhA...49l5401B, 2014NJPh...16l3020E, 2014JSMTE..12..017C, 2015NuPhB.898...78H, 2015PhRvB..92g5109W, 2013PhRvA..88d2319C, 2013PhRvA..88d2318L, 2016JHEP...09..012W, 2016PhRvB..93x5140W, 2018arXiv180904689W, 2015JSMTE..06..021D}.

The partial transpose of a density matrix can be defined through its matrix elements as
\begin{align}
    \bra{i_1 ,j_2} \rho^{T_2} \ket{k_1, l_2} = \bra{i_1 ,l_2} \rho \ket{k_1, j_2},
\end{align}
where $i_1, j_2$ etc. are basis vectors for bipartite subsystems $1$ and $2$.
The logarithmic negativity is then defined as
\begin{align}
    \mathcal{E} = \log\left(\left| \rho^{T_2}\right|_1  \right),
\end{align}
where $\left|\mathcal{O} \right|_1 = \Tr \sqrt{\mathcal{O}^{\dagger} \mathcal{O}}$ is the trace norm of $\mathcal{O}$ i.e.~the sum of the magnitudes of its eigenvalues. We can then calculate the logarithmic negativity in a variety of ways. For small lattice models, we may perform exact diagonalization. For 1+1d conformal field theories, we can use the replica trick to compute partition functions on Riemann surfaces corresponding to the moments of the partially-transposed density matrix \cite{2012PhRvL.109m0502C}. Finally, in holographic conformal field theories, we may transfer the problem to the bulk where the negativity is conjectured to be dual to a particular geometric object \cite{2019PhRvD..99j6014K}, a procedure that will be reviewed shortly.  

In addition to these motivations from many-body physics, thermalization and the scrambling of quantum information have played major roles in recent studies of black hole physics, largely because the AdS/CFT correspondence states that thermal states in the boundary conformal field theory are dual to black hole states in the bulk gravity theory. After quantum information in sufficiently excited pure states is scrambled, the states accurately mimic the thermal behavior of black holes. A successful and intensely studied probe of the relevant time scales and thermalization behavior is the out of time ordered correlator (OTOC) which diagnoses quantum chaos by its exponential growth at early times analogous to classical Lyapunov growth \cite{2014JHEP...03..067S,2015PhRvL.115m1603R,2016JHEP...08..106M,2016PhRvD..94j6002M}. The entanglement of unitary evolution operators is a complementary probe of chaotic behavior. One advantage of the operator entanglement is that it is a state independent quantity and one does not need to choose a specific operator. It has consequently been shown to be an effective diagnostic of scrambling and chaos \cite{2016JHEP...02..004H,2018arXiv181200013N}. In this paper, we propose that the \textit{operator negativity} may be a superior diagnostic of scrambling of \textit{quantum} information. 

The paper is organized as follows: In the rest of Section \ref{intro}, we review the necessary background regarding operator entanglement and logarithmic negativity in holographic field theories. In Section \ref{randU_section}, we discuss the line-tension picture for ``hydrodynamic'' information flow and propose a generalization of this picture to logarithmic negativity. Here, we discuss the apparent similarities between the minimal surfaces in holographic entanglement entropy and the line-tension picture. We study operator entanglement in random unitary circuits which serve as tractable toy models for strongly interacting quantum systems. In Section \ref{op_neg_section}, we compute the operator logarithmic negativity for free fermion, compact boson, and holographic CFTs as well as for quantum spin chains. We focus on the late-time behavior of the tripartite logarithmic negativity, finding clear distinctions of scrambling between the models. In Section \ref{discussion_section}, we discuss future directions, leaving certain details of computations and a review/expansion of the holographic negativity conjecture to the appendices.

\subsection{Review of unitary operator entanglement}
The idea of studying the scrambling ability of unitary channels by looking at the operator entanglement was first proposed in Ref.~\cite{2016JHEP...02..004H}\footnote{We note that operator space entanglement was first studied in Refs.~\cite{2001PhRvA..63d0304Z,2007PhRvA..76c2316P} and was discussed in the context of ``simulatability'' in Ref.~\cite{2017JPhA...50w4001D}.}. In particular, they considered the time evolution operator $U(t) = e^{-i H t}$ and performed a state-operator map
\begin{equation}
U(t) = \sum_{i,j}u_{ij}|i\rangle \langle j | \longrightarrow |U(t) \rangle = \mathcal{N} \sum_{i,j}u_{ij}|i\rangle_{\text{in}} \otimes|j \rangle_{\text{out}}^*
\end{equation}
where $\mathcal{N}$ is some normalization factor and $\ket{\cdot}^*$ denotes the CPT conjugate. The corresponding operator state now lives in a doubled Hilbert space. The analog in quantum field theory looks like the thermofield double state because we must regularize the high energy modes
\begin{align}
   \ket{ U_{\beta}(t) }= \mathcal{N}\sum_a e^{-itE_a}e^{-\beta E_a/2}\ket{a}_{in} \otimes \ket{a}_{out}^*,
\end{align}
where the inverse temperature, $\beta$, acts as the regulator.
One can then study the corresponding entanglement of this state and its time dependence, which only depends on the Hamiltonian of the system and not on any particular choice of a state.

A particular quantity of interest is the triparitite operator mutual information (TOMI)
\begin{equation}
I_3(A:B:C) \equiv I(A,B)+I(A,C)-I(A,B\cup C),
\end{equation}
where $A$ is an interval in the input Hilbert space while $B$ and $C$ form a bipartition of the total output Hilbert space. The first two bipartite mutual informations characterize the local correlations between $A$ and $B$, and $A$ and $C$ respectively, while the third bipartite mutual information measures the global correlation between $A$ and the entire output $B\cup C$. If information is scrambled, then it stands to reason that it is stored non-locally over the entire output rather than in local intervals in the output. Hence, the scrambling of information would imply that $I_3(A:B:C)$ is negative. This is observed in the numerical study of spin chains in Ref.~\cite{2016JHEP...02..004H}.

The authors of Ref.~\cite{2018arXiv181200013N} further studied the problem in the context of two-dimensional conformal field theories where analytic computations are tractable. In particular, they studied the Dirac fermion, compactified boson, and holographic CFTs. The amount of information scrambling is characterized by the late-time saturation value of $I_3(A:B:C)$. The non-interacting Dirac fermion showed no scrambling and the bipartite mutual information could be perfectly described by a model of quasiparticles. Next, they studied the compactified boson at both rational and irrational radii. The compactified boson at rational radii showed a small amount of scrambling, with the late-time value of the tripartite operator mutual information depending on the quantum dimension of the twist operator for a two-sheeted Riemann surface. The compactified boson at irrational radii showed a larger amount of scrambling, with the late-time value of the tripartite operator mutual information scaling as $-\log l_A/\beta$, where $l_A$ is the size of subsystem $A$. 
% The last type of CFT studied in Ref.~\cite{2018arXiv181200013N} are the holographic CFTs. These are
Finally, CFTs with a large central charge and sparse low-lying spectrum (possessing a holographic dual) were studied. It was shown that the saturation value of the tripartite operator mutual information for these CFTs behave as $-l_A/\beta$. These CFTs exhibit the maximal amount of scrambling, which is consistent with the conjecture that holographic CFTs are maximally chaotic.

We are interested in comparing these results for operator mutual information to operator logarithmic negativity to distinguish the scrambling behavior of classical and quantum information. We would therefore like to construct an analogous quantity to the tripartite mutual information for the logarithmic negativity. A tripartite generalization of \textit{entanglement negativity} has been proposed in the past for finite dimensional systems \cite{2008EPJD...48..435S}. However, to our knowledge, there is no generalization for \textit{logarithmic negativity}. In particular, there is no generalization to quantum field theories\footnote{We note that multipartite constructions for the entanglement wedge cross-section in the context of entanglement of purification have been studied in Refs.~\cite{2018arXiv180502625U, 2018arXiv180500476B}.}. Analogous to the tripartite mutual information, we define the tripartite logarithmic negativity (TOLN) as
\begin{align}
    \mathcal{E}_3(A:B:C) \equiv \mathcal{E}(A,B) + \mathcal{E}(A,C) - \mathcal{E}(A, B\cup C).
\end{align}
$\mathcal{E}_3$ can characterize the tripartite quantum entanglement in pure states where $I_3$ fails. This can be seen, for example, in the GHZ state
\begin{align}
    \ket{GHZ} = \frac{1}{\sqrt{2}} \left(\ket{000} + \ket{111} \right)
\end{align}
where $\mathcal{E}_3 = -\log 2$, but $I_3 = 0$. This state is purely classical when losing a single qubit, but is clearly entangled when one has access to all three qubits. The quantum correlations and classical correlations are mixed in a way that make $I_3$ misleading. 
Unlike $I_3$, $\mathcal{E}_3$ is not generally symmetric about the three subregions. We will examine this quantity in the following sections.

\subsection{Logarithmic negativity and entanglement wedge cross-sections}
\label{negativity_section}

A conjecture for the holographic dual of logarithmic negativity was proposed in Ref.~\cite{2019PhRvD..99j6014K}\footnote{After the completion of this work, the conjecture was derived more convincingly in $AdS_3/CFT_2$ \cite{2019PhRvL.123m1603K}}. It is based on the entanglement wedge of the bulk geometry that is dual to the CFT. Given a Cauchy slice, $\Xi$, a (possibly disconnected) subregion of the boundary, $A$, and the Ryu-Takayanagi surface \cite{2006PhRvL..96r1602R,2006JHEP...08..045R,2007JHEP...07..062H}, $\gamma_A$, the entanglement wedge of $A$ is the codimension-1 surface in $\Xi$ whose boundary is $\gamma \cup A$. With this definition in mind, the prescription states that the logarithmic negativity is proportional to the minimal cross-sectional area of the entanglement wedge in appropriate limits\footnote{For the generalization to arbitrary entangling surfaces using the backreactions of cosmic branes, see Appendix \ref{app_neg_blocks} and Ref.~\cite{2019PhRvL.123m1603K}.}
\begin{align}
    \mathcal{E} = \mathcal{X}_d \frac{E_W}{4G_N} ,
    \label{main}
\end{align}
%\begin{align}
%    \mathcal{X}_d = \left(\frac{1}{2}x_d^{d-2}\left(1 + x_d^2\right) - 1\right), \quad x_d = \frac{2}{d}\left(1 + \sqrt{1 - \frac{d}{2} + \frac{d^2}{4}}\right),
%\end{align}
where $G_N$ is the bulk Newton constant, $E_W$ is the area of the minimal entanglement wedge cross-section\footnote{Note that we have not included the $4G_N$ in the definition of $E_W$ as is sometimes done in the literature.}, and $d$ is the dimension of the CFT. For 2d CFTs, $\mathcal{X}_2 = 3/2$. This has been shown to reproduce many nontrivial CFT results: a single interval at zero and finite temperatures, adjacent intervals at zero and finite temperatures, disjoint intervals at zero temperature, the thermofield double state, and holographic perfect tensor codes. There is further evidence from geodesic Witten diagrams that describe the correlation functions for negativity described in Appendix \ref{app_neg_blocks}. Besides the tensor network derivation and the various checks in CFT, there are a few reasons why it makes sense for the negativity to be dual to an object related to the entanglement wedge cross section. Most importantly, because the negativity does not count classical correlations, its holographic dual should avoid the horizon in a black hole spacetime. This is precisely how the entanglement wedge cross section behaves, by terminating on the horizon and not picking up its volume law contributions. Furthermore, with motivations from bit threads \cite{2017CMaPh.352..407F}, the authors of Ref.~\cite{Agon:2018lwq} noted that the entanglement wedge cross section may be related to the maximum number of Bell pairs that can be distilled between the regions using only local operations and classical communications (LOCC). Logarithmic negativity has a close relationship with distillable entanglement and provides an upper bound \cite{2002PhRvA..65c2314V}.

The entanglement wedge is interesting in its own right and has played a major role in understanding bulk reconstruction and subregion-subregion duality \cite{2014arXiv1412.8465J,2016JHEP...06..004J,2015JHEP...04..163A, 2015JHEP...06..149P,2018arXiv180610560F}. Furthermore, in addition to logarithmic negativity, the minimal cross-sectional area of the entanglement wedge has recently become a quantity of great interest in the contexts of entanglement of purification, ``odd entanglement entropy\footnote{Odd entanglement entropy is defined as 
$$
S_o = -\sum_{\lambda_i > 0 } |\lambda_i | \log |\lambda_i| + \sum_{\lambda_i < 0 } |\lambda_i | \log |\lambda_i|
$$ where $\{\lambda_i\}$ are the eigenvalues of the partially transposed density matrix. 
% We discuss in Appendix \ref{app_neg_blocks} how the odd entropy appears to be a precise dual to the entanglement wedge cross-section in $AdS_3/CFT_2$ without backreaction.
},'' and ``reflected entropy'' 
\cite{2017arXiv170809393T,2018JHEP...01..098N,2018arXiv180909109T, 2019arXiv190500577D}.
% \cite{2017arXiv170809393T, 2018JHEP...04..132B,Agon:2018lwq, 2018JHEP...01..098N, 2018arXiv180502625U,2018JHEP...03..006B,2018PTEP.2018f3B03H,2018PhRvD..98b6010N,2018arXiv180500476B,2018arXiv180405855E,2018arXiv180909109T,2018arXiv181205268C,2018arXiv181101983B,2019JHEP...01..114Y,2019arXiv190406871D,2019arXiv190202369B,2019arXiv190409582J,2019arXiv190308490B,2019arXiv190210161P, 2019arXiv190500577D, 2019arXiv190605970H} 
% \textcolor{red}{(JKF: Do we need to cite every paper that talks about purification or just the originals? If its every one then I will need to do a more thorough literature search.)}.
In fact, one can show that the odd entropy is precisely computed by the entanglement wedge cross-section in $AdS_3/CFT_2$ using geodesic Witten diagrams (see Appendix \ref{app_neg_blocks}). Furthermore, a replica trick for the reflected entropy establishes its connection to the entanglement wedge cross-section in general dimensions \cite{2019arXiv190500577D}. For $AdS_3/CFT_2$, a R\'enyi version of the reflected entropy was shown to equal the logarithmic negativity \cite{2019PhRvL.123m1603K}. The geometric distinction between the holographic negativity proposal and those for the entanglement of purification, odd entropy, and reflected entropy is that the probe brane to compute the cross-section for negativity has finite tension (leading e.g.~to the constant $\mathcal{X}_d$) while the brane for the other quantities is tensionless. This is because R\'enyi entropies are dual to backreacting cosmic branes \cite{2016NatCo...712472D}. In this paper, our goals are in the dynamics of quantum information rather than gravity, so we use the tensionless entanglement wedge cross-section (\ref{main}), in the appropriate limits, as a very good approximation for the logarithmic negativity of large c CFTs. 
% We now review previous results involving negativity from CFT methods and the entanglement wedge cross section from gravitational methods.

\section{Comparing holographic entanglement entropy and the line-tension picture}
\label{randU_section}

In the spacetime scaling limit, the behavior of entanglement in chaotic systems can be understood in terms of a model dependent line-tension, $\mathcal{T}(\bar{v}, \bar{x})$, of some membrane, $\mathcal{M}$, in the spacetime \cite{2017PhRvX...7c1016N,2018arXiv180300089J,2018PhRvD..98j6025M,2018PhRvX...8c1058R,2018PhRvX...8b1013V} 
\begin{align}
    S(t) = \int_{\mathcal{M}} d^{d}x \mathcal{T}(\bar{v}, \bar{x}),
\end{align}
where $v$ is the local ``velocity'' of the minimal membrane. This is both an effective computational tool and an illuminating phenomenological explanation of entanglement production. Similar to the Ryu-Takayanagi surface, the minimal membrane is homologous to the boundary subregion. In our convention, the velocity of the line is not bounded by the butterfly velocity as it was in Ref.~\cite{2018arXiv180300089J}. This is because we are working in the doubled Hilbert space of operator entanglement and it will be useful for us when studying logarithmic negativity. This has mainly been studied in the context of random unitary circuits, but has been applied to holographic systems as well. 

The minimal membrane picture for the operator entanglement is strongly reminiscent of the holographic formulation of operator entanglement. The holographic dual of the state created by the time evolution operator is a time slice of the eternal black hole with non-trivial $H_R + H_L$ time evolution. In this set-up, the length of the wormhole grows linearly in time, leading us to the interesting observation that the radial direction in the gravity theory plays an analogous role to the temporal direction in the line-tension picture. The operator entanglement in both scenarios is computed by minimal membranes, only with different metrics.

\subsection{Mutual information in random unitary circuits}

Random unitary circuits are tractable toy models for chaotic systems that efficiently scramble quantum information. In the spacetime scaling limit, these have been shown to possess interesting universal properties such as emergent hydrodynamics of entanglement entropy and out-of-time-ordered correlators and were the original motivation for the conception of the line-tension picture \cite{2017PhRvX...7c1016N,2018PhRvX...8b1014N,2018PhRvX...8c1057K,2018arXiv180409737Z,2017PhRvB..95i4206Z,2018PhRvX...8c1058R,2018PhRvX...8b1013V}\footnote{See Ref.~\cite{2018PhRvB..98a4309Y} for interesting discussion regarding the relation between state and operator entanglement in random circuits.}. In general, random unitary circuits are difficult to simulate efficiently. Because we are interested in the spacetime scaling limit (i.e.~large system sizes), we restrict ourselves, for the time being, to a subset of local unitary operators belonging to the Clifford group \cite{2004PhRvA..70e2328A}. This allows us to simulate large circuits because we only need to keep track of the stabilizers of the states generated by the Clifford group; this scales as $\mathcal{O}(L)$, where $L$ is the system size, instead of the exponential scaling for Haar random unitaries. Computing the entanglement entropy scales as $\mathcal{O}(L^3)$. The random Clifford group is composed of phase, Hadamard, and CNOT gates. For more details, see Appendix \ref{clif_appendix}.

\begin{figure}
    \centering
    \includegraphics[width =7cm]{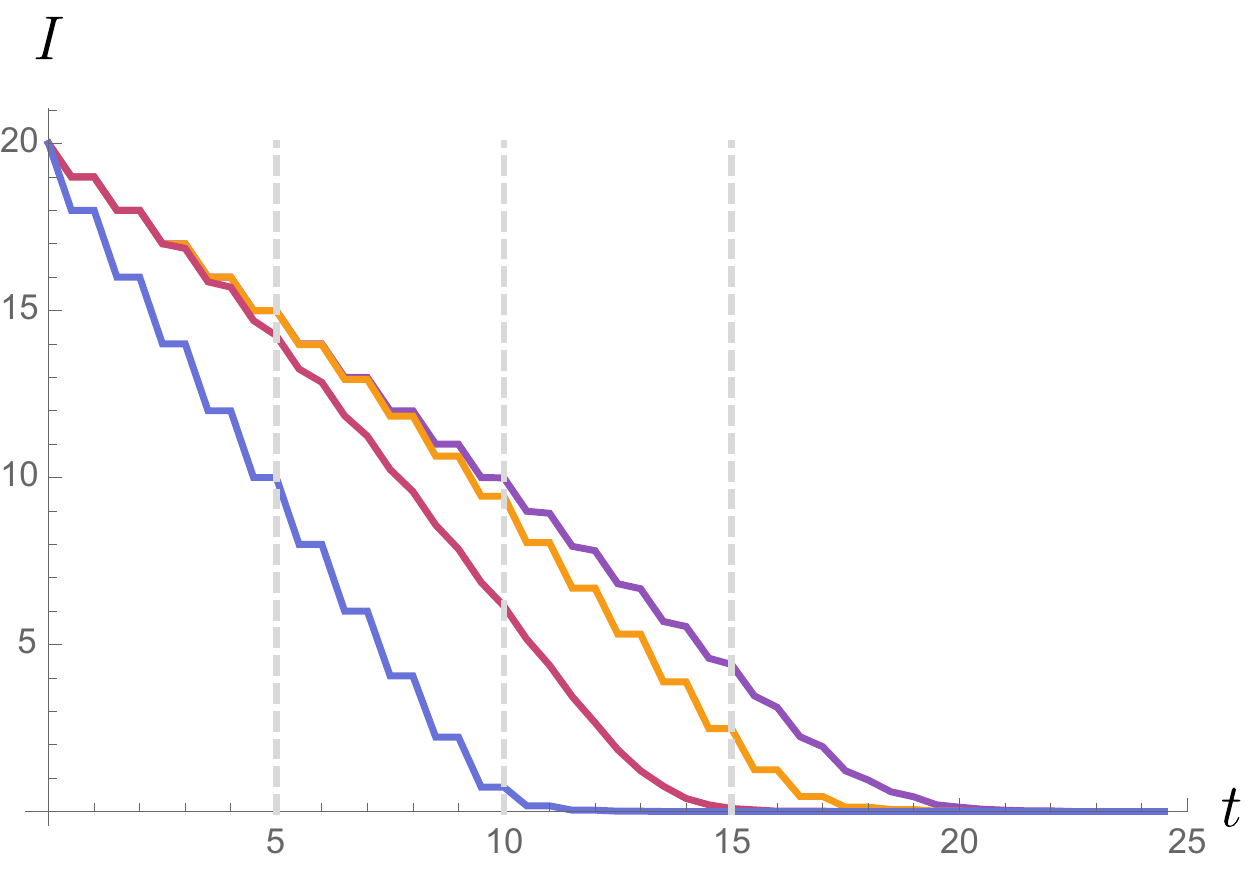}\llap{\raisebox{2.5cm}{\includegraphics[height=2cm]{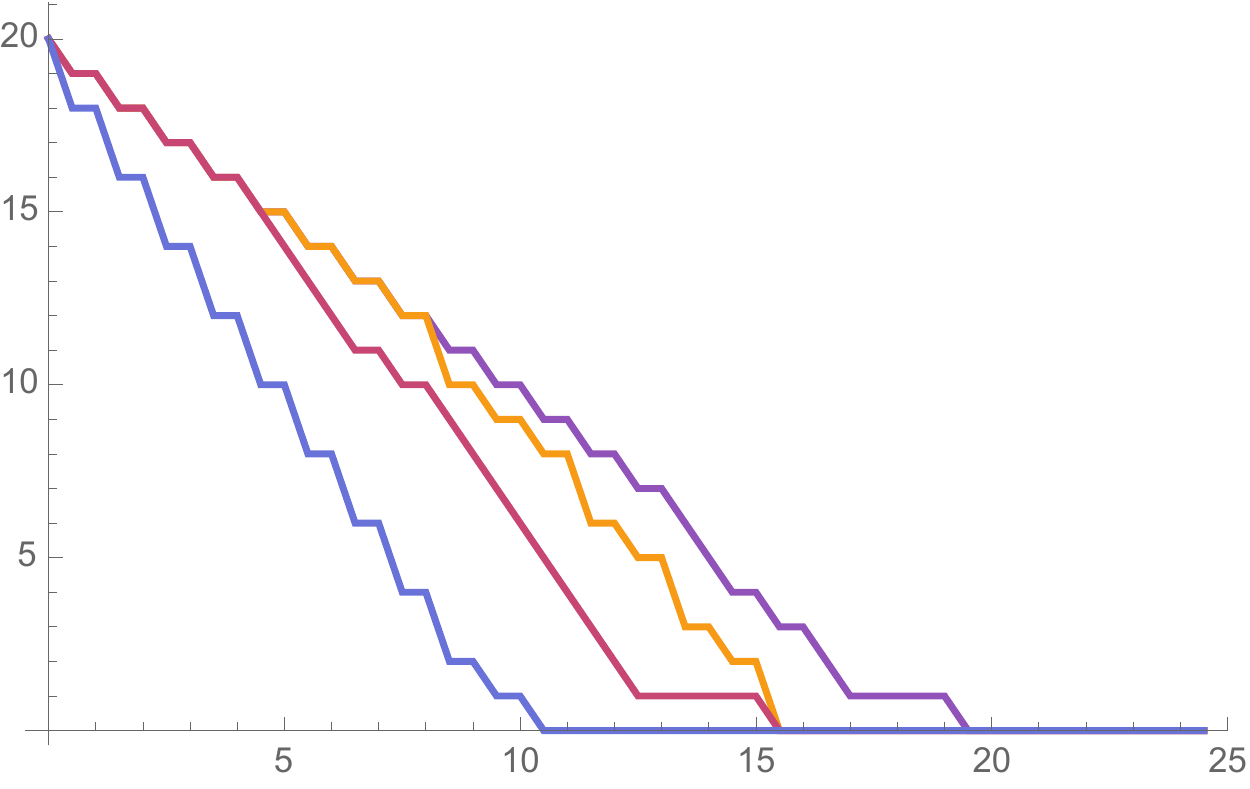}}}
    \includegraphics[width = 7cm]{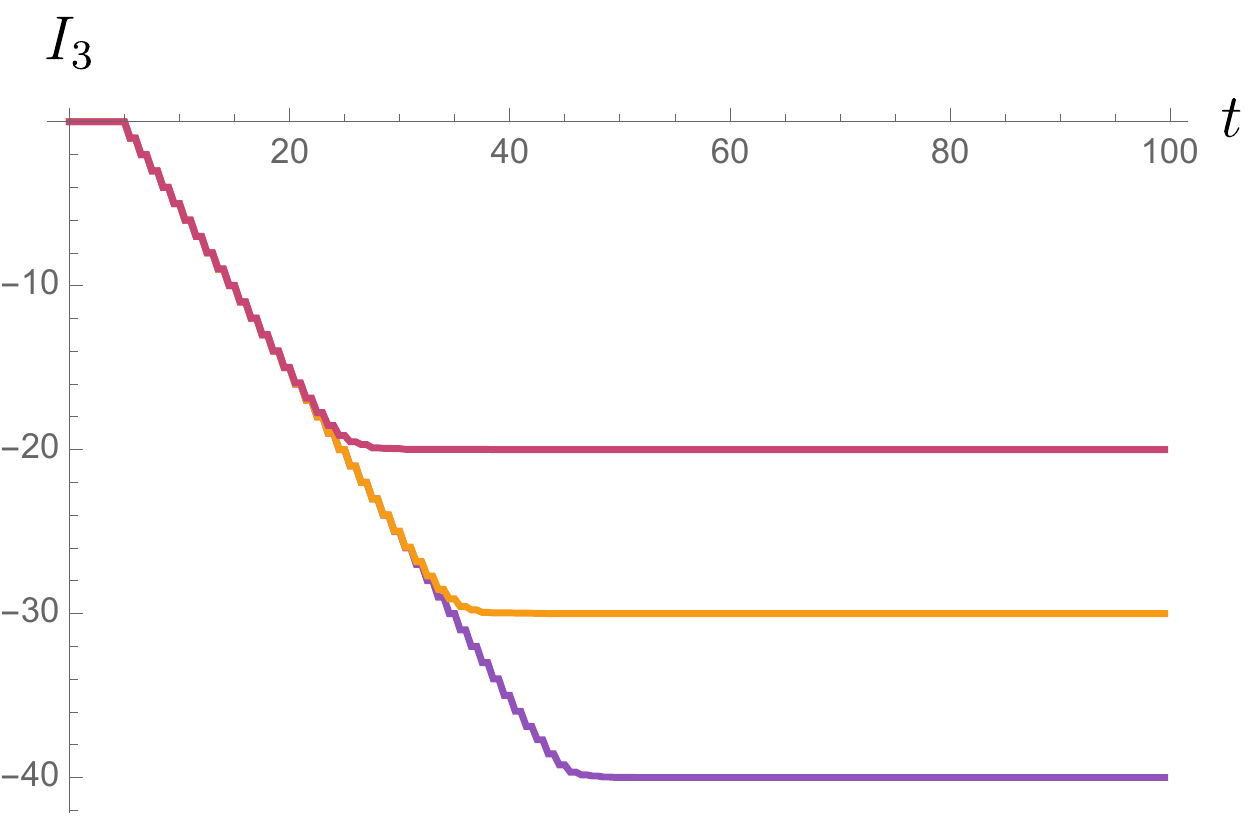}\llap{\raisebox{2.5cm}{\includegraphics[height=2cm]{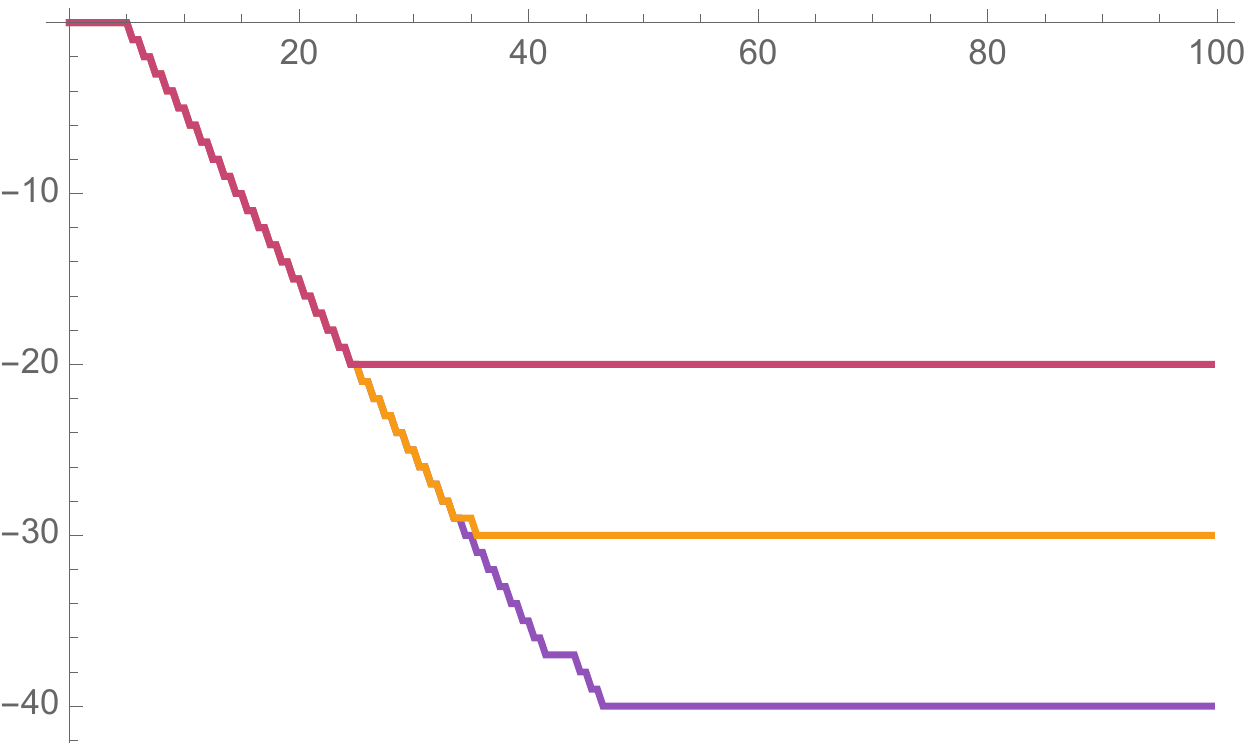}}}
    \includegraphics[width = 7cm]{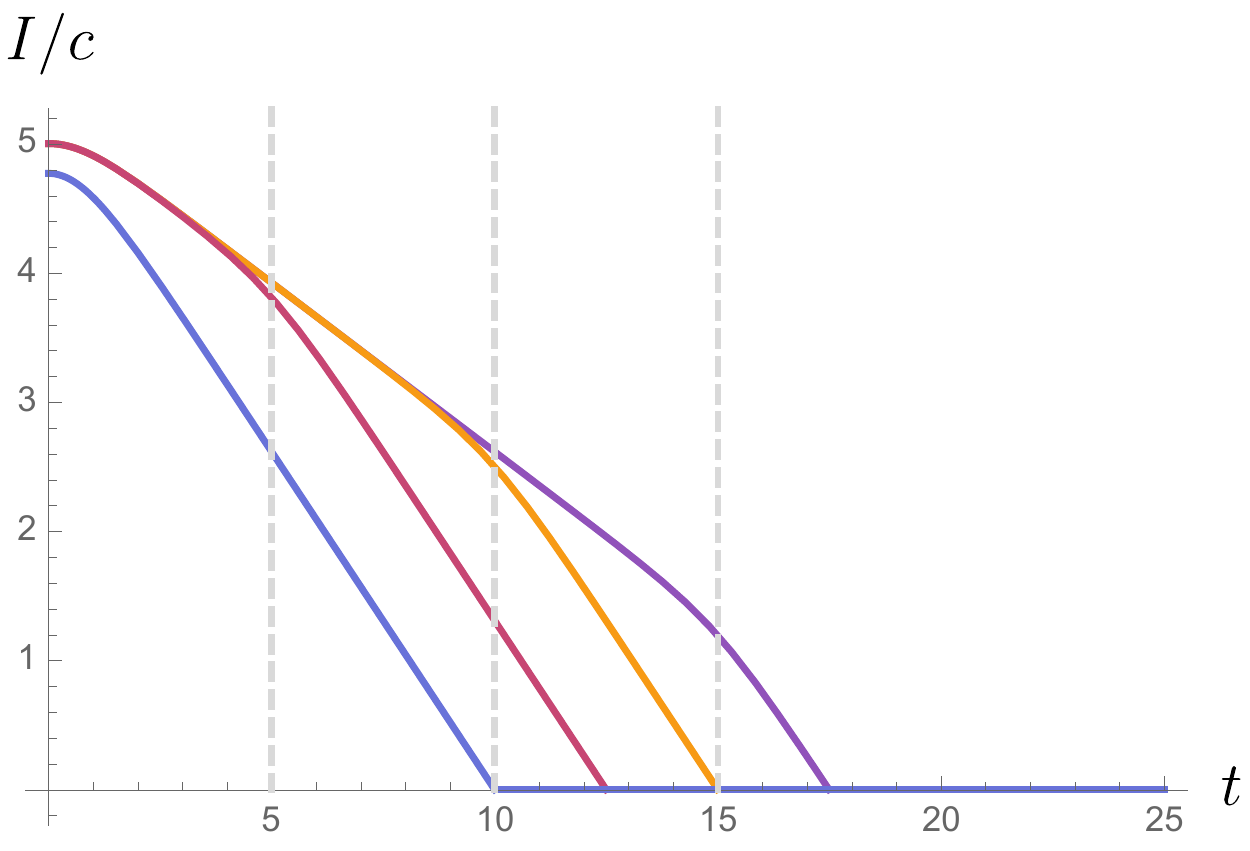}
    \includegraphics[width = 7cm]{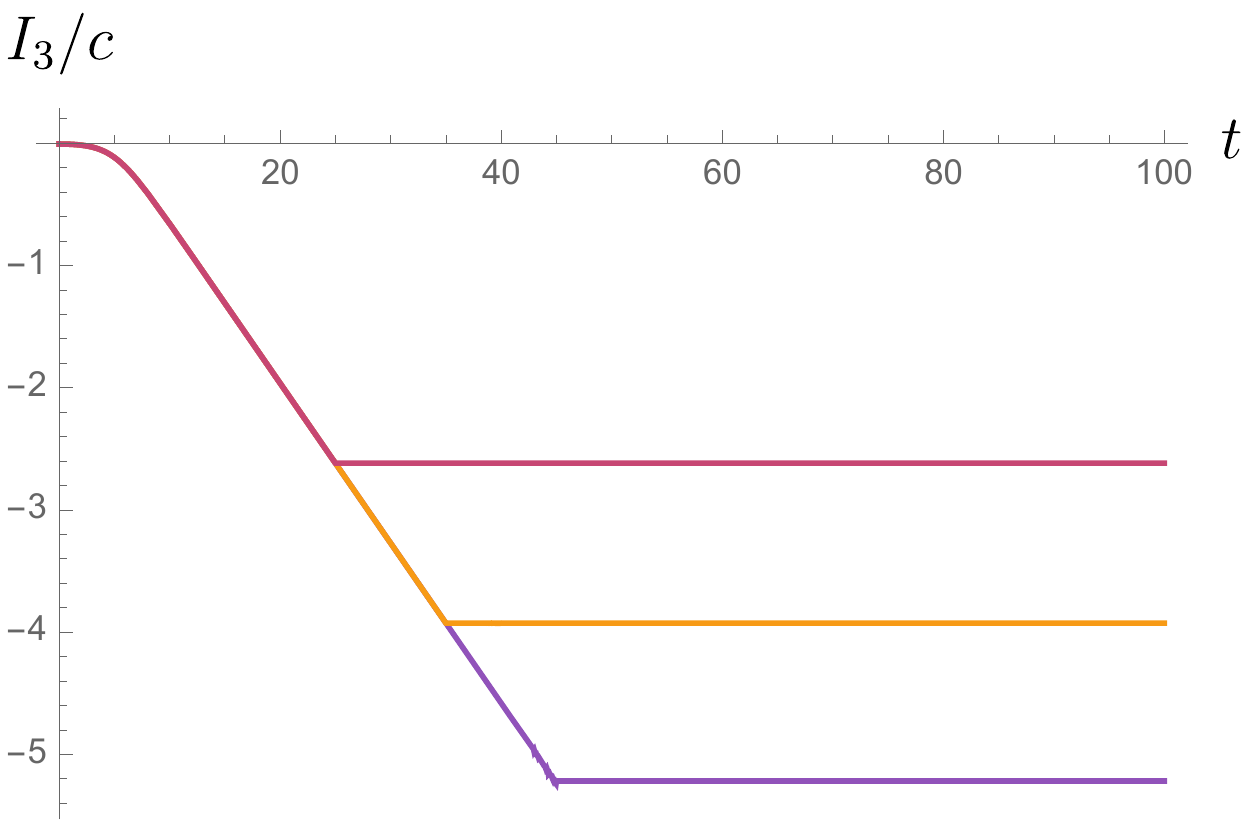}
    \caption{Numerical simulation of Clifford circuits. (top left) BOMI for asymmetric overlap with the left endpoints of the input and output intervals flush. The input interval is of length $10$ and the output varies in length as 10 (blue), 15 (magenta), 20 (yellow), 25 (purple). The dotted gray lines are the times that the line-tension picture predicts the kink to occur at. (top right) TOMI for A of length 10 (magenta), 15 (yellow), 20 (purple). We average over 100 disorder realizations. The inset shows a single disorder realization. The mutual information is in units of $\log 2$. These plots should be compared to the holographic results for the same configurations that are shown in the bottom row. For the BOMI and TOMI, we use $\beta = 2$ and $\beta =4$ respectively. The structure of the Clifford circuit is brick-layered with CNOT gates and Hadamard and phase gates applied randomly on each site with probability $0.5$ on every layer. The Hadamard and phase gates are essential for integrability breaking as we observe recurrence times of order $L$ when only using CNOT gates (see Fig \ref{fig:recurrence}).
}
    \label{fig:bomitomi}
\end{figure}

We compute the operator mutual information in random Clifford circuits and compare with the holographic CFT results from Ref.~\cite{2018arXiv181200013N}. In Fig.~\ref{fig:bomitomi}, we plot the bipartite operator mutual information (BOMI) and tripartite operator mutual information (TOMI) in both Clifford circuits and holographic CFTs. We find remarkably similar behavior between the seemingly disparate unitary channels. In particular, the BOMI for disjoint intervals is trivial. The saturation value of the TOMI is $-2 |A|$, saturating the bound from Ref.~\cite{2016JHEP...02..004H}. The Clifford circuits and the holographic channel appear to scramble information at the same speed. This is consistent with the idea that black holes are the fastest scramblers in nature \cite{2008JHEP...10..065S}.

Importantly, Clifford circuits are not perfect toy models of chaos/scrambling quantum systems. In particular, they have pathological OTOC due to the fact that they are unitary 3-designs \cite{2015arXiv151002769W,2015arXiv151002619Z}. While they indeed scramble information, their evolution is not chaotic under general definitions. One may be further concerned that entanglement is not generated because Pauli strings are mapped to Pauli strings without any superposition. However, for operator entanglement, the initial state is a tensor product of Bell pairs that is a superposition of Pauli strings. For our purposes, the Clifford circuit is then able to mimic the chaotic dynamics of entanglement spreading that is seen in holographic conformal field theories. 

% \subsection{line-tension picture}

\begin{figure}
    \centering
    \includegraphics[height = 5cm]{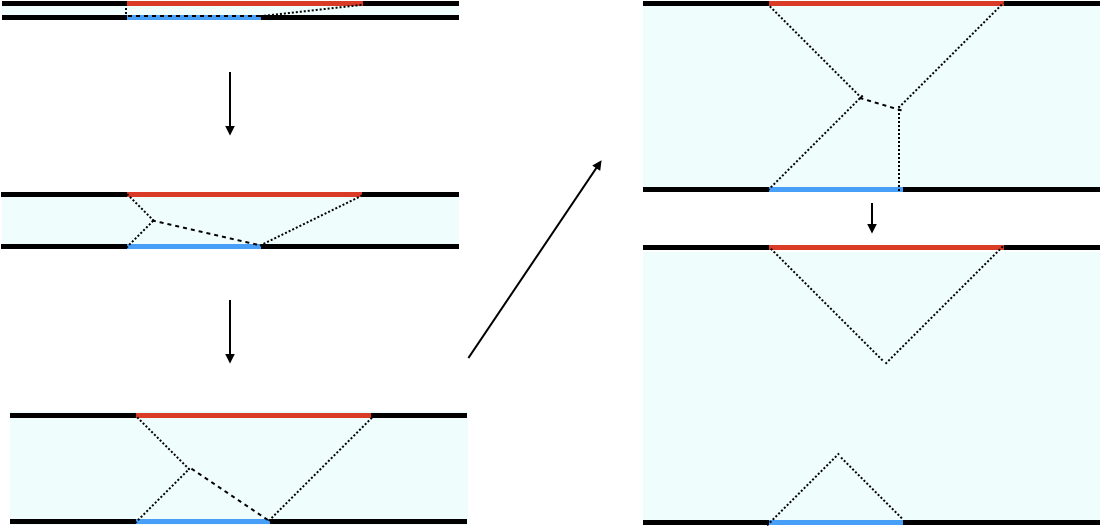}
    \caption{The cartoon shows the time evolution of the minimal membrane for operator mutual information (dotted lines) and operator logarithmic negativity (dashed lines) between input (blue) and output (red) finite intervals. In the connected regime, the membrane connecting the left ends of the intervals grows linearly, displaying the leaking of quantum information. The length of the minimal cross-section decreases linearly, showing the same leaking. The length of the right membrane remains constant for the first three time steps as the slope remains above 1; after, the length grows linearly. Similarly, the length of the minimal cross-section decreases at twice the speed after the slope decreases below 1. This demonstrates causality in that the quantum information cannot leak out the right side until the endpoints are time-like separated. At late times, the membrane reaches the disconnected regime where the BOMI and BOLN are manifestly trivial (bottom right). 
   }
    \label{membrane}
\end{figure}

In the scaling limit and limit of large bond dimension, $q$, we can find the line-tension of the brick-layered random unitary circuit by simply counting the number of bonds cut
\begin{align}
\label{linetension_eq}
    \mathcal{T}(v, x,t) =
    \begin{cases}
\log q & v < 1 , \\
v \log q & v > 1.
\end{cases}
\end{align}
 This line-tension picture is able to predict the behavior of the simulations including the previously mysterious kink.
%  We show this by first noting that the minimal membrane has a constant velocity for spacetime translationally invariant systems and that our random circuit is effectively translationally invariant in the scaling limit.
 We partition the Hilbert space into the union of two finite intervals, one in the input Hilbert space and the other in the output Hilbert, with its complement. There are two configurations for the minimal membrane, the connected regime where the membrane stretches from the input to output system and the disconnected regime, where each interval has a membrane homologous to itself. Generically, there will be a phase transition at a critical time when the disconnected regime becomes dominant. At this point, the operator mutual information is manifestly zero. We demonstrate how the membrane predicts the kink in the BOMI in Fig.~\ref{membrane}. The kink occurs because information can only leak out of the left side of the interval initially, but after a time equal to the distance between the two right endpoints, the information can leak out of both sides. This causes the operator mutual information to decrease at twice the speed i.e.~the slope doubles. Precisely the same phenomenon was found for large c conformal field theories in Ref.~\cite{2018arXiv181200013N} where the same slope and causality arguments apply.

The holographic results have the same origin as in the random unitary circuit model, only it is more complicated to analyze due to the increased complexity of hyperbolic space over the simple ``metric'' defined by the line-tension for random unitary circuits (\ref{linetension_eq}). Heuristically, this may be seen by viewing the time slices of the growth of the wormhole as the MERA tensor network presented in Ref.~\cite{2013JHEP...05..014H}. There, the MERA wormhole geometry is equivalent to the geometry of a brick-layered random unitary circuit at scales larger than $\beta$ which is a UV cutoff in the BOMI. Therefore, the agreement should come as no surprise. We again stress that we have traded the time direction for the radial direction\footnote{One may also seek a connection to computational complexity. In the random unitary circuit, the complexity may be well defined; it grows linearly with time and is proportional to the volume of the circuit, consistent with the ``Complexity = Volume'' conjecture where this picture was originally discussed \cite{2014PhRvD..90l6007S}.}. 

We can take this analogy between random unitary circuits and holographic CFTs further by comparing the rates at which the information leaks. For holographic CFTs, the BOMI decreases by $\frac{c \pi}{3 \beta}$ each time step, while in random unitary circuits, the BOMI decreases by $\log q$. We can equate these and heuristically identify the local Hilbert space dimension in holographic CFTs
\begin{align}
    q \sim e^{\frac{c\pi}{3\beta}},
\end{align}
which precisely corresponds to the entropy density in the Cardy regime \cite{CARDY1986186}
\begin{align}
    \frac{S_{Cardy}}{L} = \frac{c\pi }{3\beta}.
\end{align}

\subsection{Line-tension for logarithmic negativity}
\label{linetension_LN}
Drawing motivation from the discussion above and Ref.~\cite{2019PhRvD..99j6014K} as reviewed in section \ref{negativity_section} and Appendix \ref{app_neg_blocks}, we consider a membrane picture for the logarithmic negativity in random circuits involving an effective membrane cross-section. Suggestively, we denote the subregion of the circuit that is bounded by the minimal membranes as the \textit{entanglement wedge}. An important aspect of this proposal is that the entanglement wedge is generally degenerate due to the many configurations of the membranes that have the same total energy given by the line-tension. We must choose a preferred configuration. This will be the membrane that admits the cross section of minimal length. We conjecture that the logarithmic negativity is the energy of the minimal membrane that acts as a cross-section of the entanglement wedge\footnote{In holography, there are complications to this procedure due to backreaction from the membrane. Because we are working with random unitary circuits at large $q$, it is natural that there is no backreaction because the entanglement spectrum is flat. This is analogous to the quantum-error correcting code derivation from \cite{2019PhRvD..99j6014K}.}
\begin{align}
    \mathcal{E}(t) = \int_{E_W} d^{d}x \mathcal{T}(\bar{v}, \bar{x}).
    \label{ln_line_tension}
\end{align}
In Fig.~\ref{membrane}, we demonstrate how the minimal cross-section also accurately predicts the causality of quantum information. The area of the minimal cross-section decreases linearly at early times and decreases twice as fast when the time is larger than the distance between the right sides of the intervals, predicting a kink in the BOLN of chaotic systems, analogous to what was observed in the BOMI. Unfortunately, the logarithmic negativity of stabilizer states, such as states generated by Clifford circuits, is currently intractable, so we are unable to numerically test this conjecture for large system sizes. However, we will find in the following section that the behavior of the holographic logarithmic negativity is consistent with the predictions from this line-tension picture. This deepens the connection between the logarithmic negativity and entanglement wedge cross-sections. It would be interesting to have a more rigorous understanding of this relationship in generic quantum networks. 

\section{Operator negativity}
\label{op_neg_section}

We are technically limited to studying operator negativity in in 2d conformal field theory. Higher dimensional calculations may be tractable by use of the holographic prescription, but we leave this to future work.
\subsection{Universal behavior in CFT}

\label{cft_op_section}
To compute the operator logarithmic negativity (OLN) of $A_1 = (X_2, X_1)$ and $A_2 = (Y_2, Y_1)$, we need to calculate the negativity of disjoint intervals in a thermofield double state. The moments of the partial transpose of the reduced density matrix are calculated using a thermal four-point function of twist fields\footnote{Note that we have have switched twist and anti-twist operators compared to the calculation for OMI in Ref.~\cite{2018arXiv181200013N}. Also note that there are no topological subtleties for this thermal correlator as there are for the bipartite negativity discussed in Ref.~\cite{2015JPhA...48a5006C}. This is because we are considering disjoint intervals.}
\begin{align}
     \Tr \left( \rho_{A_1, A_2}^{T_{A_2}} \right)^{n_e}  =\langle \sigma_{n_e}(X_2) \bar{\sigma}_{n_e}(X_1) \sigma_{n_e}(Y_2, \tau) \bar{\sigma}_{n_e}(Y_1, \tau) \rangle_{\beta}, 
\end{align}
where $\beta$ is the inverse temperature.
The logarithmic negativity is the logarithm of this function analytically continued to $n_e \rightarrow 1$. We first perform a conformal transformation from the cylinder to the complex plane $ w( z) = e^{2 \pi z /\beta}$
% \begin{align}
%   \quad \bar{w}( \bar{z}) = e^{2 \pi \bar{z} /\beta}.
% \end{align}
\begin{align}
    \Tr \left( \rho_{A_1, A_2}^{T_{A_2}} \right)^{n_e} = \prod_i^4 \left[\frac{dw}{dz} \right]^{h_{\sigma_{n_e}}}_{w = w_i}\left[\frac{d\bar{w}}{d\bar{z}} \right]^{\bar{h}_{\sigma_{n_e}}}_{\bar{w} = \bar{w}_i}\langle \sigma_{n_e}(e^{\frac{2\pi}{\beta}X_2}) \bar{\sigma}_{n_e}(e^{\frac{2\pi}{\beta}X_1}) \sigma_{n_e}(e^{\frac{2\pi}{\beta}(Y_2 + i \tau)}) \bar{\sigma}_{n_e}(e^{\frac{2\pi}{\beta}(Y_1 + i \tau)}) \rangle_{\mathbb{C}},
\end{align}
where the conformal dimensions of the twist fields are
\begin{align}
    h_{\sigma_{n_e}} = \bar{h}_{\sigma_{n_e}} = \frac{c}{24}\left(n_e - \frac{1}{n_e} \right).
\end{align}
We note that all conformal factors will drop out of the calculation for logarithmic negativity when we take the replica limit.

We now do the standard conformal transformation, sending the four operator insertions to $\infty$, $0$, $1$, and the cross-ratio $x$
\begin{align}
    \Tr \left( \rho_{A_1, A_2}^{T_{A_2}} \right)^{n_e} = c_n^2 \left[ \frac{\left(e^{\frac{2\pi}{\beta}(Y_2 + i \tau)}-e^{\frac{2\pi}{\beta}X_2} \right)\left(e^{\frac{2\pi}{\beta}(Y_1 + i \tau)} -e^{\frac{2\pi}{\beta}(X_1 }\right)}{\left(e^{\frac{2\pi}{\beta}X_1}-e^{\frac{2\pi}{\beta}X_2} \right)\left(e^{\frac{2\pi}{\beta}(Y_1 + i \tau)}-e^{\frac{2\pi}{\beta}(Y_2 + i \tau)} \right)} \right]^{c/6 (n-1/n)} \\ \nonumber \times \left[ \left(e^{\frac{2\pi}{\beta}(Y_1 + i \tau)}-e^{\frac{2\pi}{\beta}X_2} \right)\left(e^{\frac{2\pi}{\beta}(Y_2 + i \tau)}-e^{\frac{2\pi}{\beta}X_1} \right) \right]^{c/6 (n-1/n)} \mathcal{F}_n(x)\bar{\mathcal{F}}_n(\bar{x}).
\end{align}
The cross ratios (after proper analytic continuation to Lorentzian time) are not in the usual range of $(0,1)$ but rather in $(-\infty,0)$
\begin{align}
    x = \frac{\sinh\left(\frac{\pi}{\beta} (X_2 - X_1)\right)\sinh\left(\frac{\pi}{\beta} (Y_2 - Y_1)\right)}{\cosh\left(\frac{\pi}{\beta} (X_2 - Y_2-t)\right)\cosh\left(\frac{\pi}{\beta} (X_1 - Y_1-t)\right)}, \\ 
    \bar{x} = \frac{\sinh\left(\frac{\pi}{\beta} (X_2 - X_1)\right)\sinh\left(\frac{\pi}{\beta} (Y_2 - Y_1)\right)}{\cosh\left(\frac{\pi}{\beta} (X_2 - Y_2+t)\right)\cosh\left(\frac{\pi}{\beta} (X_1 - Y_1+t)\right)} .
\end{align}
We define new cross ratios \cite{2013JSMTE..02..008C}
\begin{align}
    y &= \frac{x}{x-1}, \quad 
    \bar{y} =  \frac{\bar{x}}{\bar{x}-1},
    \\ \mathcal{G}_{n_e}(y) &= (1-y)^{c/3(n-1/n)}\mathcal{F}_{n_e}(\frac{y}{y-1}), \quad \bar{\mathcal{G}}_{n_e}(\bar{y}) = (1-\bar{y})^{c/3(n-1/n)}\bar{\mathcal{F}}_{n_e}(\frac{\bar{y}}{\bar{y}-1})
\end{align}
% We now do the standard conformal transformation sending the four points to $\infty$, $0$, $1$, and the cross-ratio $x$
% \begin{align}
%     v(w) = \frac{(w - w(Y_1+i \tau ))(w(X_2 ) - wX_1)}{(w - wX_1)(w(Y_1+i \tau ) - wX_2)}, \quad \bar{v}(\bar{w}) = \frac{(\bar{w} - \bar{w}(Y_1-i \tau ))(\bar{w}(X_2 ) - \bar{w}X_1)}{(\bar{w} - \bar{w}X_1)(\bar{w}(Y_1-i \tau ) - \bar{w}X_2)}
% \end{align}
% giving the cross-ratio
% \begin{align}
%     x = \frac{(e^{2 \pi X_2 /\beta}-e^{2 \pi X_1 /\beta})(e^{2 \pi (Y_2+i \tau) /\beta}-e^{2 \pi (Y_1+i \tau) /\beta})}{(e^{2 \pi (Y_2+i \tau) /\beta} - e^{2 \pi X_1 /\beta})(e^{2 \pi (Y_1+i \tau) /\beta} - e^{2 \pi X_2 /\beta})}, \quad \bar{x} = x^*.
%     \label{cross_ratio}
% \end{align}
so that $y \in  (0,1)$. The operator logarithmic negativity becomes
\begin{align}
    \mathcal{E}(y, \bar{y}) = \log \lim_{n_e \rightarrow 1} \left[ \mathcal{G}_{n_e}(y) \bar{\mathcal{G}}_{n_e}(\bar{y}) \right].
    \label{neg_G_form}
\end{align}
% The correlation function can be expanded in terms of conformal blocks
% \begin{align}
%     \langle \sigma_{n_e}(\infty)\bar{\sigma}_{n_e}(1)\bar{\sigma}_{n_e}(x, \bar{x}){\sigma}_{n_e}(0) \rangle_{\mathbb{C}} = \sum_p a_p \mathcal{F}(c, h_p, h_i,x)\bar{\mathcal{F}}(c, \bar{h}_p, \bar{h}_i,\bar{x})
% \end{align}

\begin{figure}
    \centering
    \includegraphics[width = \textwidth]{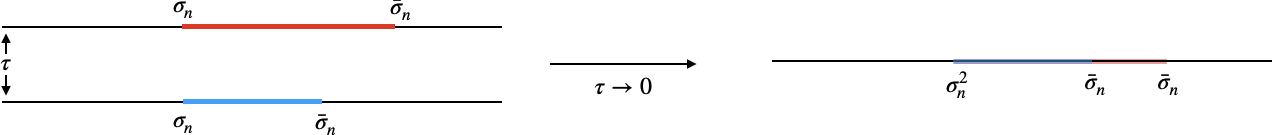}
    \caption{We display the positions of the twist-fields used for the universal limit with the blue region corresponding to $A_1$ and the red $A_2$. At very early times, we can take the $\sigma_n \times \sigma_n = \sigma^2_n$ OPE to obtain a 3-point function similar to the one for negativity of adjacent intervals \cite{2012PhRvL.109m0502C}.}
    \label{universal_OPE_cartoon}
\end{figure}

\begin{figure}
    \centering
    \includegraphics[height = 3.25cm]{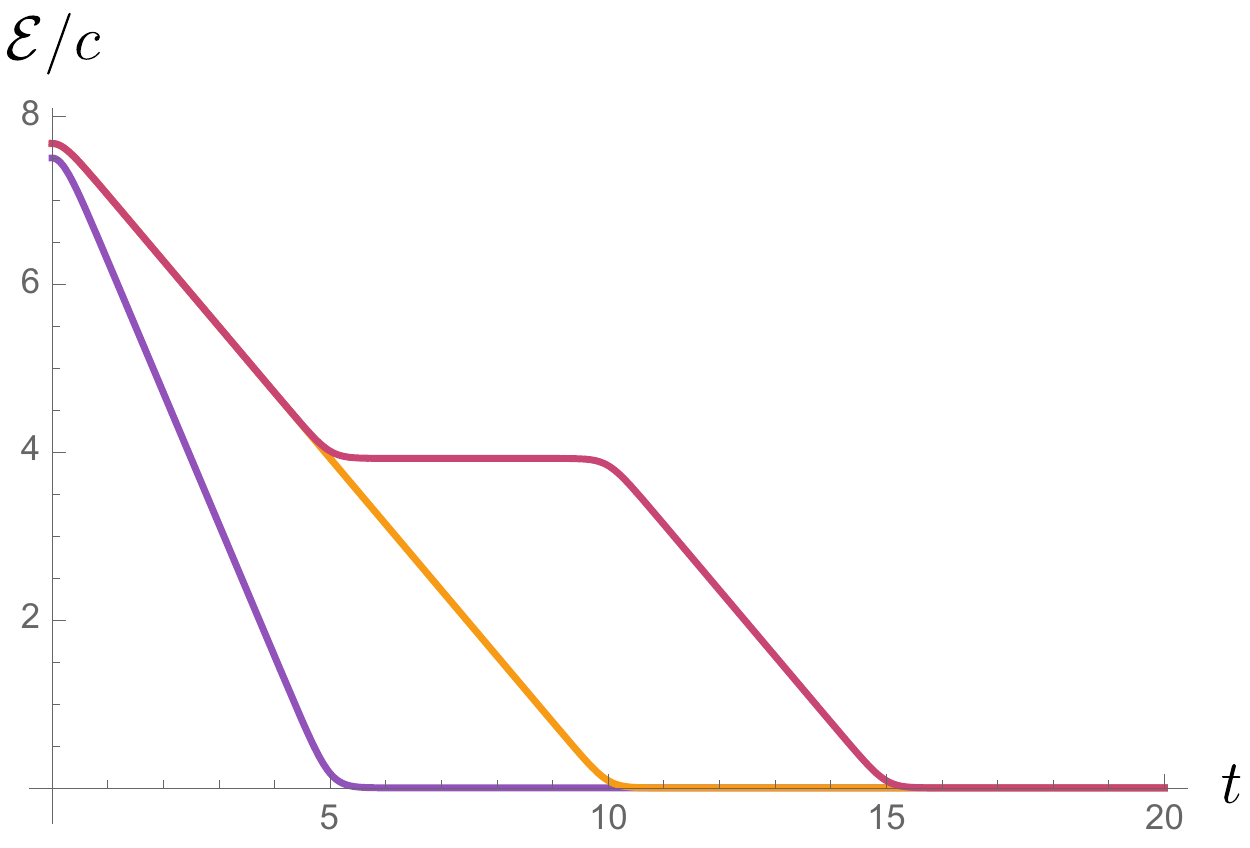} 
    % \quad
    \includegraphics[height = 3.25cm]{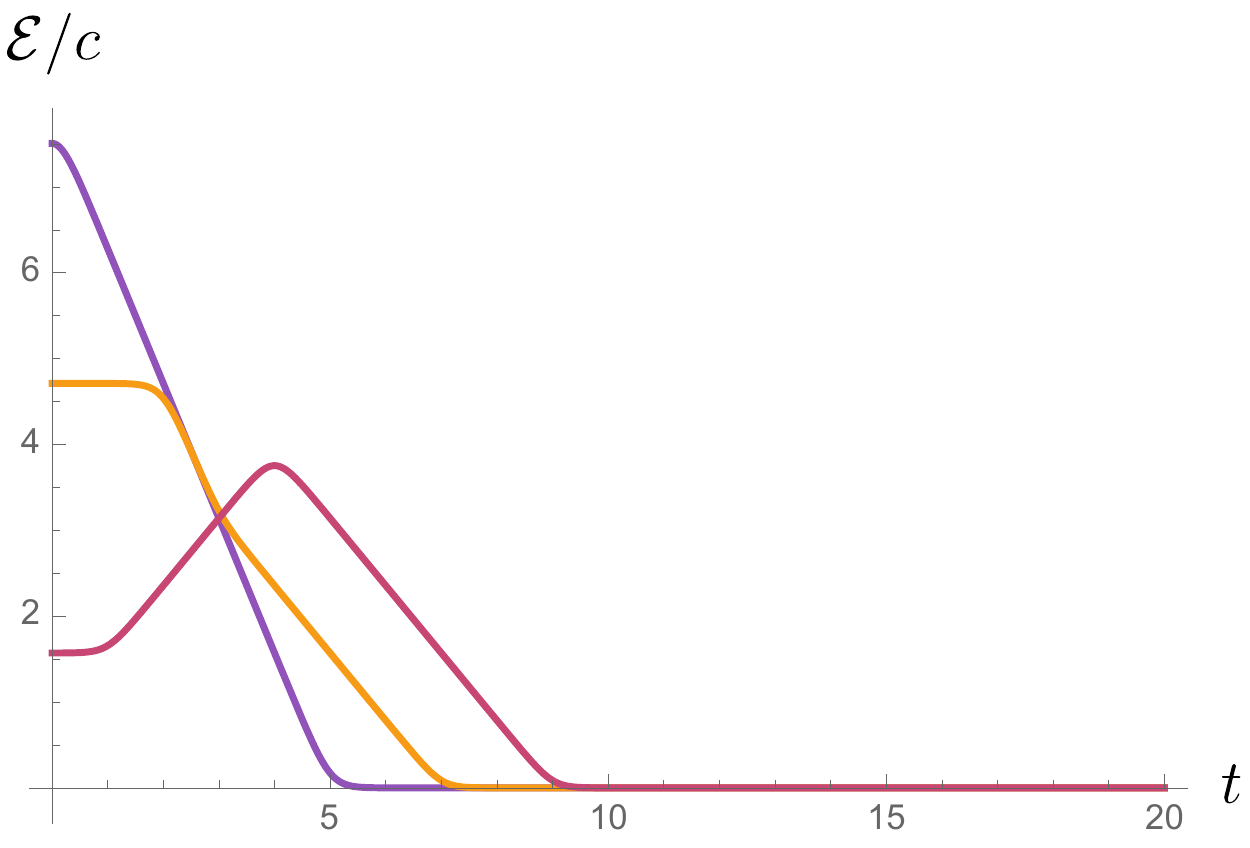} 
    % \quad
    \includegraphics[height = 3.25cm]{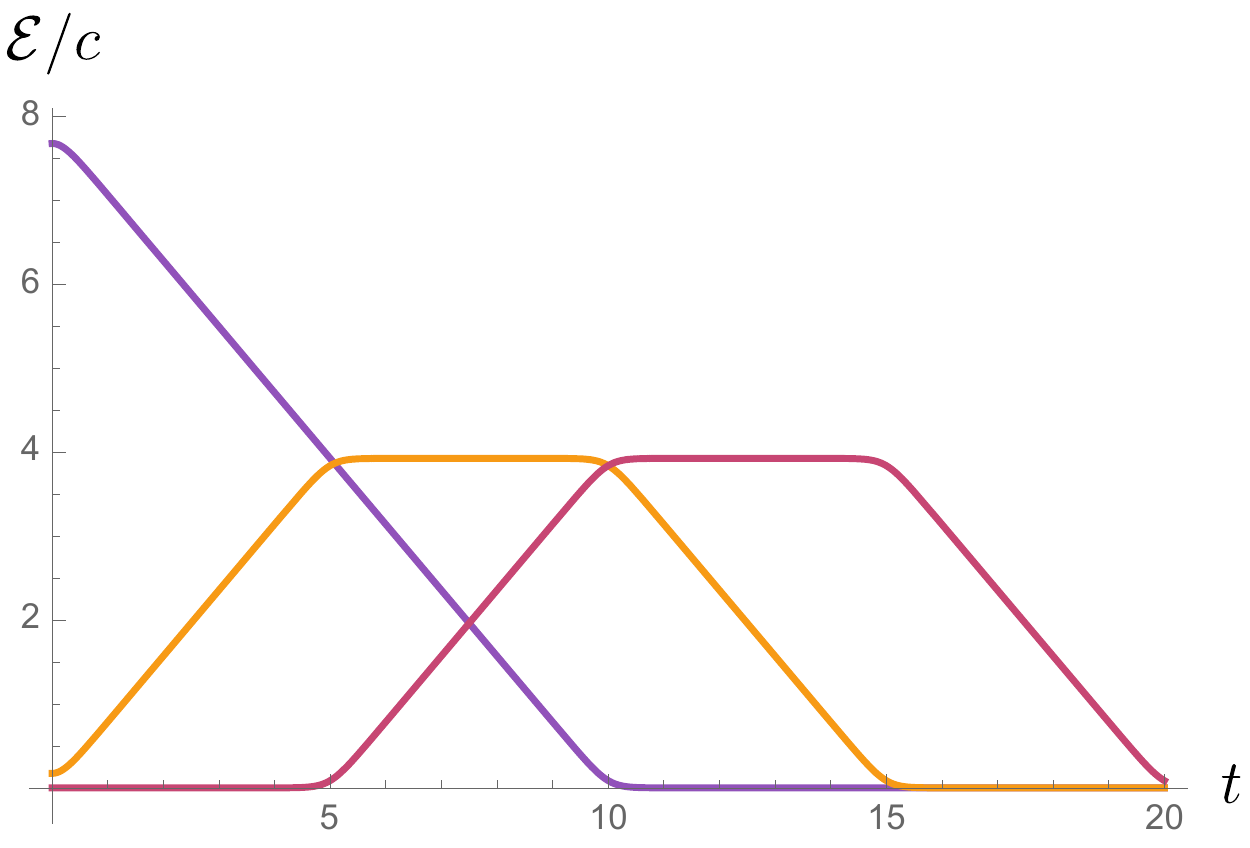} 
    \caption{The universal contribution to the negativity (\ref{univ_OLN}). (left) OLN with $l_1 = 5$, $l_2 = $ 5 (purple), 10 (yellow), 15 (magenta), and both intervals starting at the same position. (center) OLN with $l_1 = l_2 = 5$, and the distance between starting positions varying as 0 (purple), 2 (yellow), 4 (magenta). (right). OLN with $l_1 = 5$, $l_2 = 10$, fully overlapped (purple), adjacent (yellow), and separated by $5$ (magenta). We take $\beta = 2$. It is clear that the qualitative behavior of the negativity follows the quasi-particle picture. 
    }
    \label{CFT_neg}
\end{figure}

To find the universal contribution to the operator negativity, we consider $X_2 = Y_2$. Then, at early times ($\tau \ll 1$), the correlator becomes (see Fig.~\ref{universal_OPE_cartoon})
\begin{align}
    \Tr \left(\rho_A^{T_2} \right)^{n_e} &= \langle \sigma^2_{n_e}(e^{\frac{2\pi}{\beta}X_2}) \bar{\sigma}_{n_e}(e^{\frac{2\pi}{\beta}X_1}) \bar{\sigma}_{n_e}(e^{\frac{2\pi}{\beta}Y_1}) \rangle,
\end{align}
where $\sigma_{n_e}^2$ is the double twist operator with conformal dimensions
\begin{align}
    h_{\sigma^2_{n_e}}=\bar{h}_{\sigma^2_{n_e}} = \frac{c}{12}\left(\frac{n_e}{2} - \frac{2}{n_e} \right).
\end{align}
Therefore, the negativity diverges universally at early times as
\begin{align}
   \mathcal{E}(t\rightarrow 0) = \frac{c}{4} \log \left(\frac{(e^{\frac{2\pi}{\beta}X_1}-e^{\frac{2\pi}{\beta}X_2})(e^{\frac{2\pi}{\beta}Y_1}-e^{\frac{2\pi}{\beta}Y_2})}{\epsilon \left| e^{\frac{2\pi}{\beta}Y_1} - e^{\frac{2\pi}{\beta}X_1} \right|} \right)
    \label{t0limit}
\end{align}
where we have introduced a regulator, $\beta$. Using this early-time limit, we can find the behavior of $\mathcal{G}_n(y)$ as $y \rightarrow 1$ because (\ref{neg_G_form}) only reduces to (\ref{t0limit}) if
\begin{align}
    \bar{\mathcal{G}}_n(y)  \simeq \mathcal{G}_n(y) \simeq g_n (1- y)^{\Delta_{\sigma_n^2}/2}.
\end{align}
where $g_n$ is some constant only depending on the replica number. This leads us to the early-time evaluation of the negativity
\begin{align}
    \mathcal{E}(y \rightarrow 1) = -\frac{c}{8} \log \left[ (1- y) (1-\bar{y})\right].
    \label{univ_OLN}
\end{align}
At late times ($y\rightarrow 0$), the negativity vanishes faster than any power \cite{2013JSMTE..02..008C}
\begin{align}
    \mathcal{E}(y \rightarrow 0) = 0.
\end{align}
Thus, we have solved for the universal early and late-time limits. Extrapolating these results to $0 < y < 1$, we find this behavior to follow the quasi-particle picture for integrable CFTs quite similarly to what was found for OMI in Ref.~\cite{2018arXiv181200013N}. We plot this in Fig.~\ref{CFT_neg} and note that our results display the quasi-particle picture even after relaxing the condition that $X_2 = Y_2$. This is expected because the negativity is a conformally invariant quantity. 

The tripartite negativity is identically zero because the universal contribution follows the quasi-particle picture in which all information remains coherent and localized
\begin{align}
    \mathcal{E}_3(t) = 0.
\end{align}

\subsection{Free fermion CFT}

\begin{figure}
    \centering
    \includegraphics[height = 3.25cm]{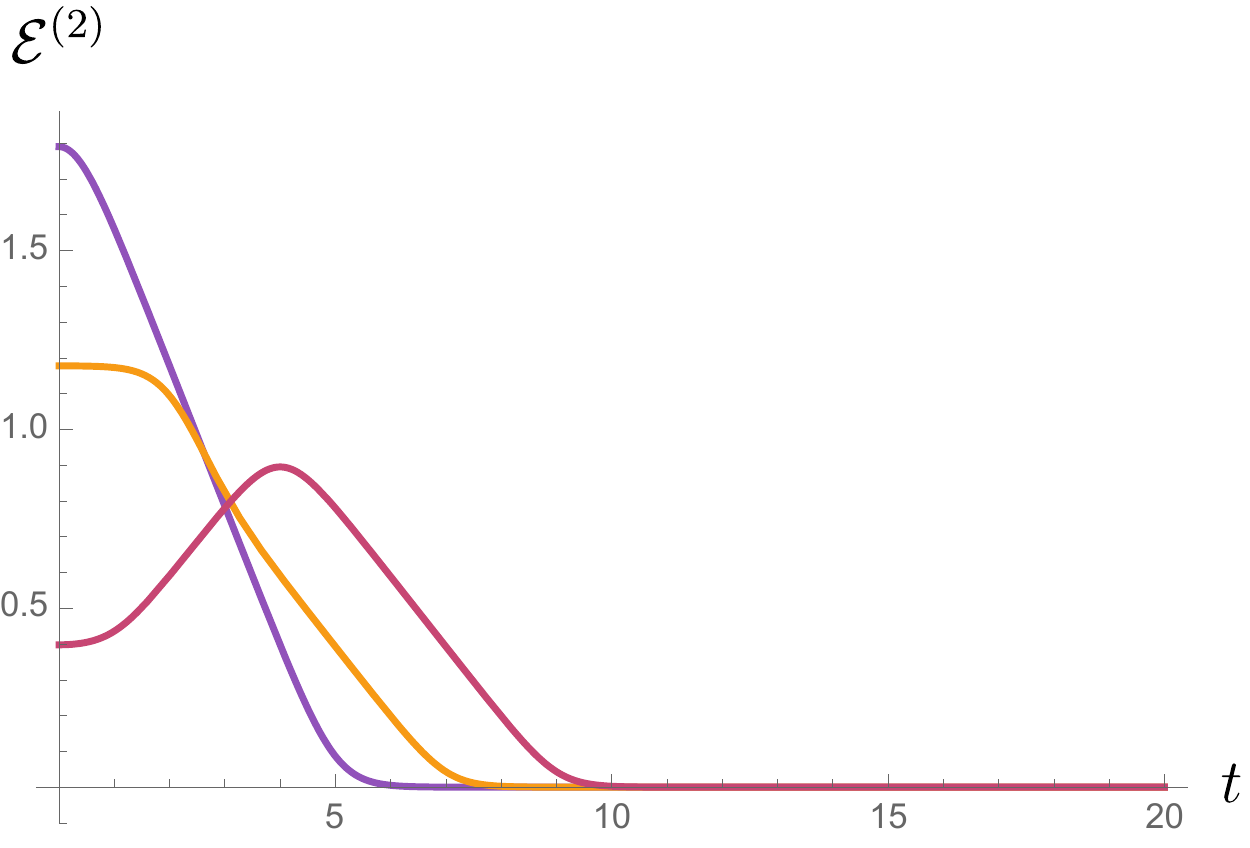}
    % \quad 
    \includegraphics[height = 3.25cm ]{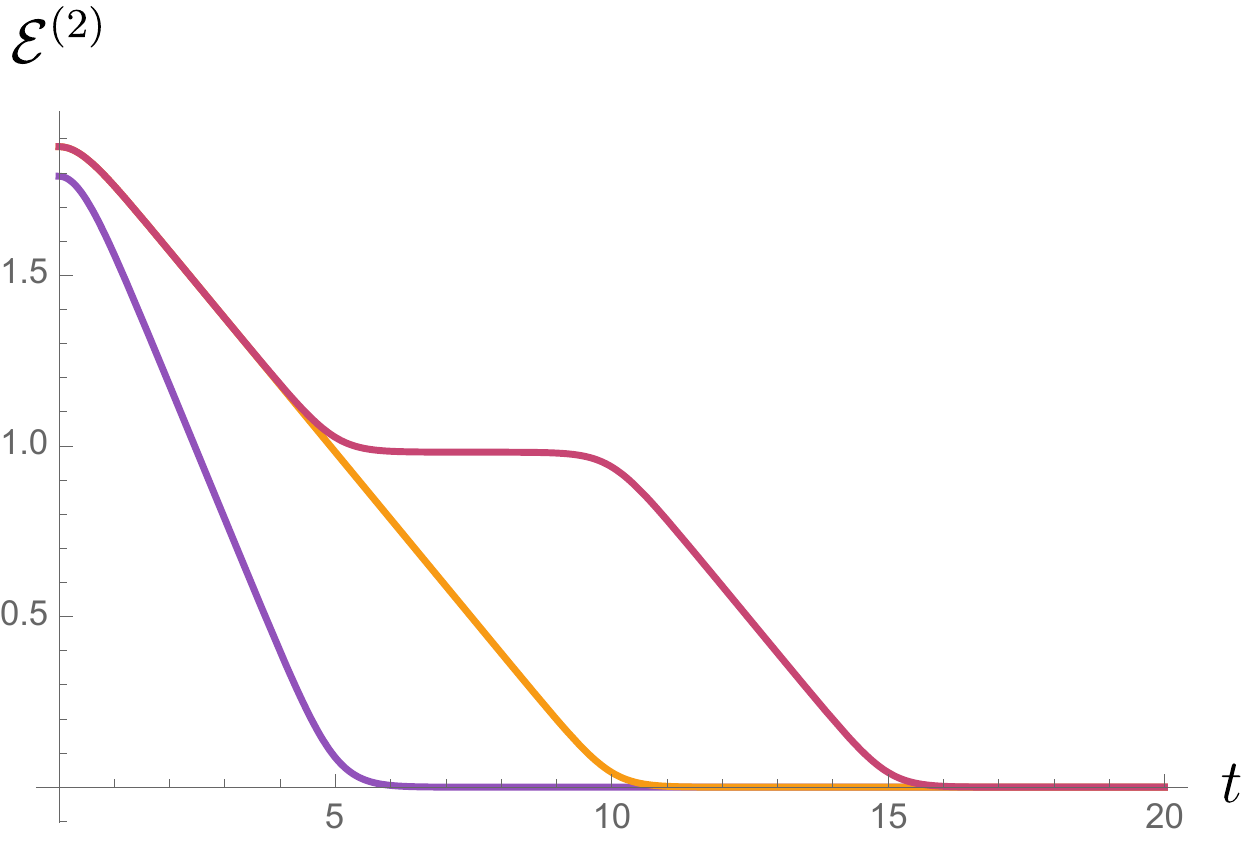} 
    % \quad 
    \includegraphics[height = 3.25cm ]{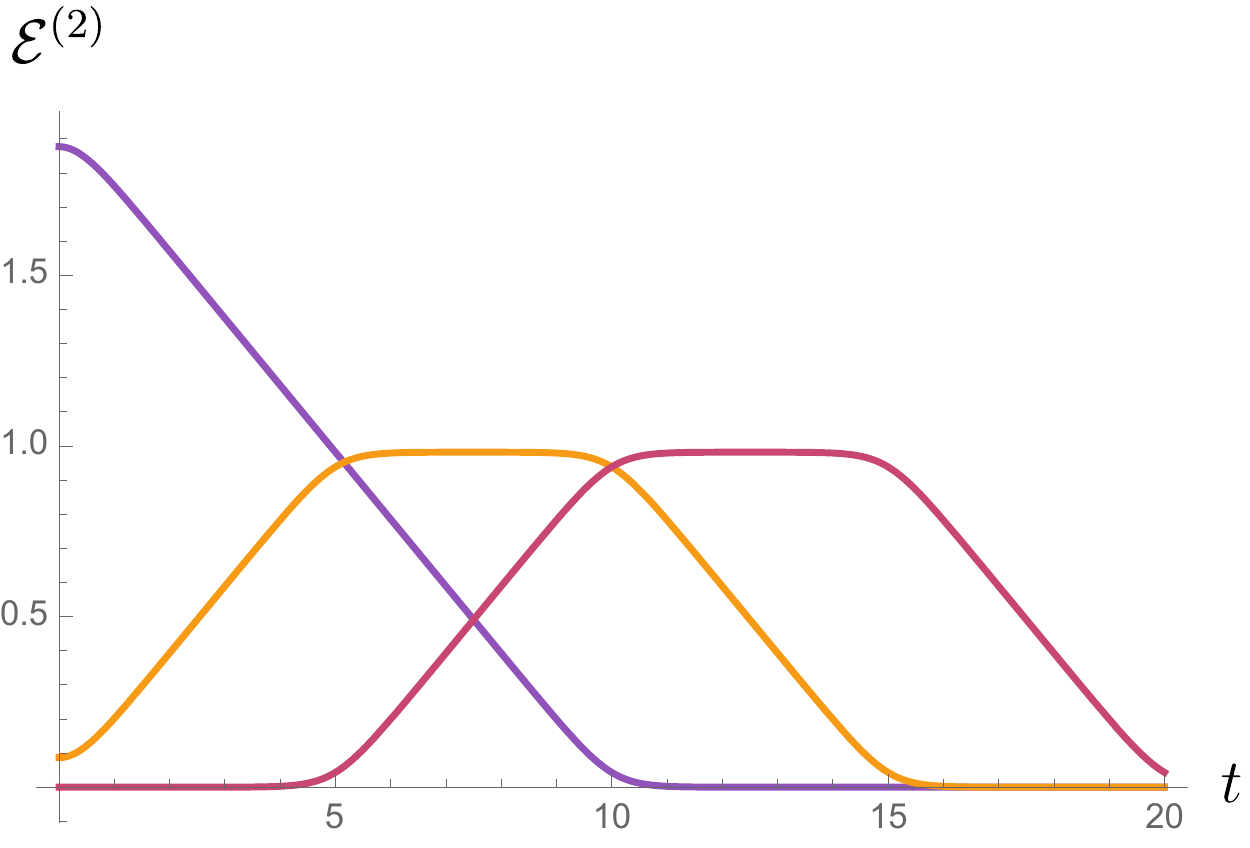} \\
    \includegraphics[height = 3.25cm]{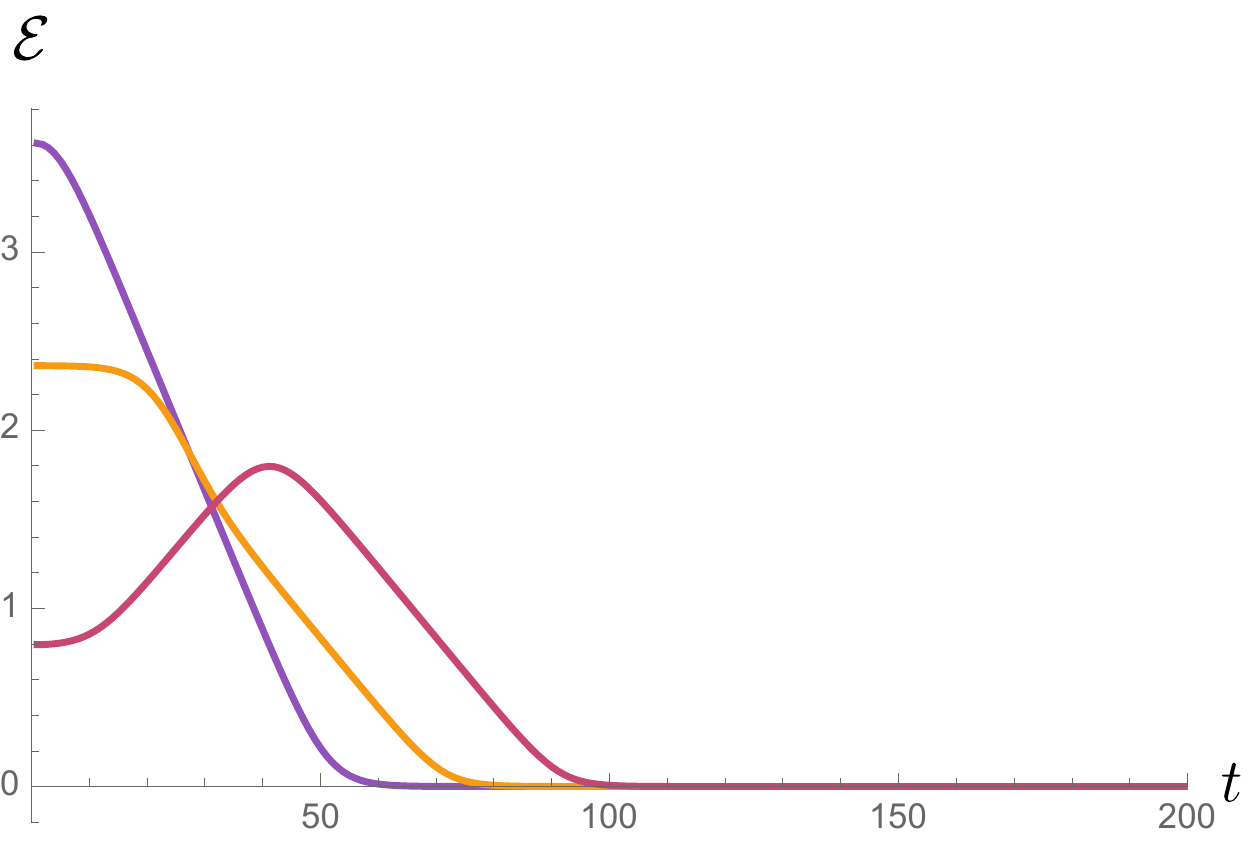} 
    % \quad 
    \includegraphics[height = 3.25cm ]{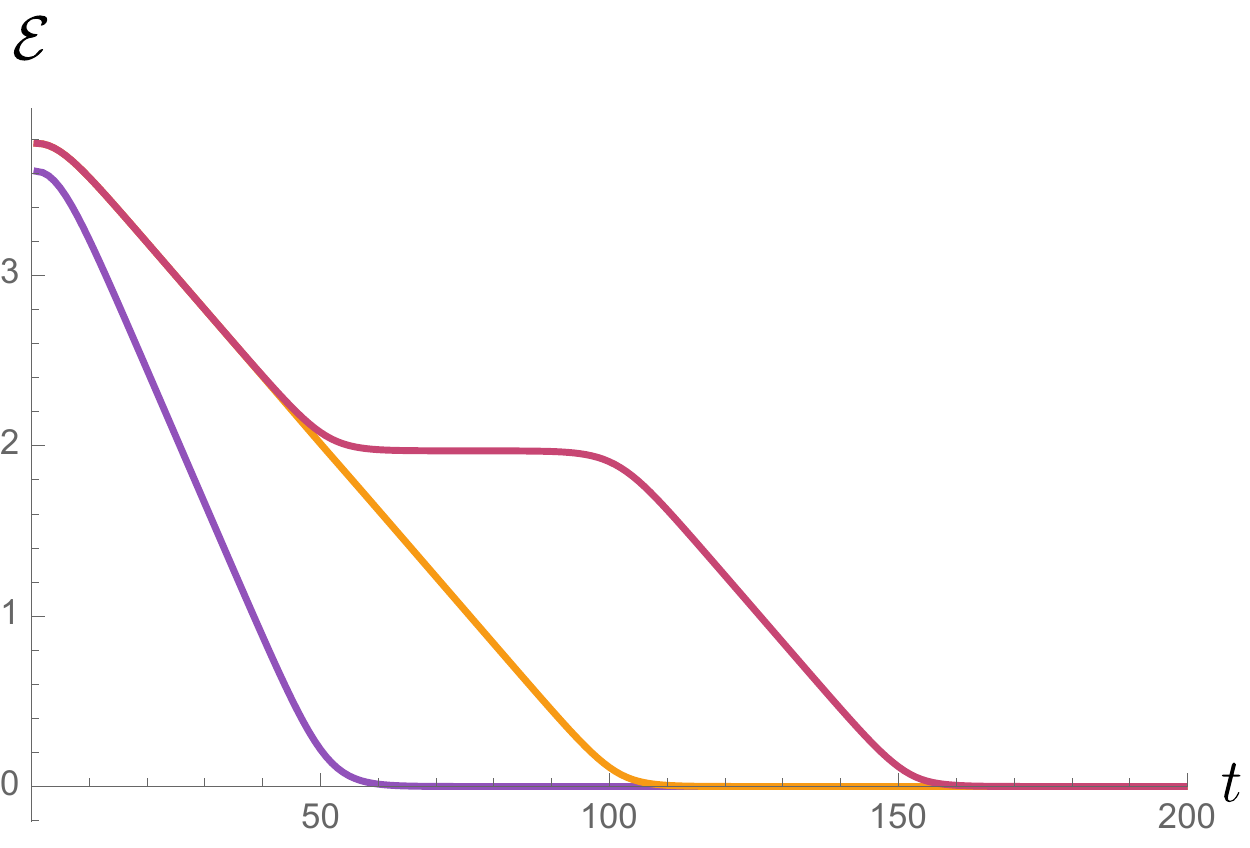}
    % \quad
    \includegraphics[height = 3.25cm ]{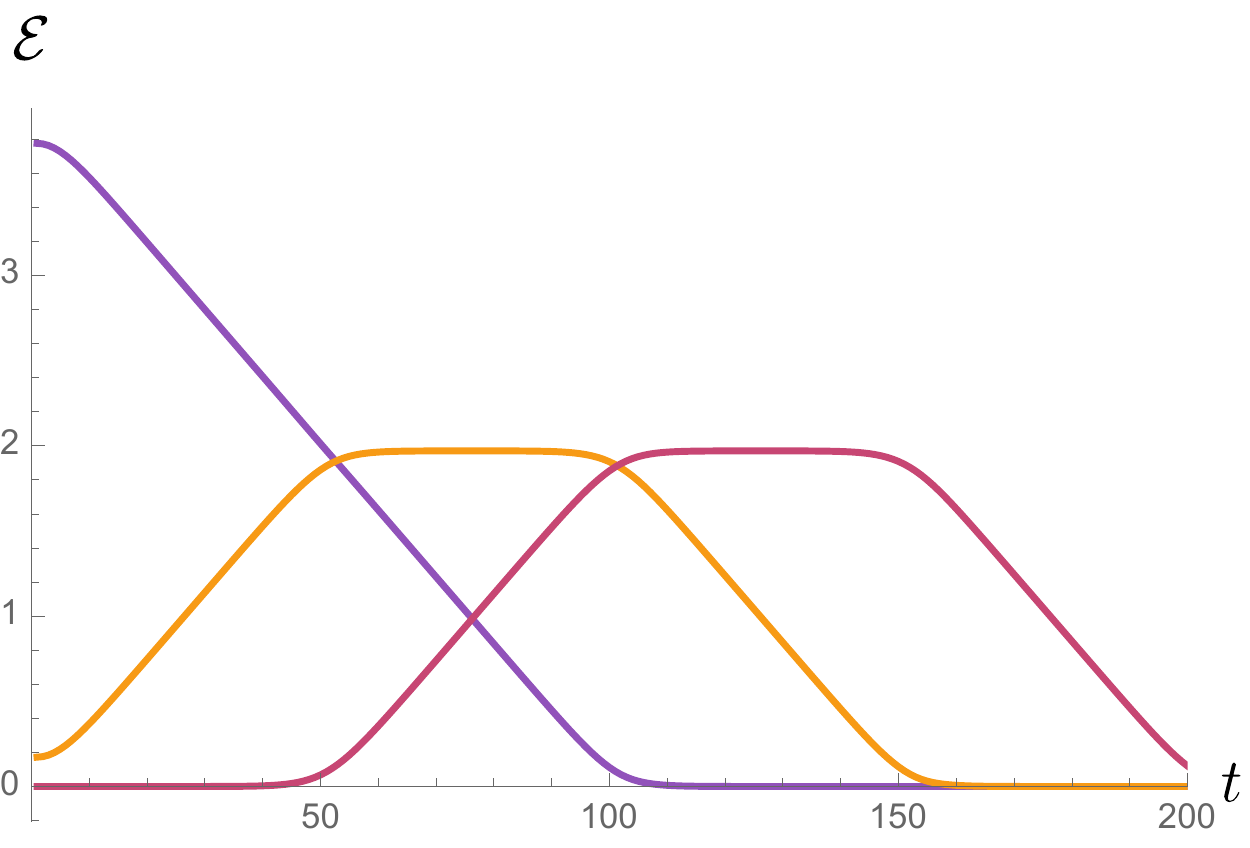}\\
    \includegraphics[height = 3.25cm]{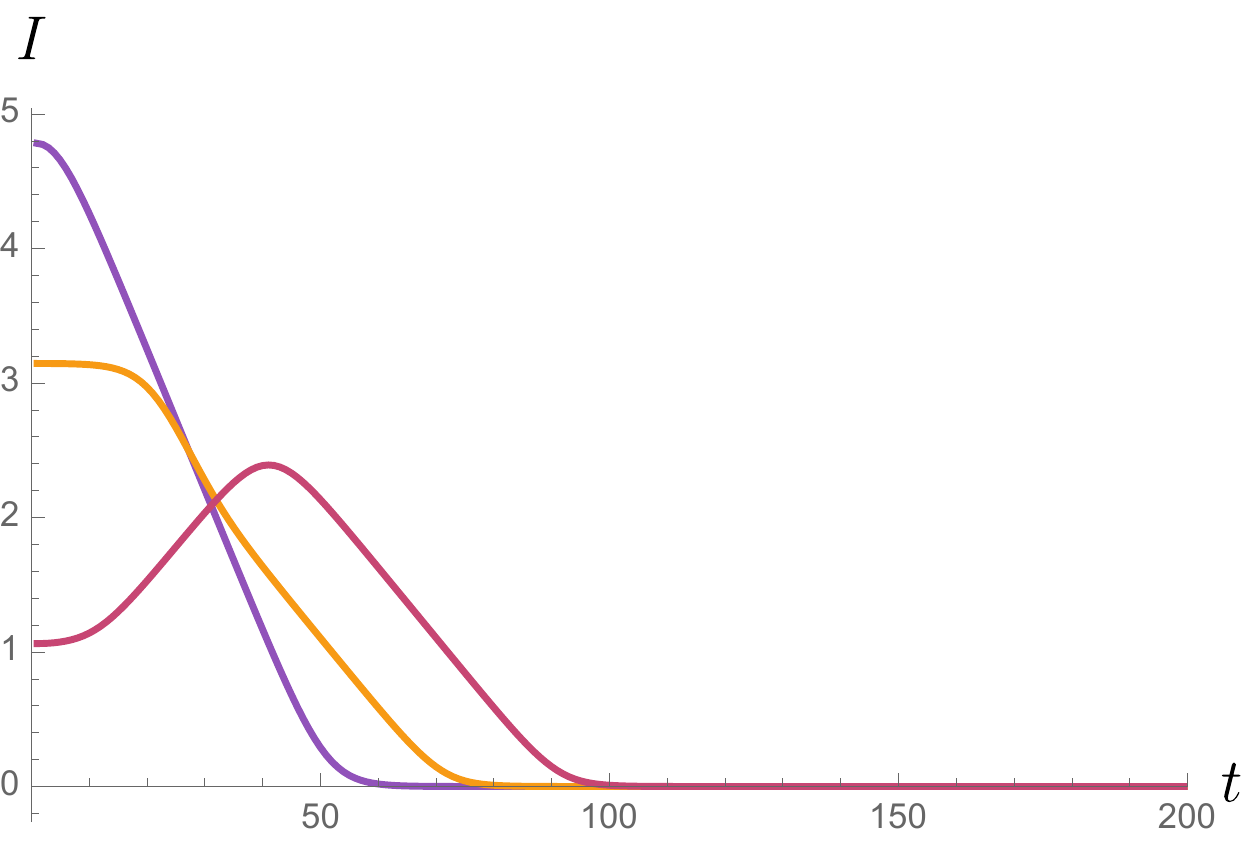} 
    % \quad 
    \includegraphics[height = 3.25cm ]{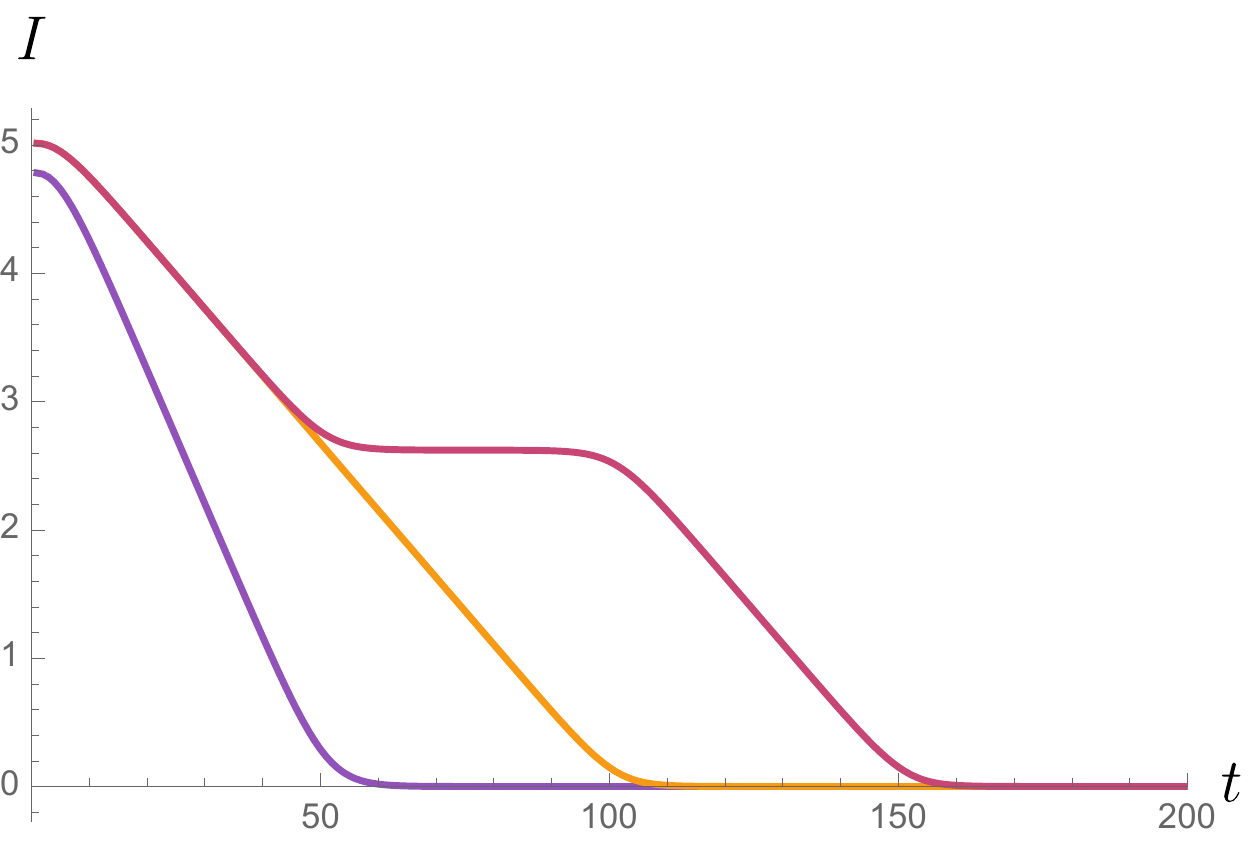}
    % \quad
    \includegraphics[height = 3.25cm ]{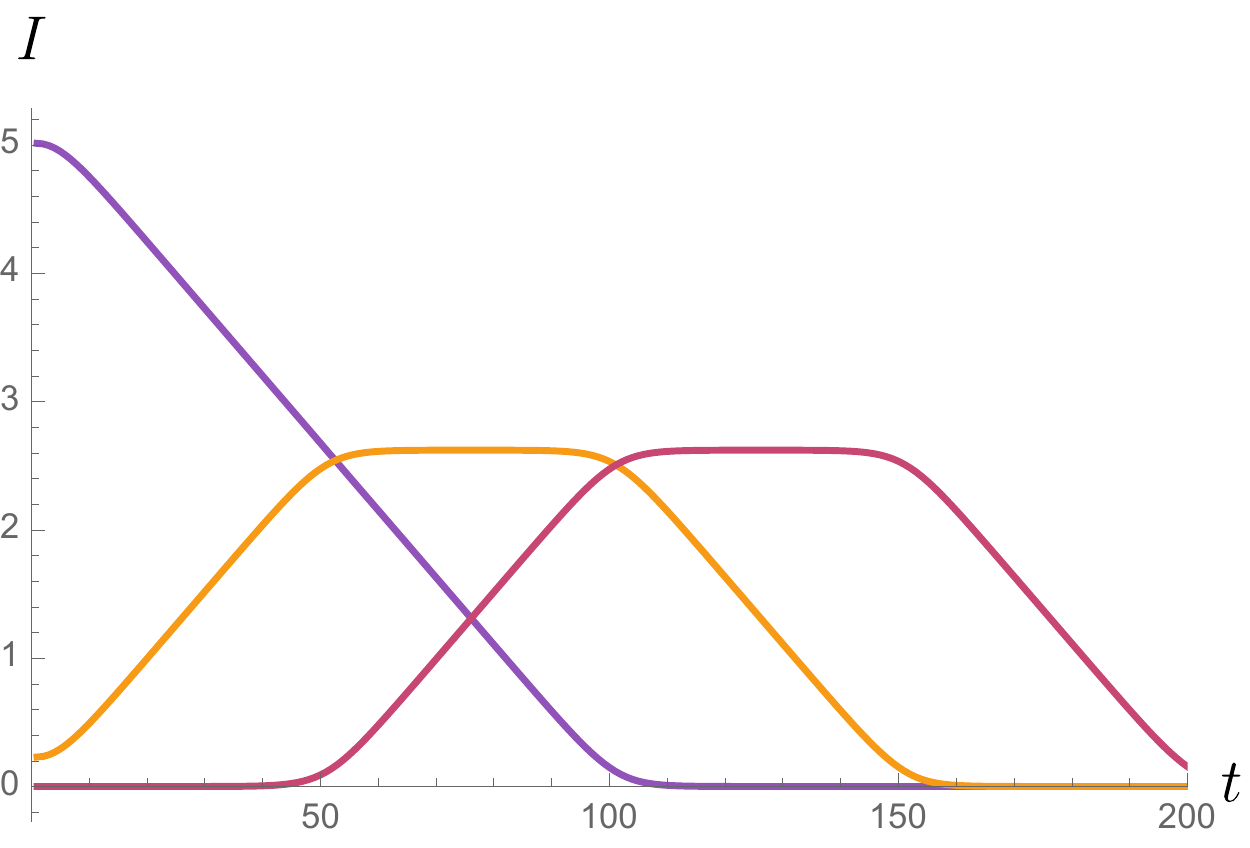}
    \caption{Operator negativity and mutual information for free fermions. The top row is computed by (\ref{ff_2nd_OLN_eq}). (top left) $\Delta \mathcal{E}^{(2)}$ for free fermions with $l_1 = 5$, $l_2 = \{5, 10, 15\}$, and both intervals starting at the same position. The quantity is normalized such that the conformal factor is absent. Furthermore, it is multiplied by $\epsilon$. (top middle) $\mathcal{E}^{(2)}$ with $l_1 = l_2 = 5$, and the distance between starting positions varying as $\{0,2,4 \}$. (top right) $\mathcal{E}^{(2)}$ with $l_1 = 5$, $l_2 = 10$, fully overlapped (purple), adjacent (yellow), and separated by $5$ (magenta). We take $\beta = 2$. It is clear that the qualitative behavior of the negativity follows the quasi-particle picture. (middle row) Numerically computed using the same configurations except all linear dimensions are scaled $\times 10$ and we take the replica limit using the correlator method for a system of 450 lattice sites. (bottom row) Numerically computed using the same configurations except now for the BOMI using the correlator method.}
    \label{free_fermion_OLN}
\end{figure}

Let us now study the $2^{nd}$ R\'enyi operator logarithmic negativity ($2^{nd}$ OLN) for free fermion CFTs bcause the replica limit is presently out of reach.
% between the input subsystem $A$ and  the output subsystem $B$, which are defined by 
% \be
% \begin{split}
% &A=\left\{X| X_2\le X \le X_1\right\}, \quad B=\left\{X,\tau_1|Y_2\le X \le Y_1\right\}, \quad X_2<X_1<Y_2<Y_1.
% \end{split}
% \ee 
By acting with a partial transpose
% $\cdot^{T_B}$
on the subsystem $B$, the second R\'enyi operator logarithmic negativity of free fermion channel $U(t)$ is defined by
\be \label{2ndoln}
\mathcal{E}^{(2)}=\log{\left[\Tr_{A_1, A_2}\left(\rho_{A_1, A_2}^{T_{A_2}}\right)^{2}\right]}.
\ee
We lay out the details of this calculation in App. \ref{CB_FF_details_App} and present the result
\begin{align}
\label{ff_2nd_OLN_eq}
\mathcal{E}^{(2)}
&=\f{1}{8}\log{\left[\f{\pi}{2\beta}\right]}+\f{1}{16}\log{\left[\f{1}{\sinh^2{\left[\f{\pi}{\beta}(Y_1-Y_2)\right]}\sinh^2{\left[\f{\pi}{\beta}(X_1-X_2)\right]}}\right]}
\\ \nonumber
&+ \f{1}{16}\log{\left[\f{\cosh{\left[\f{\pi}{\beta}(X_1-Y_2+t)\right]}\cosh{\left[\f{\pi}{\beta}(X_1-Y_2-t)\right]}\cosh{\left[\f{\pi}{\beta}(X_2-Y_1+t)\right]}\cosh{\left[\f{\pi}{\beta}(X_2-Y_1-t)\right]}}{\cosh{\left[\f{\pi}{\beta}(X_1-Y_1+t)\right]}\cosh{\left[\f{\pi}{\beta}(X_1-Y_1-t)\right]}\cosh{\left[\f{\pi}{\beta}(X_2-Y_2+t)\right]}\cosh{\left[\f{\pi}{\beta}(X_2-Y_2-t)\right]}}\right]}.
\end{align}
The second term is large and negative. The late-time tripartite OLN $\mathcal{E}_3^{(2)}$ is given by
\be
\mathcal{E}_3^{(2)} \approx -\f{\pi}{16\epsilon}(X_1-X_2)
\ee
We should not be fooled into thinking that the free fermion scrambles quantum information scaling extensively with system size. This value is given by the unphysical conformal factor term that disappears in the $n_e \rightarrow 1$ limit. Indeed, for the $2^{nd}$ R\'enyi negativity, we should really study the change from the initial time. This necessity is most explicitly seen when studying the operator logarithmic negativity between disjoint intervals. Of course, the quantum correlations are initially zero, but the conformal factor spuriously shows nontrivial contribution. From here, we can simply fix the normalization to see the expected quasi-particle picture for free fermions with
\begin{align}
    \Delta \mathcal{E}_3^{(2)}(t)  = 0.
\end{align}
While the second moment of the negativity is generally not very illuminating, we are able to find numerically that $\mathcal{E}^{(2)}$ is simply proportional to the replica limit, thus has physical meaning. We have performed finite scaling analysis for the TOLN in the replica limit and have found the TOLN to vanish in the spacetime scaling limit for free fermions, matching the universal term which can be described by the quasi-particle picture
\begin{align}
    \mathcal{E}_3(t) = 0.
\end{align}
See Fig.~\ref{free_fermion_OLN} for the remarkable agreement between $\mathcal{E}^{(2)}$, the replica limit logarithmic negativity, and the mutual information. 
We note that we must use the proper fermionic partial transpose when numerically computing the negativity \cite{2015NJPh...17e3048E,2015JSMTE..08..005C,2016JSMTE..03.3116C,2016JSMTE..05.3109C,2017PhRvB..95p5101S,2019PhRvA..99b2310S,2018arXiv180709808S}.
More precisely, we use 
partial time-reversal, which properly implements 
(s)pin structures of spacetime.

% \begin{figure}
%     \centering
%     \includegraphics[height = 5cm]{toln_scaling.pdf}
%     \caption{Finite scaling analysis for the TOLN of the free fermion. The power law fit has exponent $\sim -0.81$. $t_f = 2 l = L/5$.}
%     \label{finite_scaling}
% \end{figure}

\subsection{Compact boson CFT}

\begin{figure}
    \centering
    \includegraphics[height = 2.75cm]{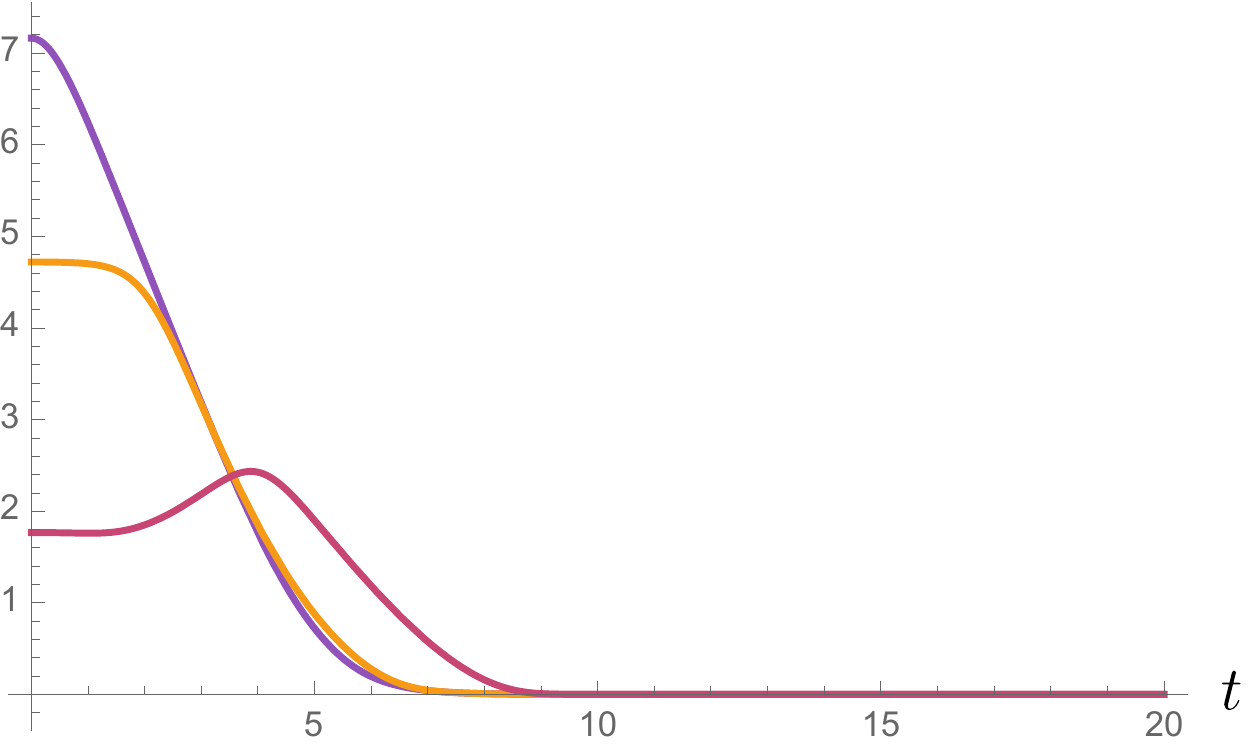} 
    \includegraphics[height = 2.75cm ]{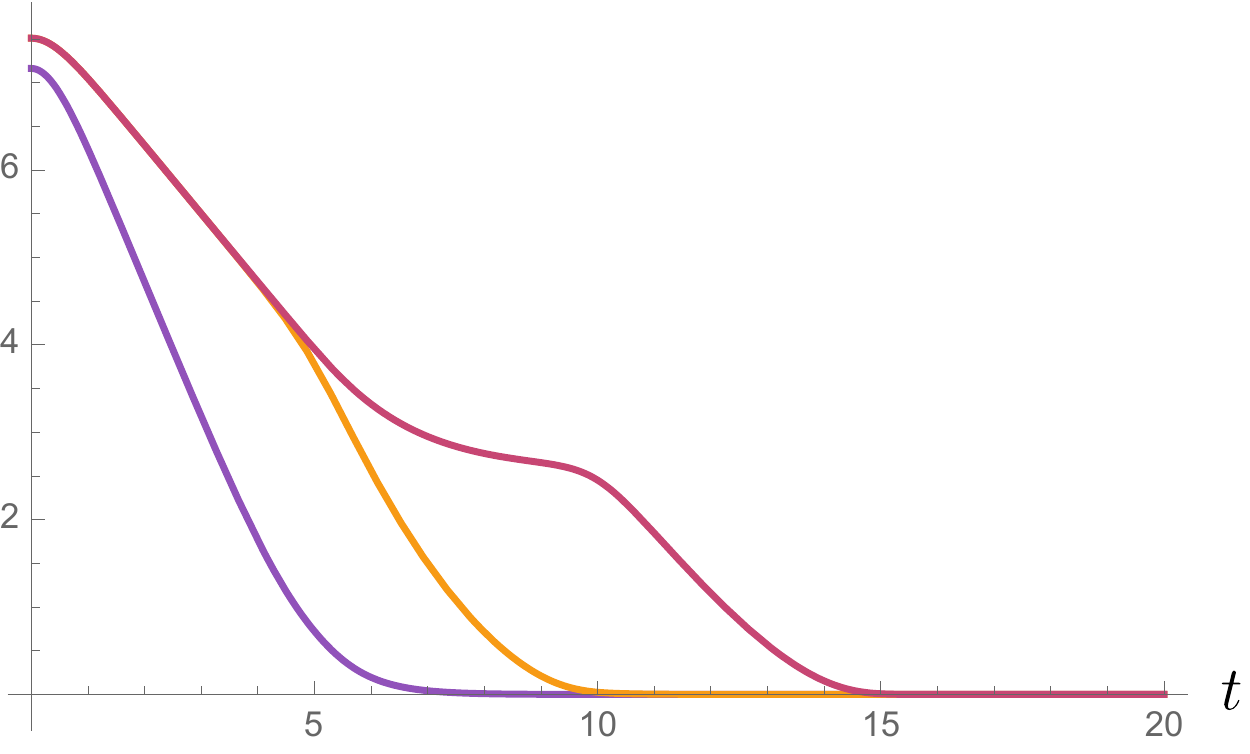}  
    \includegraphics[height = 2.75cm ]{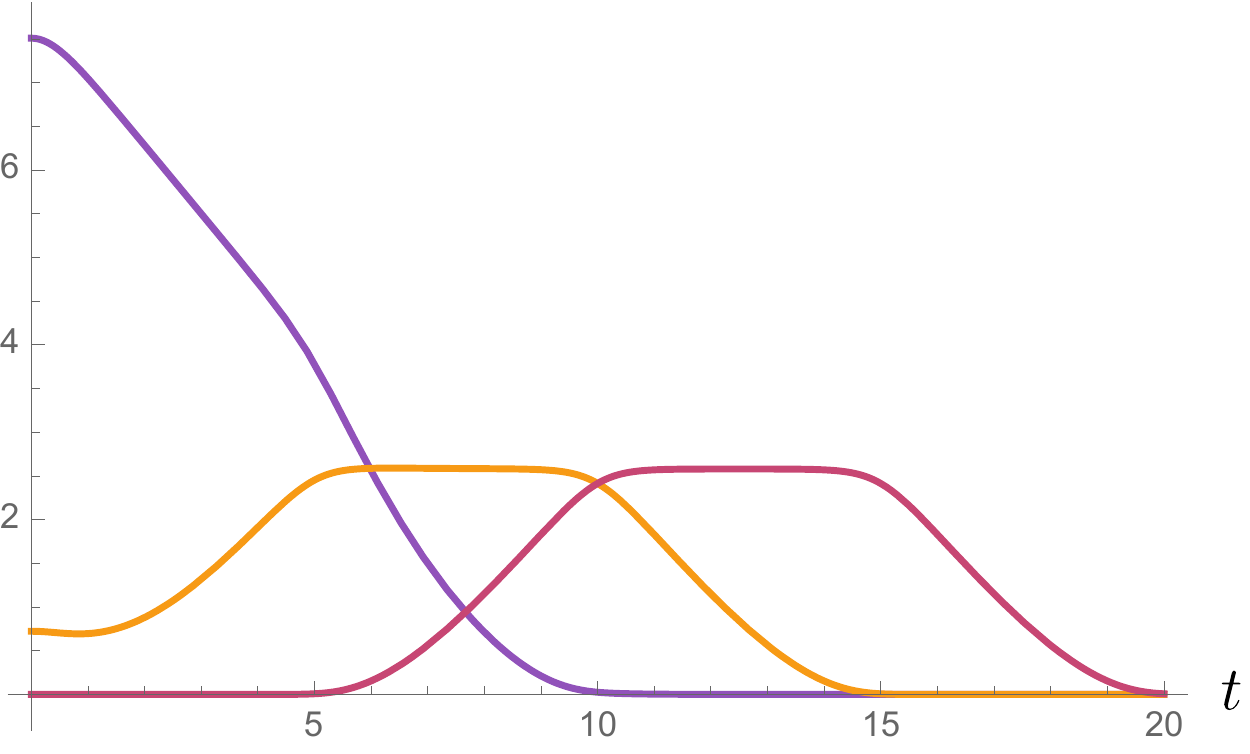}
    \caption{The time dependent piece of the second operator R\'enyi negativity (or equivalently the bipartite second R\'enyi operator mutual information) is shown for the compact boson (\ref{CB_OLN}). As usual, we have (left) OLN with $l_1 = l_2 = 5$, and the distance between starting positions varying as $\{0,2,4 \}$. (center) $l_1 = 5$, $l_2 = \{5, 10, 15\}$ with both intervals starting at the same position. (right) $l_1 = 5$, $l_2 = 10$, fully overlapped (purple), adjacent (yellow), and separated by $5$ (magenta). We use a compactification radius of $\eta = 2$ and cutoff $\beta  = 4$. The quasi-particle picture is certainly present though less pronounced than for the free fermion. }
    \label{CB2_OLN}
\end{figure}

We progress to the $c = 2$ free boson compactified on $S^1 \times S^1$ with radius $R$. We parametrize the radius as
\begin{align}
    \eta = R^2.
\end{align}
The theory is rational when $\eta$ is rational and irrational otherwise. In our convention, the self-dual radius corresponds to $\eta = 1$.  
The OMI and OTOC behaviors are something between free quasi-particles and maximal scrambling/chaos \cite{2018arXiv181200013N,2017PhRvD..96d6020C}, so we expect some deviation from the free fermion. The R\'enyi operator logarithmic negativities are
\begin{equation}
\mathcal{E}^{(n)} = \log \left[ \left(\frac{\pi}{\beta} \right)^{8h_n} \frac{F_n(x,\bar{x})}{\left[\sinh\frac{\pi(X_1-X_2)}{\beta}\sinh\frac{\pi(Y_1-Y_2)}{\beta}\right]^{4h_n}(1-x)^{2h_n}(1-\bar{x})^{2h_n}} \right]
\end{equation}
where 
\begin{align}
    F_n(x,\bar{x}) = \frac{\Theta(0|T)^2}{\prod_{k=1}^{n-1}f_{k/n}(x)f_{k/n}(\bar{x})}.
\end{align}
Analogous to the free fermion, we drop the conformal factors so that the negativity begins at zero for disjoint intervals, thus respecting causality
\begin{equation}
\label{CB_OLN}
\Delta \mathcal{E}^{(n)}(A,B) = \log \left[  \frac{F_n(x,\bar{x})}{(1-x)^{2h_n}(1-\bar{x})^{2h_n}} \right].
\end{equation}
Again, we relegate the details of the calculation for different replica numbers to App. \ref{CB_FF_details_App}. 
We plot the second R\'enyi operator negativity in Fig.~\ref{CB2_OLN}.

\begin{figure}
    \centering
    \includegraphics[height = 2.75cm]{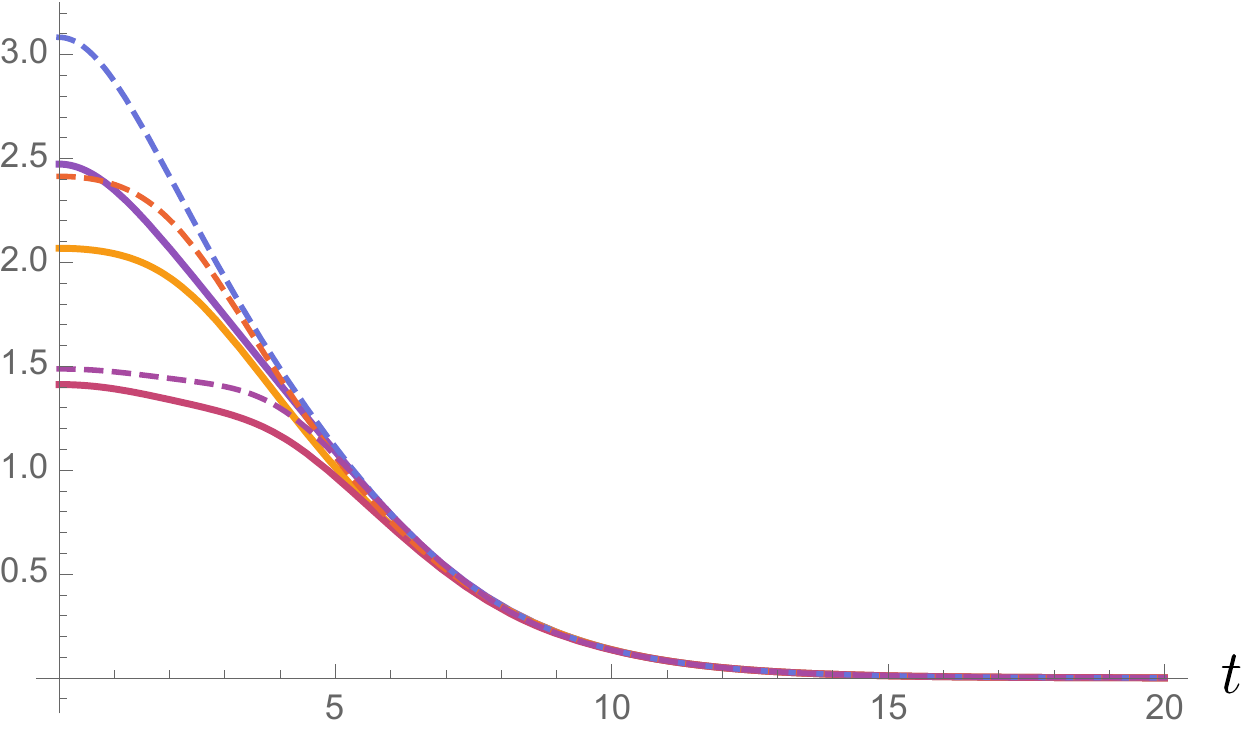} 
    \includegraphics[height = 2.75cm ]{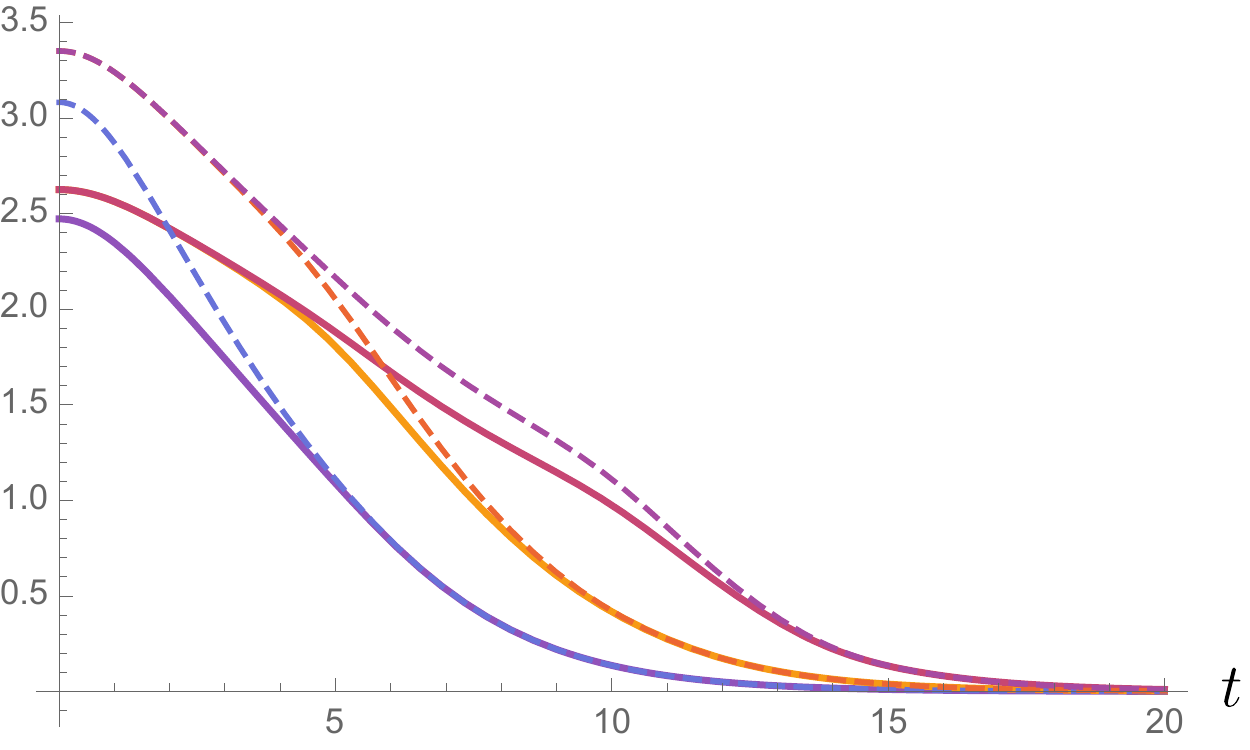}  
    \includegraphics[height = 2.75cm ]{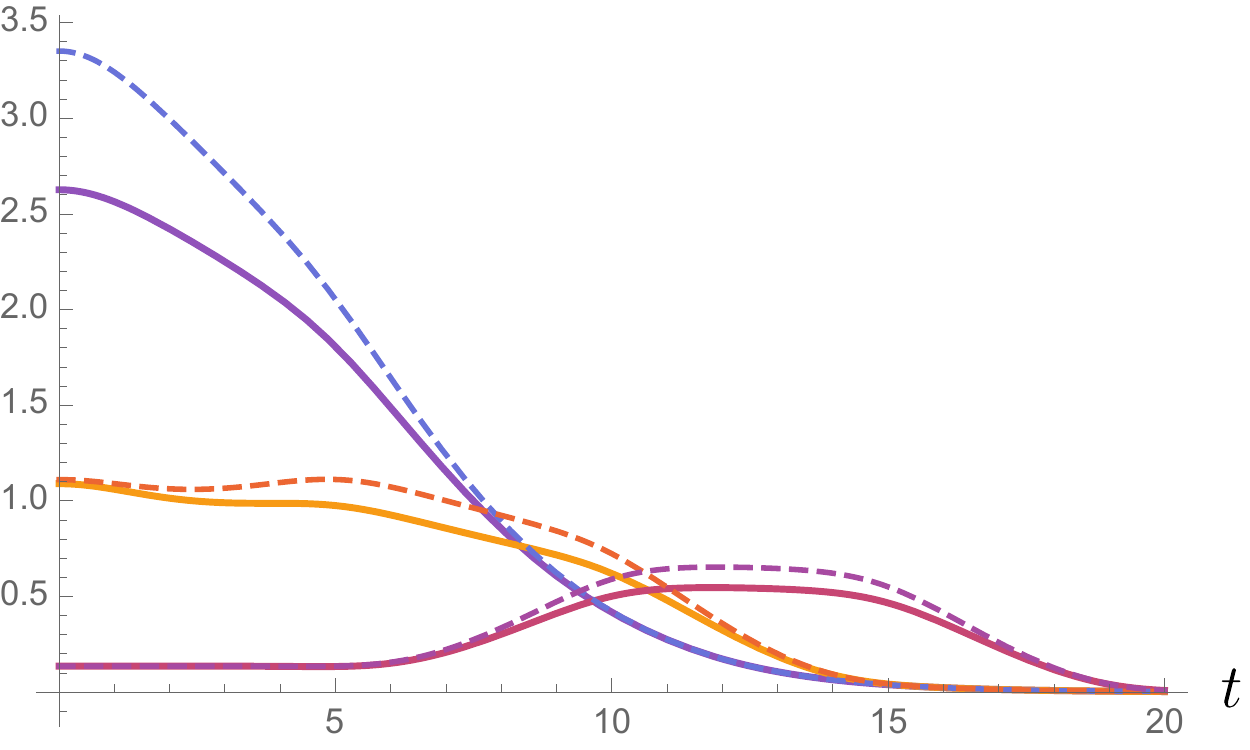}
    \caption{The time dependent piece of $\mathcal{E}^{(3)}$ (solid) and $I^{(3)}$ (dashed) are shown for the same parameters as Fig.~\ref{CB2_OLN} using (\ref{CB_OLN}). The convergence of the two quantities at late times is noted. A preliminary interpretation may be that the classical information is lost prior to the quantum information.}
    \label{fig:E3_cboson}
\end{figure}

In order to better compare the moments of the negativity and those of the mutual information for odd integers, we use the following normalization reminiscent of the odd entropy \cite{2018arXiv180909109T}
\begin{align}
    \tilde{\mathcal{E}}^{(n_o)} = \frac{1}{n_o-1} \mathcal{E}^{(n_o)}.
\end{align}
We plot $\tilde{\mathcal{E}}^{(3)}$ along with its tripartite generalization in Fig.~\ref{fig:E3_cboson}.
Ultimately, we are interested in the analytic continuation from the even integers to one to find the logarithmic negativity, though this is a highly nontrivial technical task. We instead have provided analytic formulas for general integer replica numbers ($>1$). A parity of the moments may be seen in the fact that only the even integers contain hypergeometric functions with $k/n = 1/2$ that are described by elliptic functions (see App. \ref{CB_FF_details_App}). It is presently unclear to us how to perform the proper analytic continuation. 

% \subsection{Bipartite Logarithmic Negativity Plots}
% \begin{figure}
%   \includegraphics[width=50mm]{BosonNegativity_Eta2_LA10_LB10_Symm}
%   \includegraphics[width=50mm]{BosonNegativity_Eta2_LA10_LB10_ASymm}
%   \includegraphics[width=50mm]{BosonNegativity_Eta2_LA10_LB10_d10}
% \includegraphics[width=50mm]{BosonNegativity_Eta2_LA10_LB20_d10}
% \includegraphics[width=50mm]{BosonNegativity_Eta2_LA20_LB10_d10}
% \caption{Plots of 2\textsuperscript{nd} R\'{e}nyi bipartite operator mutual information (blue curve)  and 2\textsuperscript{nd} R\'{e}nyi bipartite logarithmic negativity (orange curve) for $\epsilon = 1.1$ with $\eta = 2$. The intervals are \textbf{Top Left:} $(X_2,X_1) = (-10,0)$ and $(Y_2,Y_1) = (-10,0)$. \textbf{Top Right:} $(X_2,X_1) = (-10,0)$ and $(Y_2,Y_1) = (-5,5)$. \textbf{Middle Left:} $(X_2,X_1) = (-10,0)$ and $(Y_2,Y_1) = (10,20)$. \textbf{Middle Right:} $(X_2,X_1) = (-10,0)$ and $(Y_2,Y_1) = (10,30)$. \textbf{Bottom Left:} $(X_2,X_1) = (-20,0)$ and $(Y_2,Y_1) = (10,20)$.}
% \end{figure}

\begin{figure}
\centering
    \includegraphics[width = 6cm]{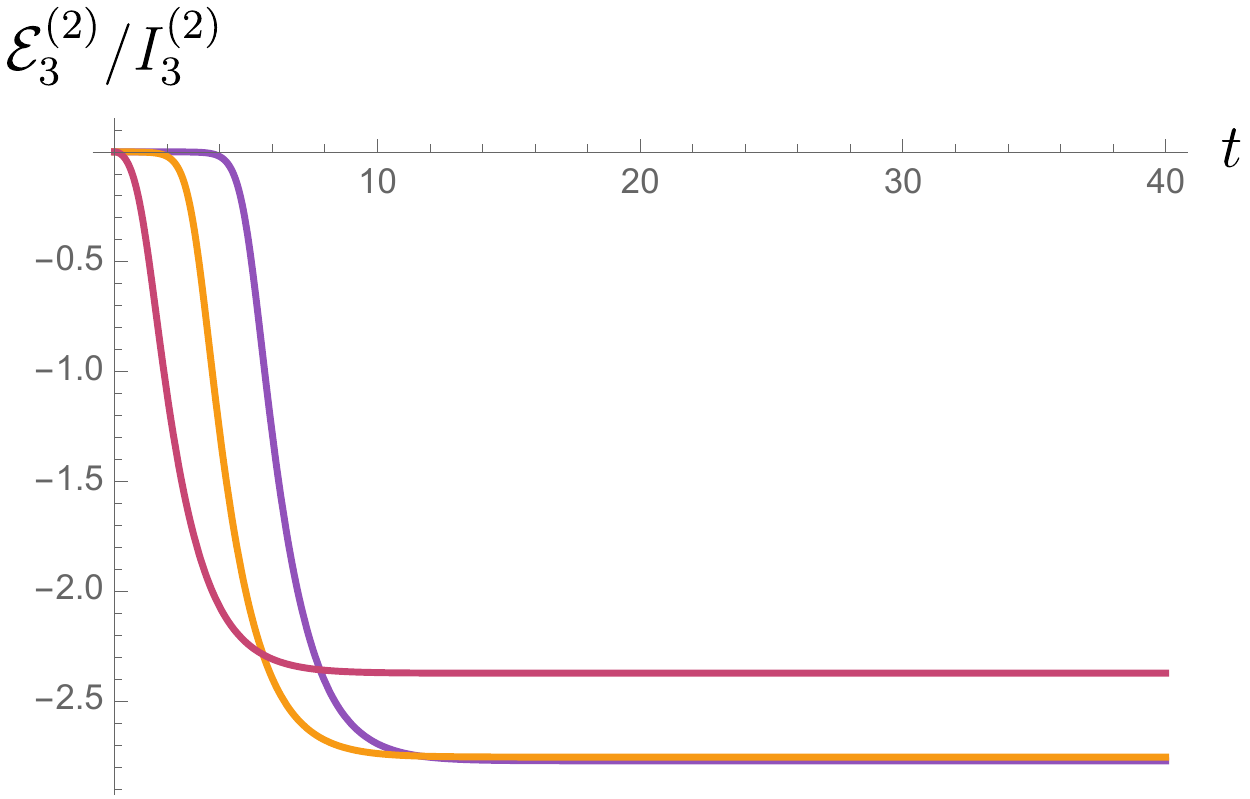} 
     \includegraphics[width = 6cm]{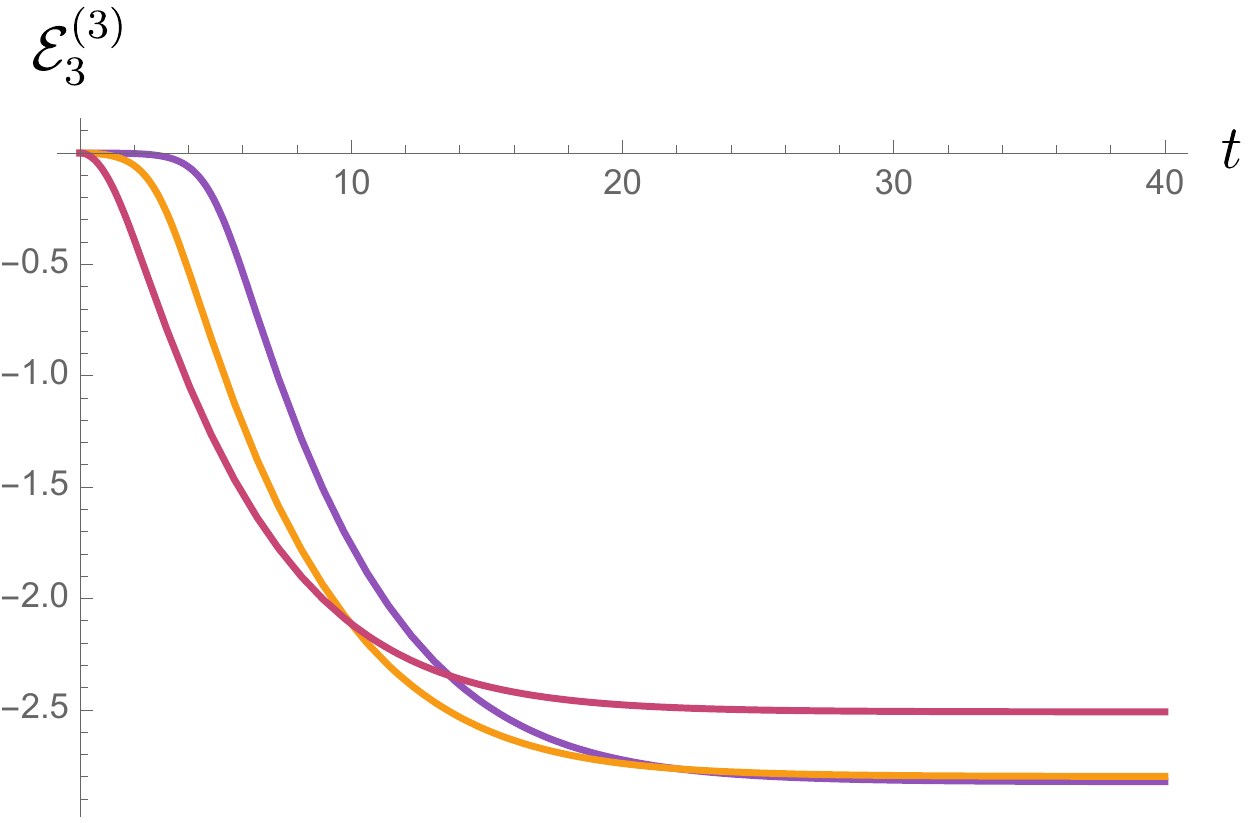}
   \caption{(left) The change in tripartite $\mathcal{E}^{(2)}_3$ 
   and $I^{(3)}_3$ for the compactified boson (\ref{CB_OLN}) with $\eta = 2$, $\beta = 2$, $(X_2,X_1) = (-1,1)$ (magenta)$, (-3,3)$ (yellow)$, (-5,5)$ (purple) and $Y_2 = 0$. (right) The third tripartite R\'enyi negativity for the same intervals and $\eta = 3$, $\beta =4$.}
   \label{boson3_fig}
\end{figure}

\subsection{Holographic calculation}
\label{holo_quench}
For holographic conformal field theories, we rely on the bulk dual description of the logarithmic negativity reviewed in section \ref{negativity_section} and Appendix \ref{app_neg_blocks}.

To compute the operator logarithmic negativity, we adjust the holographic global quench set up introduced in Ref.~\cite{2013JHEP...05..014H}. This uses the holographic dual of the thermofield double state, the eternal black hole \cite{2003JHEP...04..021M}. The Killing vector of this geometry time evolves forward on one of the copies of the CFT and backwards on the other. To introduce nontrivial time dependence and model the operator entanglement, we use a Hamiltonian that evolves both CFT's forward in time. We are interested in finite intervals on opposite copies of the CFT. Using a transformation from the BTZ black string geometry to the pure $AdS_3$ Poincar\'e patch, we can compute the lengths of extremal geodesics. The boundary positions of the intervals are defined by their boundary points
\begin{align}
    P_1 &= (\sinh \frac{4\pi t}{\beta}, \cosh \frac{4\pi t}{\beta}), \quad P_2 = e^{\frac{2 \pi }{\beta}l_1} (\sinh \frac{4\pi t}{\beta}, \cosh \frac{4\pi t}{\beta}) \\
    \nonumber P_5 &= e^{\frac{2 \pi }{\beta}(d+l_1+ l_2)}(\sinh \frac{4 \pi t}{\beta} , -\cosh \frac{4\pi t}{\beta}), \quad P_6 = e^{\frac{2 \pi }{\beta}(d+l_1)}(\sinh \frac{4 \pi t}{\beta}, -\cosh \frac{4 \pi t}{\beta}),
\end{align}
where $l_1$ and $l_2$ are the lengths of the two intervals and $d$ is the distance between them. There are two regimes; the disconnected regime does not connect the two intervals through the bulk while the connected regime does. We calculate the OMI
\begin{align}
    \label{disjoint_MI_holo_eq}
    I_{dis} &= 0, \\ \nonumber \quad I_{con} &= \frac{c}{6} \log \left[ \frac{\mathcal{L}_{12} \mathcal{L}_{56}}{\mathcal{L}_{52} \mathcal{L}_{16}}\right],
\end{align}
where $\mathcal{L}_{ij}$ are the geodesic lengths between $P_i$ and $P_j$
\begin{align}
    \mathcal{L}_{12} &= \log\left[\frac{\left(e^{\frac{2 \pi }{\beta}l_1}-1 \right)^2}{\epsilon^2 e^{\frac{2 \pi }{\beta}l_1}} \right], \\ 
    \mathcal{L}_{56} &= \log\left[\frac{\left(e^{\frac{2 \pi }{\beta}(d+l_1+ l_2)}-e^{\frac{2 \pi }{\beta}(d+l_1)} \right)^2}{\epsilon^2 e^{\frac{2 \pi }{\beta}(2 d + 2l_1 + l_2)}} \right], \\ 
    \mathcal{L}_{52} &= \log\left[\frac{-\left(e^{\frac{2 \pi }{\beta}(d+l_1+ l_2)}-e^{\frac{2 \pi }{\beta}l_1}\right)^2\sinh^2 \frac{4 \pi t}{\beta}+\left(e^{\frac{2 \pi }{\beta}(d+l_1+ l_2)}+e^{\frac{2 \pi }{\beta}l_1}\right)^2\cosh^2 \frac{4 \pi t}{\beta}}{\epsilon^2 e^{\frac{2 \pi }{\beta}(d+l_1)}} \right], \\
    \mathcal{L}_{16} &= \log\left[\frac{-\left(e^{\frac{2 \pi }{\beta}(d+l_1)}-1\right)^2\sinh^2 \frac{4 \pi t}{\beta}+\left(e^{\frac{2 \pi }{\beta}(d+l_1)}+1\right)^2\cosh^2 \frac{4 \pi t}{\beta}}{\epsilon^2 e^{\frac{2 \pi }{\beta}(d+l_1)}} \right]. 
\end{align}
We now calculate the entanglement wedge cross-sectional to find the OLN. The disconnected regime is trivial 
\begin{align}
    \mathcal{E}_{dis} = 0,
\end{align}
while the connected regime is a function of only the cross-ratio \cite{2017arXiv170809393T}
\begin{align}
    \mathcal{E}_{con} &= \frac{c}{4}\log(1 + 2 z + 2 \sqrt{z (z+1)}), \label{disjoin_LN_holo_eq} \\ \nonumber
    z &= \frac{g(P_5, P_6)~g(P_1, P_2)}{g(P_1, P_6)~g(P_5, P_2)},
\end{align}
where we use the metric $g$ on the $Mink_2$ boundary. We list the elements of the cross ratio below
\begin{align}
    \label{cross_elements}
    g(P_5, P_6) &= \left(e^{\frac{2 \pi }{\beta}(d+ l_1+l_2)}-e^{\frac{2 \pi }{\beta}(d+l_1)}\right), \\ 
    g(P_1, P_2) &= \left(e^{\frac{2 \pi }{\beta}l_1}-1 \right), \\ 
    g(P_1, P_6) &= \left(-\left(e^{\frac{2 \pi }{\beta}(d+l_1)}-1 \right)^2 \sinh^2 \frac{4 \pi t}{\beta} + \left(e^{\frac{2 \pi }{\beta}(d+l_1)}+1 \right)^2 \cosh^2 \frac{4 \pi t}{\beta}\right)^{1/2} ,\\ 
    g(P_5, P_2) &= \left(-\left(e^{\frac{2 \pi }{\beta}(d+l_1 + l_2)}-e^{\frac{2 \pi }{\beta}l_1} \right)^2 \sinh^2 \frac{4 \pi t}{\beta} + \left(e^{\frac{2 \pi }{\beta}(d+l_1 + l_2)}+e^{\frac{2 \pi }{\beta}l_1} \right)^2 \cosh^2 \frac{4 \pi t}{\beta}\right)^{1/2}.
\end{align}

\begin{figure}
    \centering
    \includegraphics[height= 3.25cm]{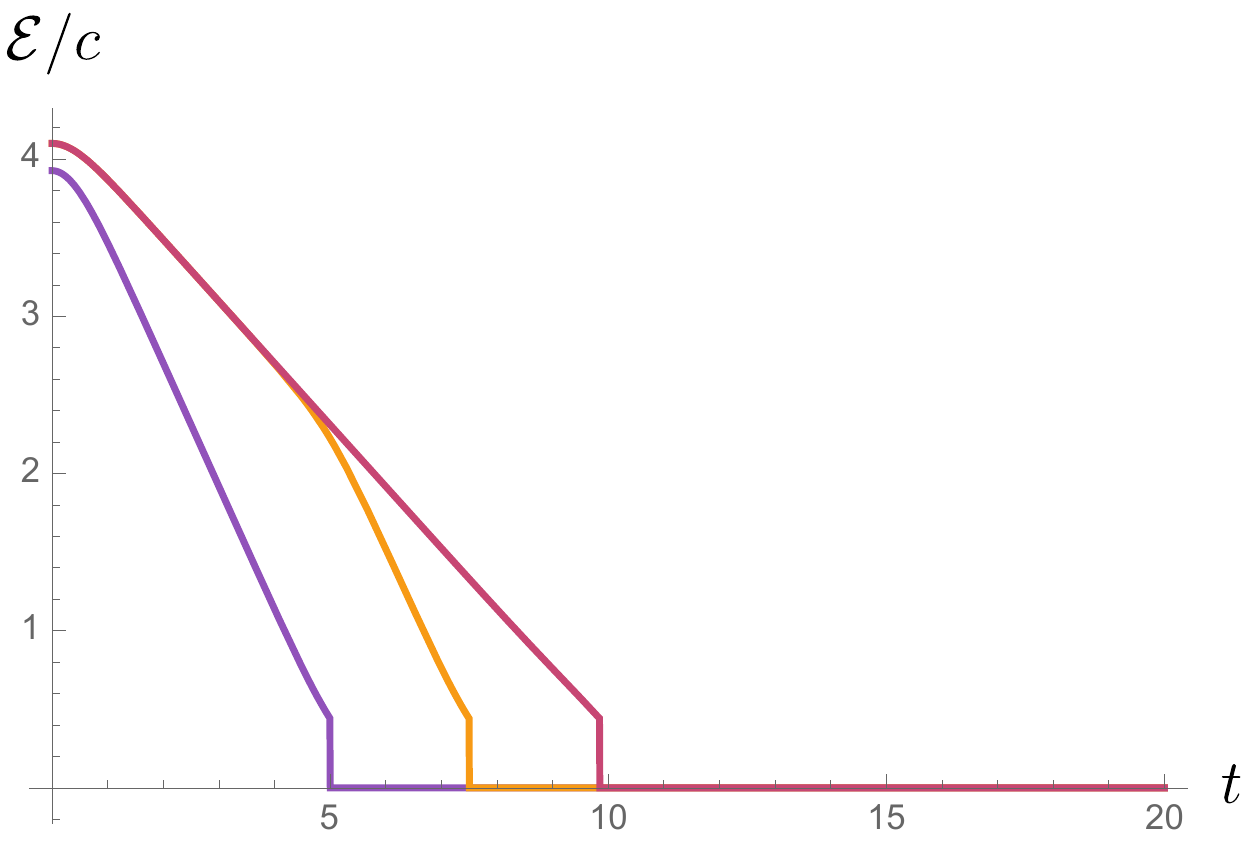} 
    % \quad
    \includegraphics[height= 3.25cm]{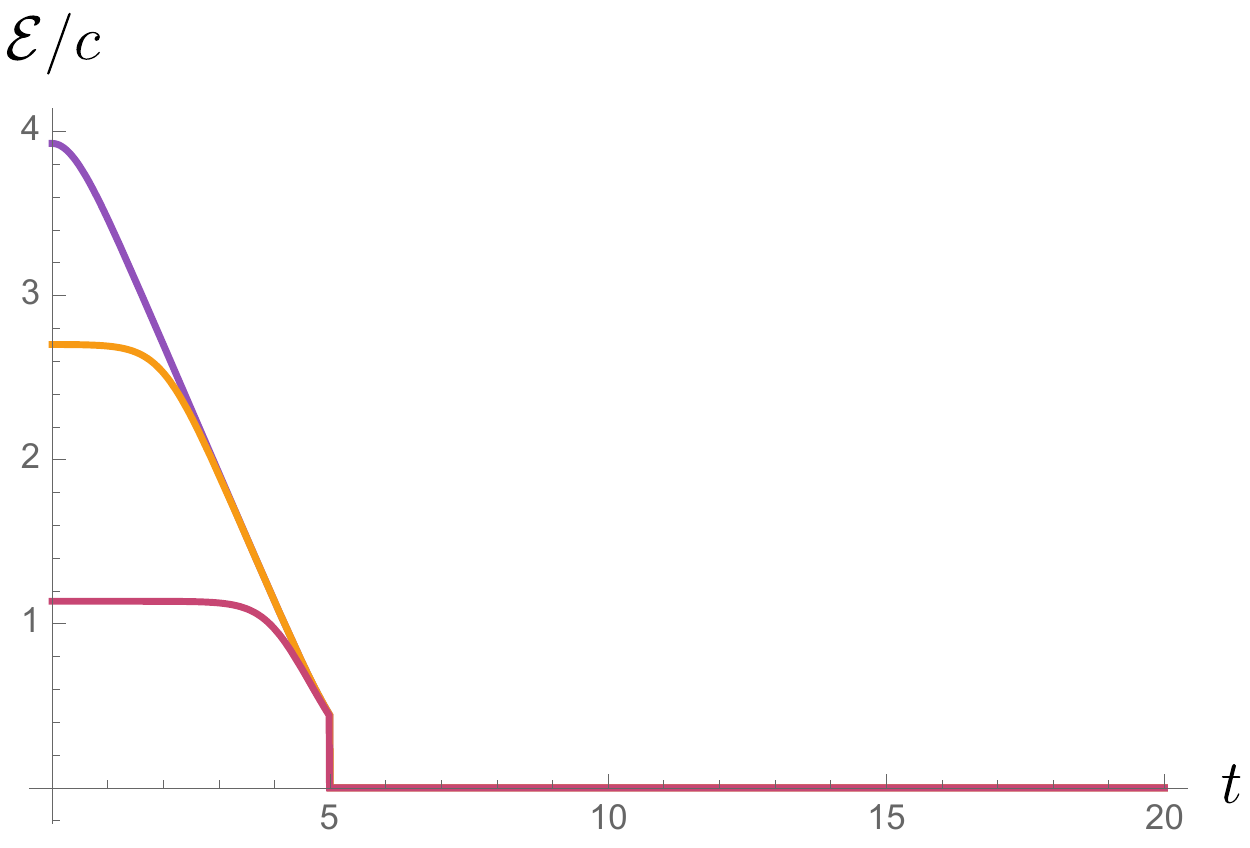} 
    % \quad
    \includegraphics[height= 3.25cm]{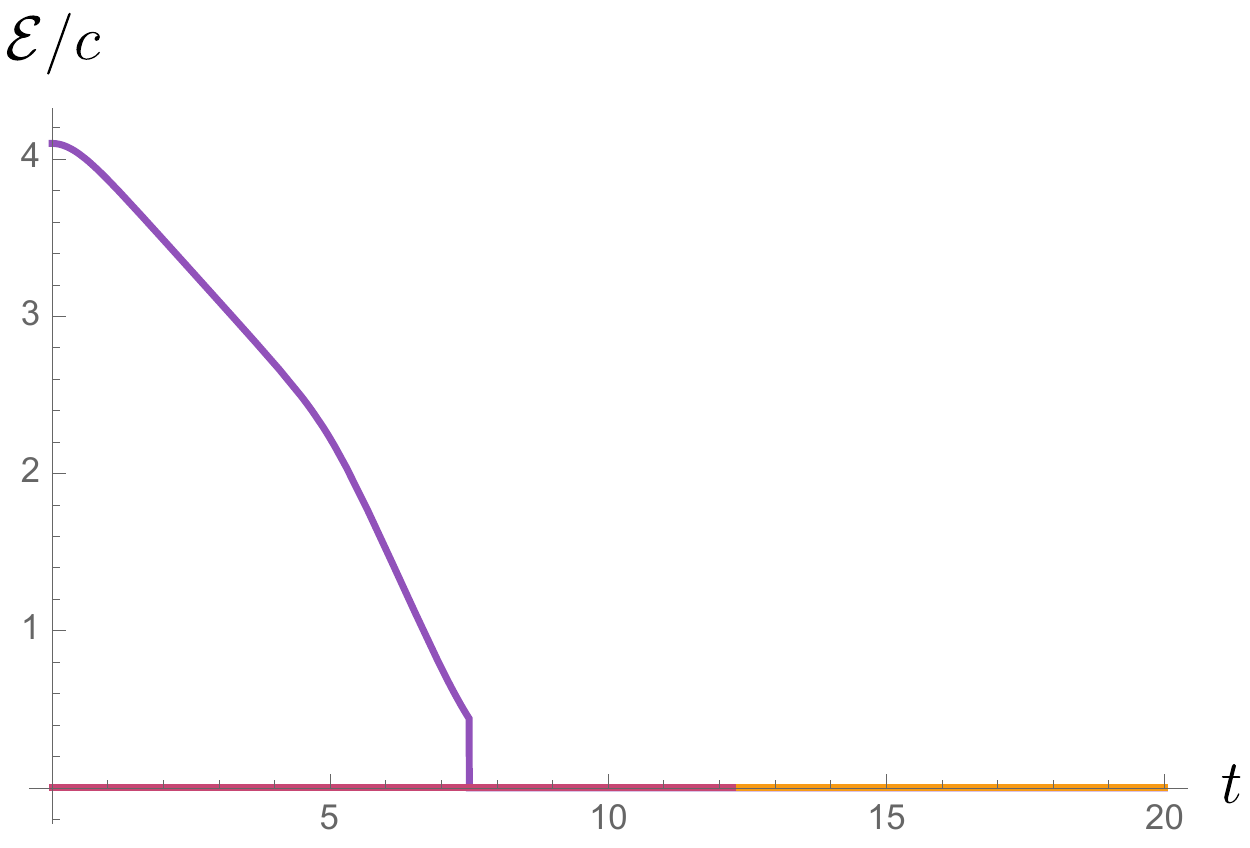} 
    % \quad
    \includegraphics[height= 3.25cm]{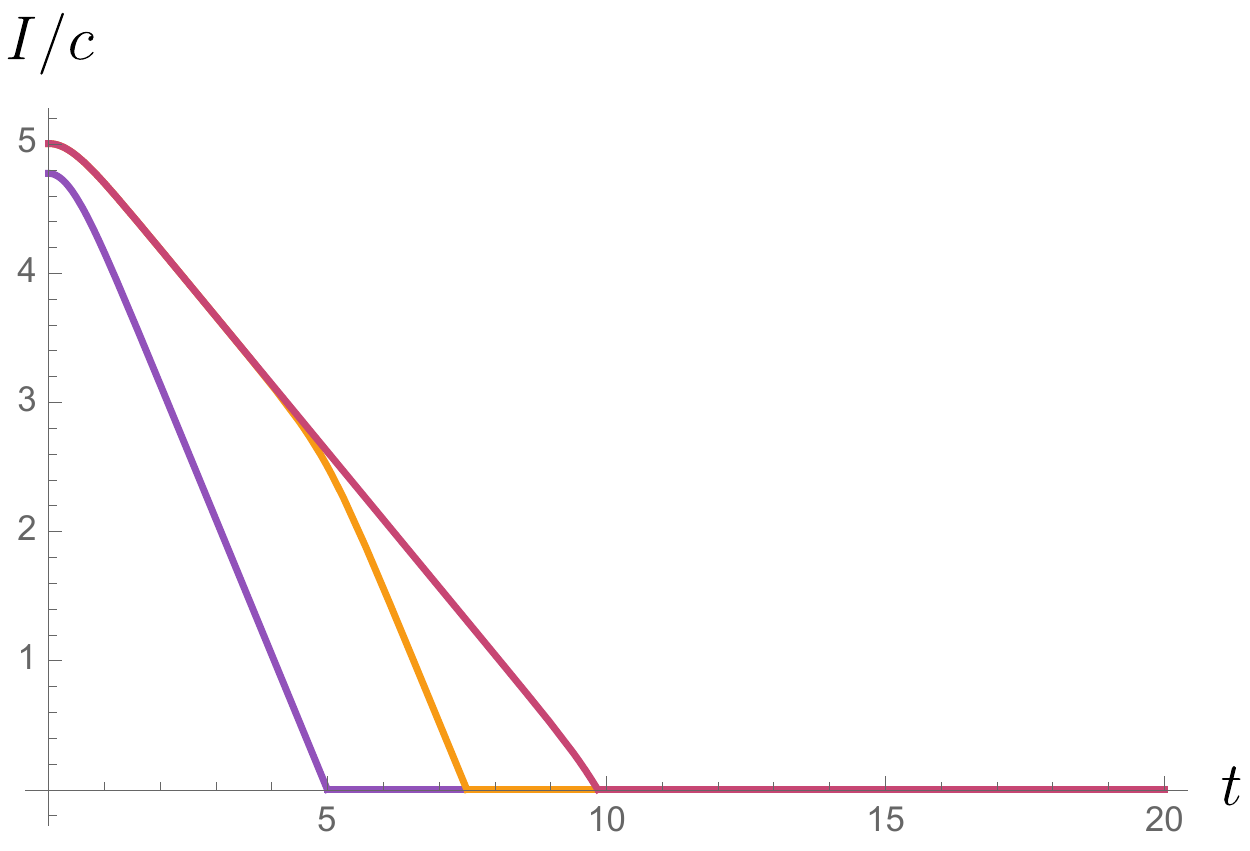} 
    % \quad 
    \includegraphics[height= 3.25cm]{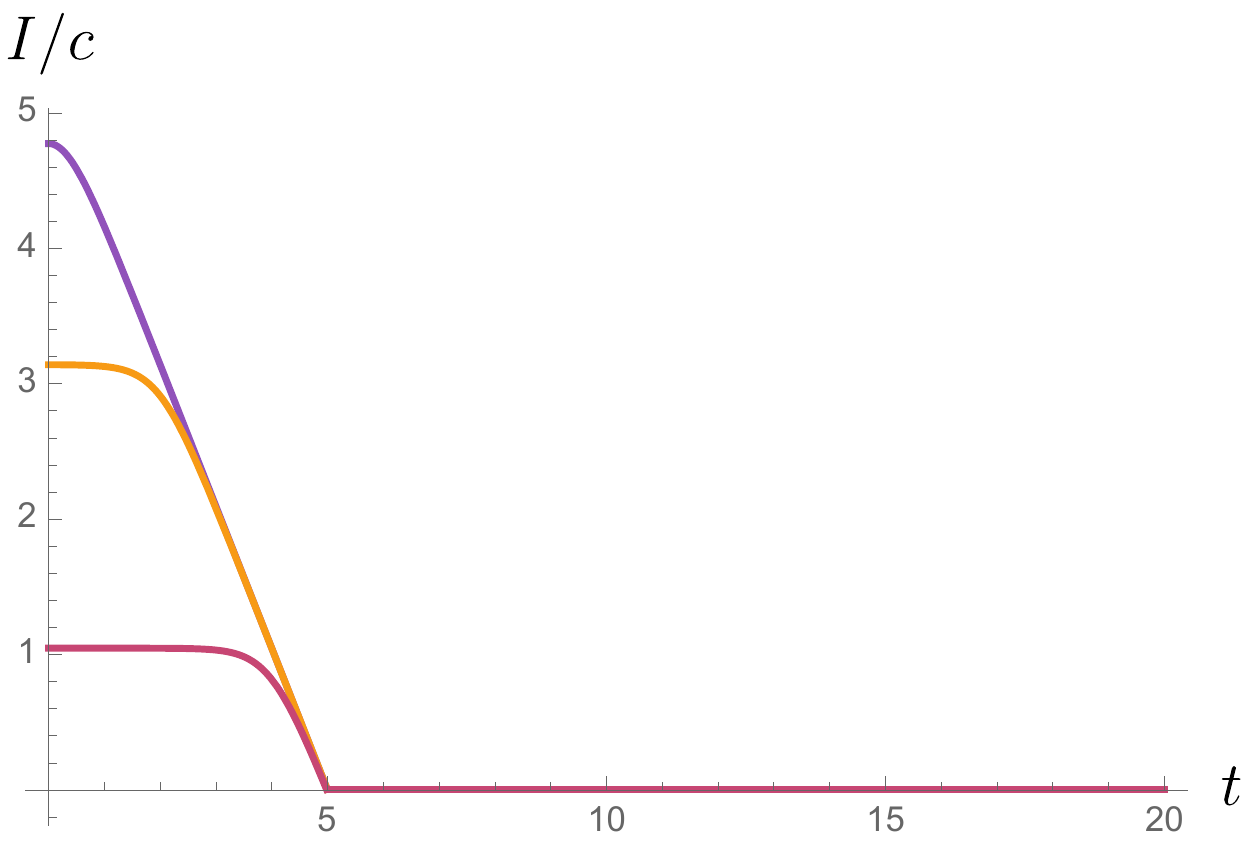}
    % \quad
    \includegraphics[height= 3.25cm]{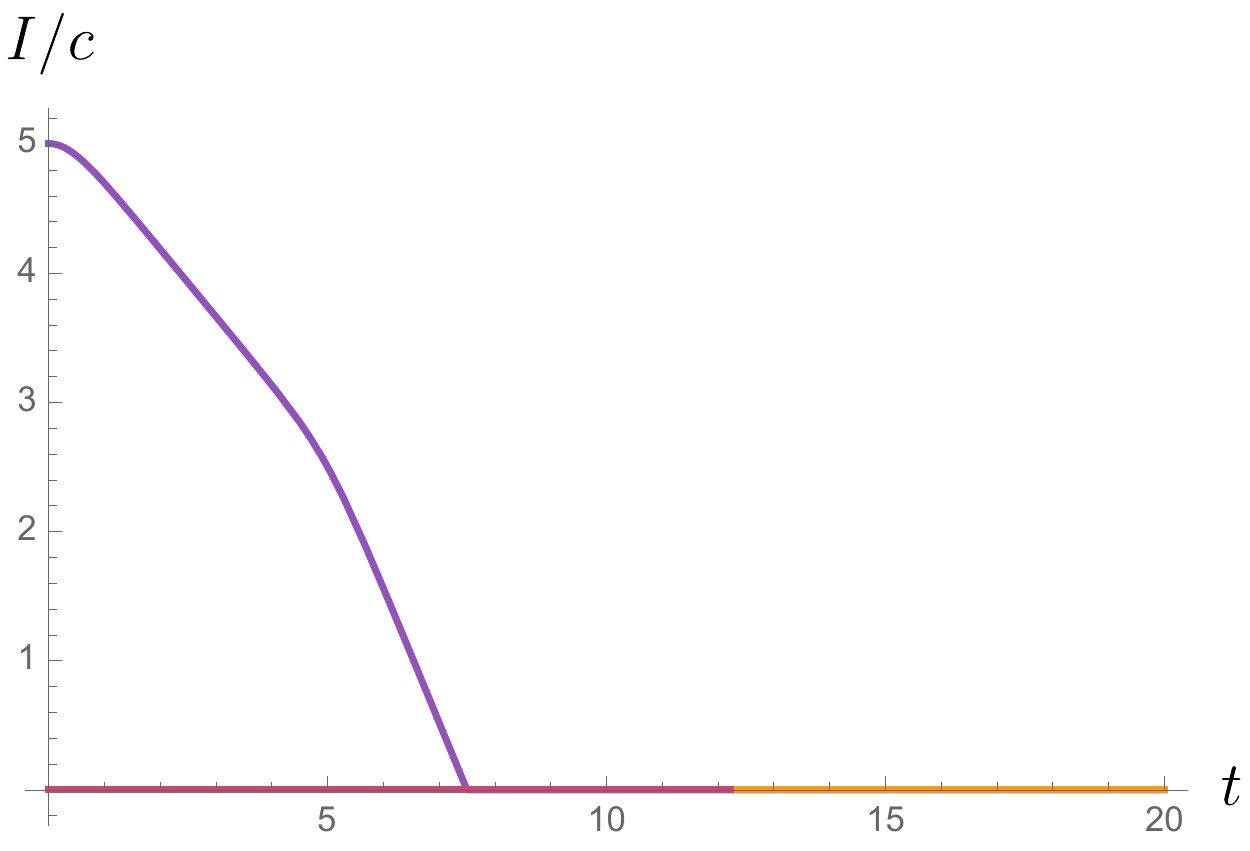}
    \caption{The large-c OLN (\ref{disjoin_LN_holo_eq}) is plotted in the top row and the OMI (\ref{disjoint_MI_holo_eq}) in the bottom row. We use the same configurations as in Figs.~\ref{CFT_neg}-\ref{boson3_fig} and set $\beta =2$.}
    \label{OMI_OLN_BOMI}
\end{figure}

We plot the BOMI and BOLN in Fig.~\ref{OMI_OLN_BOMI} and find that there are small but significant differences between the behavior of mutual information and negativity for operator entanglement. Mainly, the negativity is discontinuous when it transitions to zero. This may be interpreted as locking of the distillable entanglement, a feature not seen in entanglement entropies but seen in logarithmic negativity \cite{2005PhRvL..94t0501H}. Interestingly, the amplitude of the discontinuity is universal, depending only on the central charge
\begin{align}
    \lim_{\beta \rightarrow 0} \Delta \mathcal{E} = \frac{c}{4}\log\left[3  + 2 \sqrt{2} \right].
\end{align}
Because this discontinuity is cutoff and system size independent, it is unnoticeable in the spacetime scaling limit. It would be interesting to have a better information theoretic understanding of this discontinuity in the context of locking of distillable entanglement \cite{2005PhRvL..94t0501H}.

It is important to note that if we take non-overlapping input and output intervals, we will always be in the disconnected regime such that the OLN and OMI are trivial. This is consistent with the prediction made in Ref.~\cite{2018arXiv180909119A} that the negativity of disjoint intervals for chaotic systems in the spacetime scaling limit should be trivial. Furthermore, our holographic results (save for the discontinuity) match the behavior predicted by the line-tension picture in Section \ref{linetension_LN}. The linear decline of OLN and the kink at the time determined by causality are precisely predicted by (\ref{ln_line_tension}). 

\begin{figure}
    \centering
    \includegraphics[height = 4.5cm]{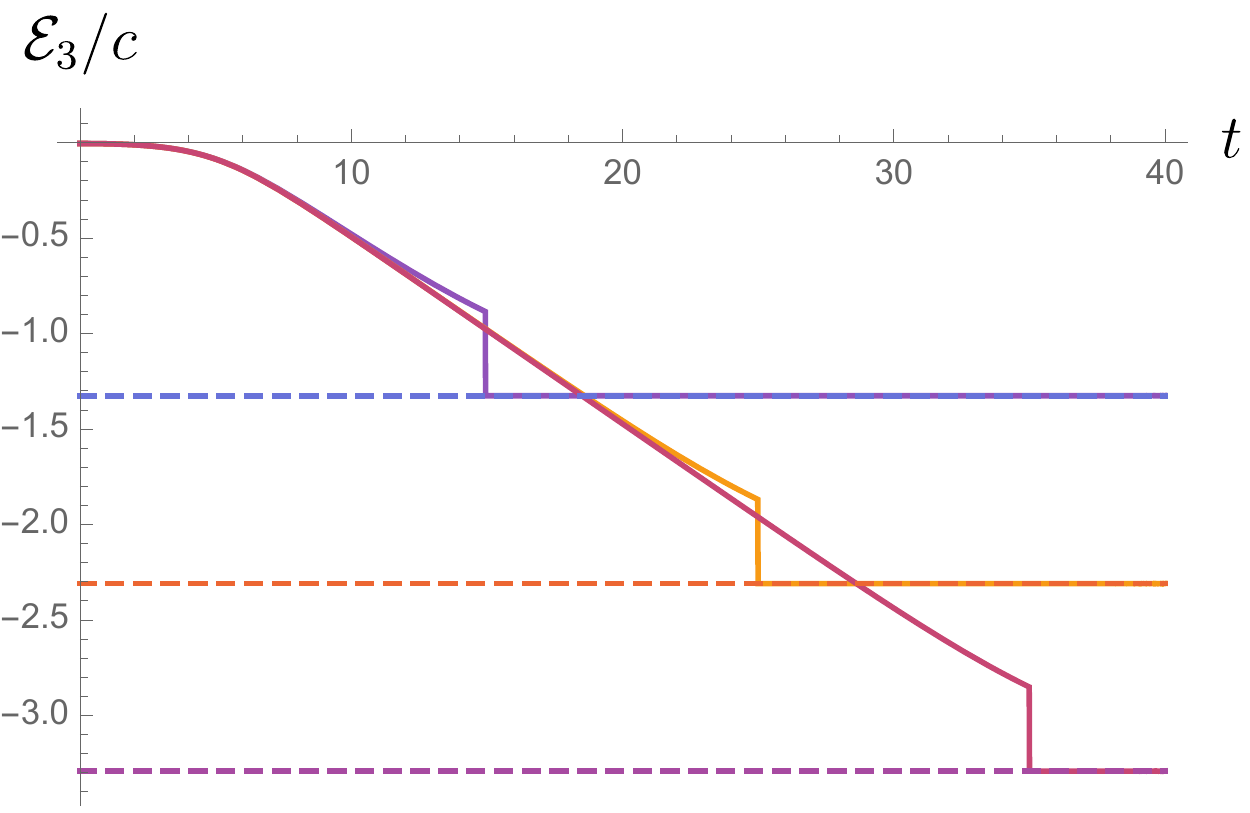}  
    \includegraphics[height = 4.5cm]{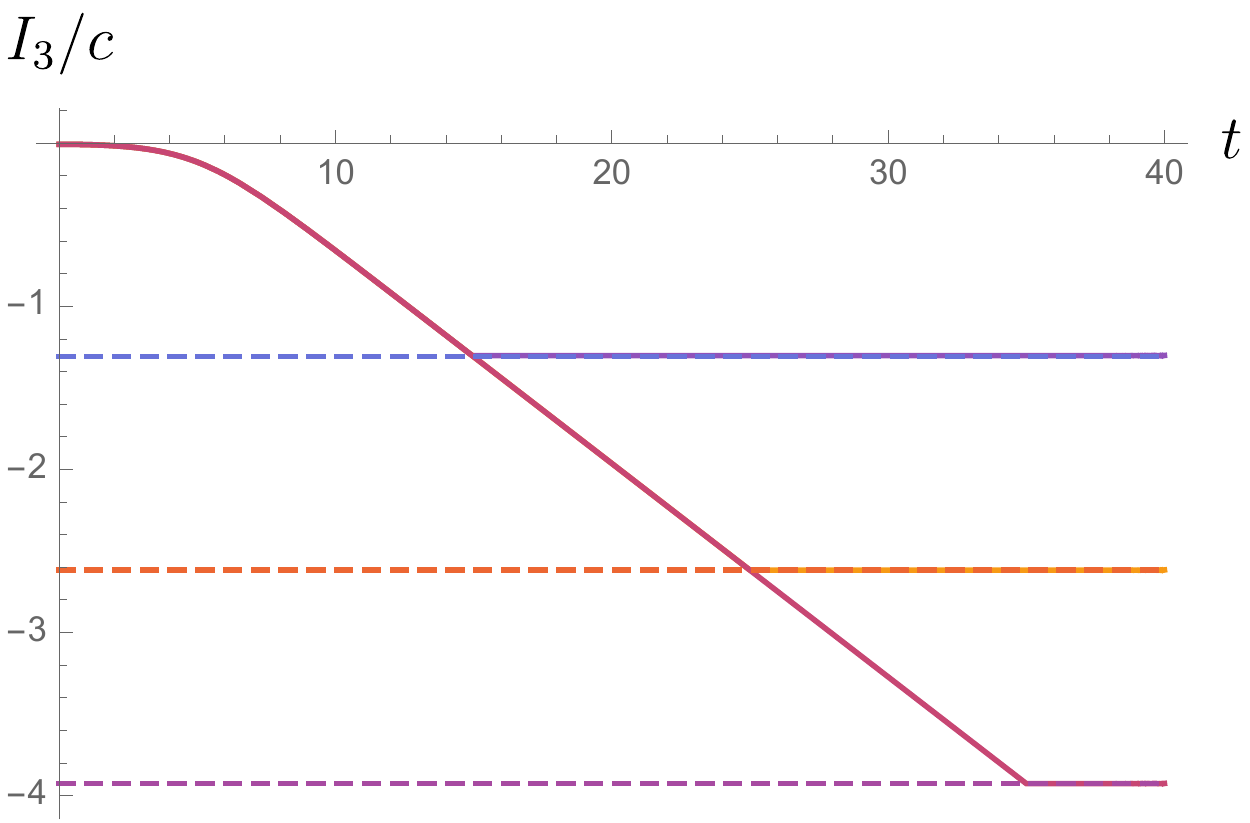}
    \caption{(left) The tripartite operator negativity (\ref{disjoin_LN_holo_eq}) is plotted for an input interval, $A$, of lengths 5 (magenta), 10 (yellow), 15 (purple). (right) TOMI (\ref{disjoint_MI_holo_eq}) for the same configurations. The dotted lines are the lower bounds given by \eqref{TOLN_bound}.}
    \label{OMI_OLN_TOMI}
\end{figure}

Multipartite generalizations of the entanglement wedge cross section were studied in Refs.~\cite{2018JHEP...03..006B,2018arXiv180502625U,2018arXiv180500476B,2018arXiv181101983B,2019arXiv190605970H} as holographic duals to multipartite entanglement of purification though these are distinct from the tripartite negativity that we study. We calculate the tripartite negativity in holographic theories. To do so, we take the $l_2 \rightarrow \infty$ limit in (\ref{cross_elements}). This gives
\begin{align}
    g(P_5, P_6)_{\infty} &= e^{\frac{2 \pi}{\beta}(d + l_1 + l_2)}, \\ \nonumber
    g(P_1, P_2)_{\infty} &=  e^{\frac{2 \pi}{\beta}l_1} - 1 , \\ \nonumber
    g(P_1, P_6)_{\infty} &= \left(-\left(e^{\frac{2 \pi }{\beta}(d+l_1)}-1 \right)^2 \sinh^2 \frac{4 \pi t}{\beta} + \left(e^{\frac{2 \pi }{\beta}(d+l_1)}+1 \right)^2 \cosh^2 \frac{4 \pi t}{\beta}\right)^{1/2} ,\\ \nonumber
    g(P_5, P_2)_{\infty} &= e^{\frac{2 \pi}{\beta}(d + l_1 + l_2)}.
\end{align}
Therefore,
\begin{align}
    z_{\infty} = \frac{g(P_1, P_2)_{\infty}}{g(P_1, P_6)_{\infty}},
\end{align}
which is independent of $l_2$.
We can simply determine a lower bound on the tripartite logarithmic negativity. 
\begin{align}
	\mathcal{E}_{3(A: B_1, B_2)}(t) = \mathcal{E}_{A, B_1}(t) + \mathcal{E}_{A, B_2}(t) - \mathcal{E}_{A, B_1 \cup B_2} \geq - \mathcal{E}_{A, B_1 \cup B_2}
	\label{TOLN_bound}
\end{align}
by the positivity of logarithmic negativity.
% In summary, we have the following bounds
% \begin{align}
% 	- LN(A, B_1 \cup B_2) \leq LN_3(A: B_1, B_2) \leq2 S_{1/2} (A).
% \end{align}
We can solve for the time-independent lower bound using (\ref{disjoin_LN_holo_eq}) with 
\begin{align}
    z = e^{2 \pi l_1/\beta} - 1.
\end{align}
Because $\beta$ is effectively a cutoff,
\begin{align}
    \mathcal{E}_{3(A: B_1, B_2),{min}} \simeq -\frac{c \pi l_1}{2 \beta} \equiv -2S_{1/2}^{(R)}(A) = -2\mathcal{E}^{(R)}(A,\bar{A}),
\end{align}
where ${\cdot}^{(R)}$ denotes the regularized entropy/negativity that subtracts the UV cutoff dependence
\begin{align}
    S^{(R)}(A) = \frac{1}{1-n}\log\left[ \frac{\langle\sigma_n(\omega_1, \bar{\omega}_1)\bar{\sigma}_n(\omega_2, \bar{\omega}_2) \rangle}{\left|dz/d\omega \right|_{\omega=0}^{4h_n}}\right],
\end{align}
where $z = e^{2\pi \omega/\beta}$. 
This simple lower bound may seem not restrictive enough, however we find that it is generically saturated at late times. It is equal to the regularized R\'enyi entropy at index $1/2$ of region $A$.
% We also can find $LN(A, B_2)$ with
% \begin{align}
%     z = \frac{e^{\frac{2 \pi}{\beta}l_1} - 1}{\left(-\left(e^{\frac{2 \pi }{\beta}(d+l_1)}-1 \right)^2 \sinh^2 \frac{2 \pi t}{\beta} + \left(e^{\frac{2 \pi }{\beta}(d+l_1)}+1 \right)^2 \cosh^2 \frac{2 \pi t}{\beta}\right)^{1/2}}.
% \end{align}
The resulting behavior is shown for various sizes of input interval $A$ in Fig.\ \ref{OMI_OLN_TOMI}. It behaves similarly to the TOMI whose saturation value is
\begin{align}
    I_3(t \rightarrow \infty) = -\frac{2c\pi l_1}{3\beta} = -2 S^{(R)}(A),
\end{align} 
which is extensive in the input system size and equal to twice the regularized von Neumann entropy of region $A$.

\subsection{Spin chains}

We now investigate the OLN for spins chains with the following two parameter Hamiltonian
\begin{align}
    H = \sum_i(- Z_iZ_{i+1} + g X_i + hZ_i).
    \label{sc_H}
\end{align}
We look at the integrable case corresponding to the transverse field Ising model with $g=1$ and $h=0$ and the chaotic case corresponding to $g=-1.05$ and $h=0.5$. As argued above, the negativity directly probes the scrambling properties of the chaotic Hamiltonian. For the integrable system, there is a finite recurrence time, while the chaotic system relaxes to a value close to the Haar-scrambled value.

\begin{figure}
    \centering
    \includegraphics[height = 3.25cm]{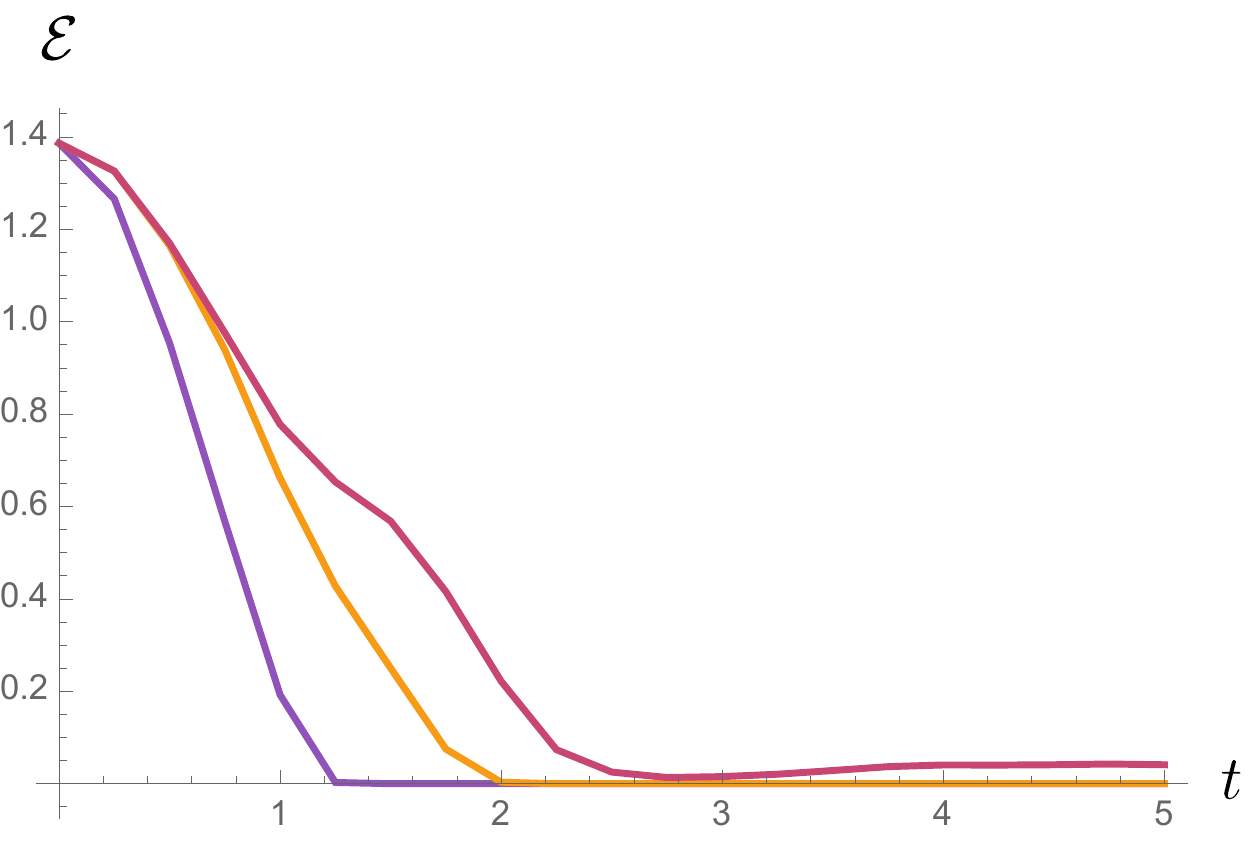}
    \includegraphics[height = 3.25cm]{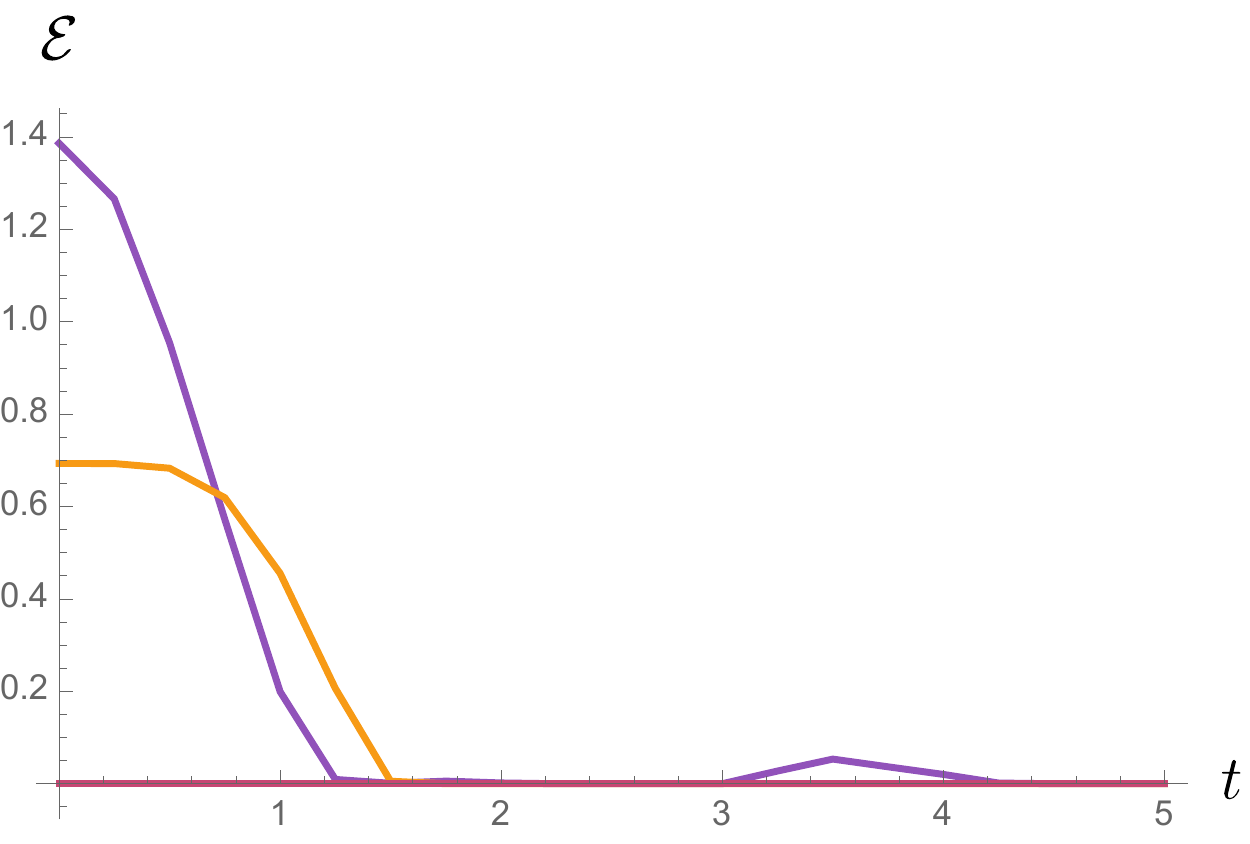}
    \includegraphics[height = 3.25cm]{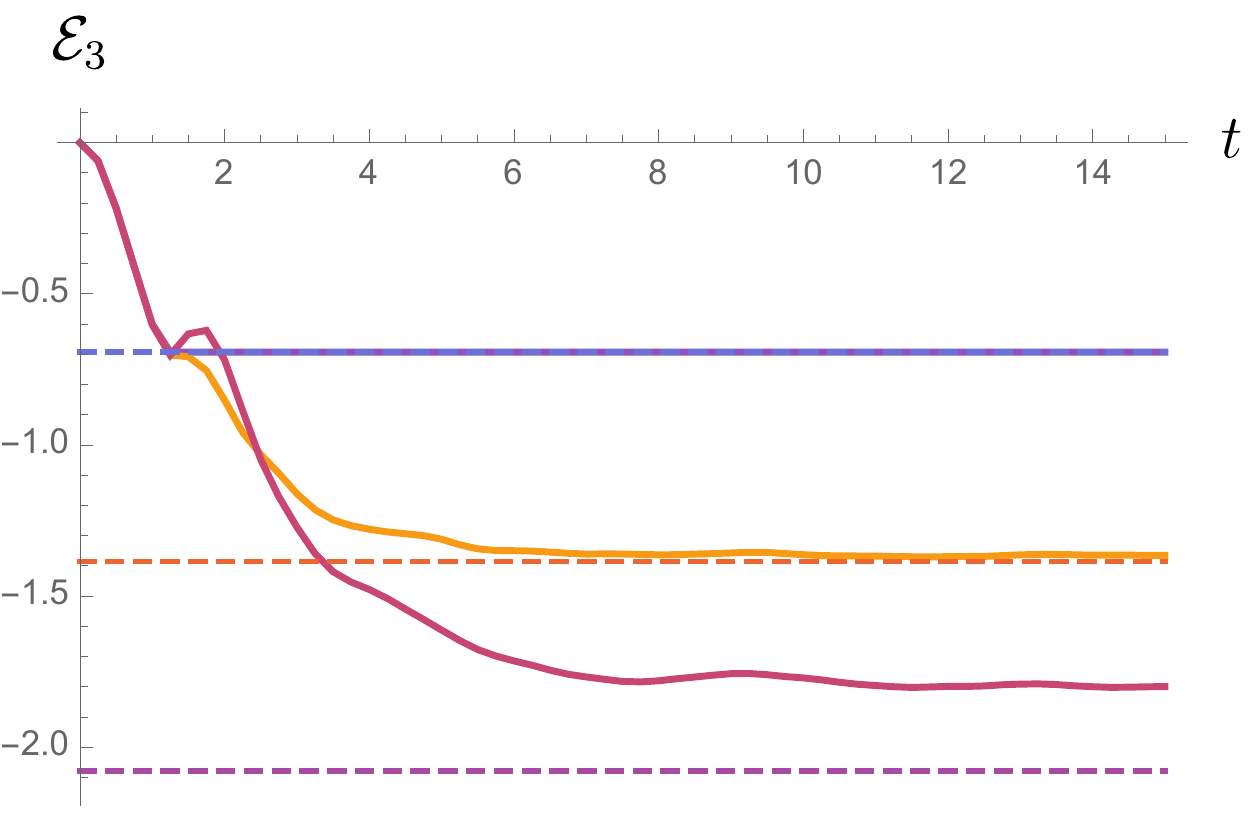}\\
    \includegraphics[height = 3.25cm]{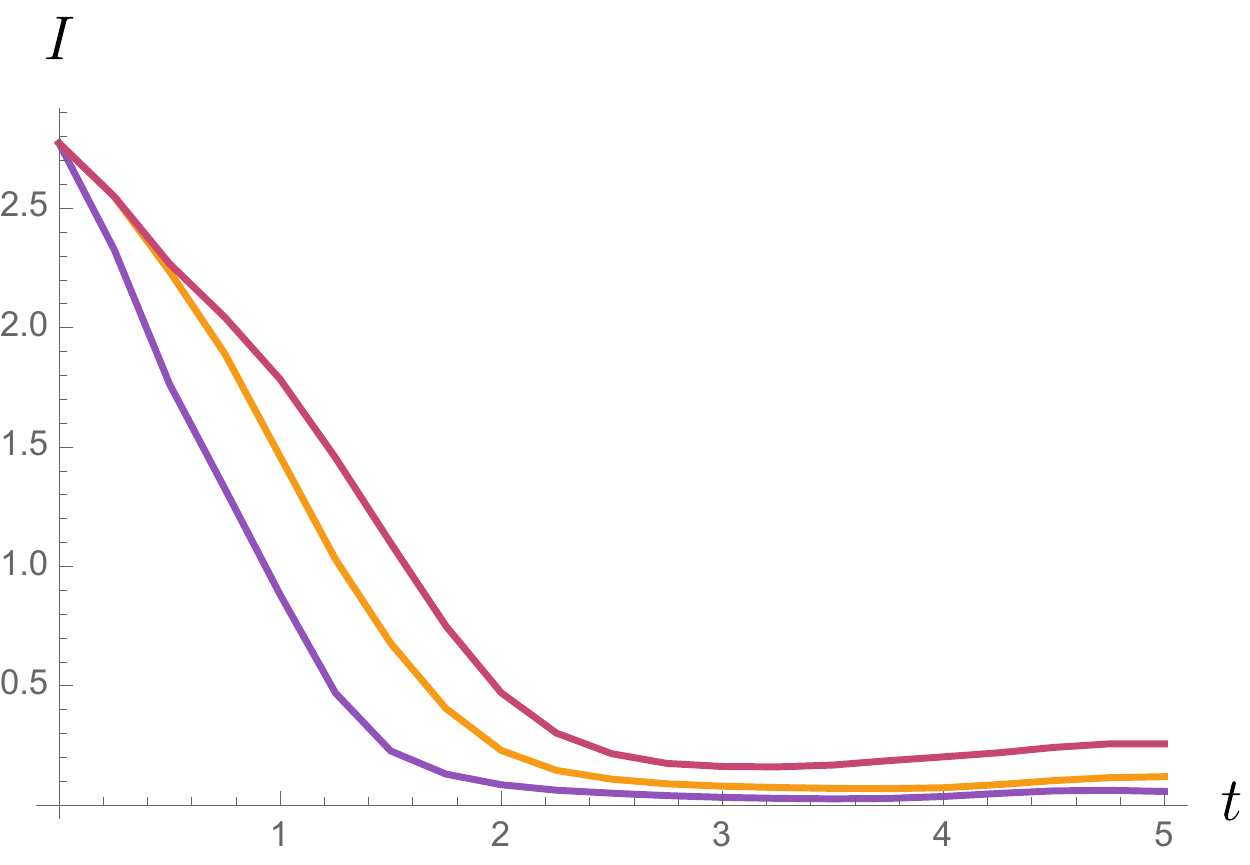}
    \includegraphics[height = 3.25cm]{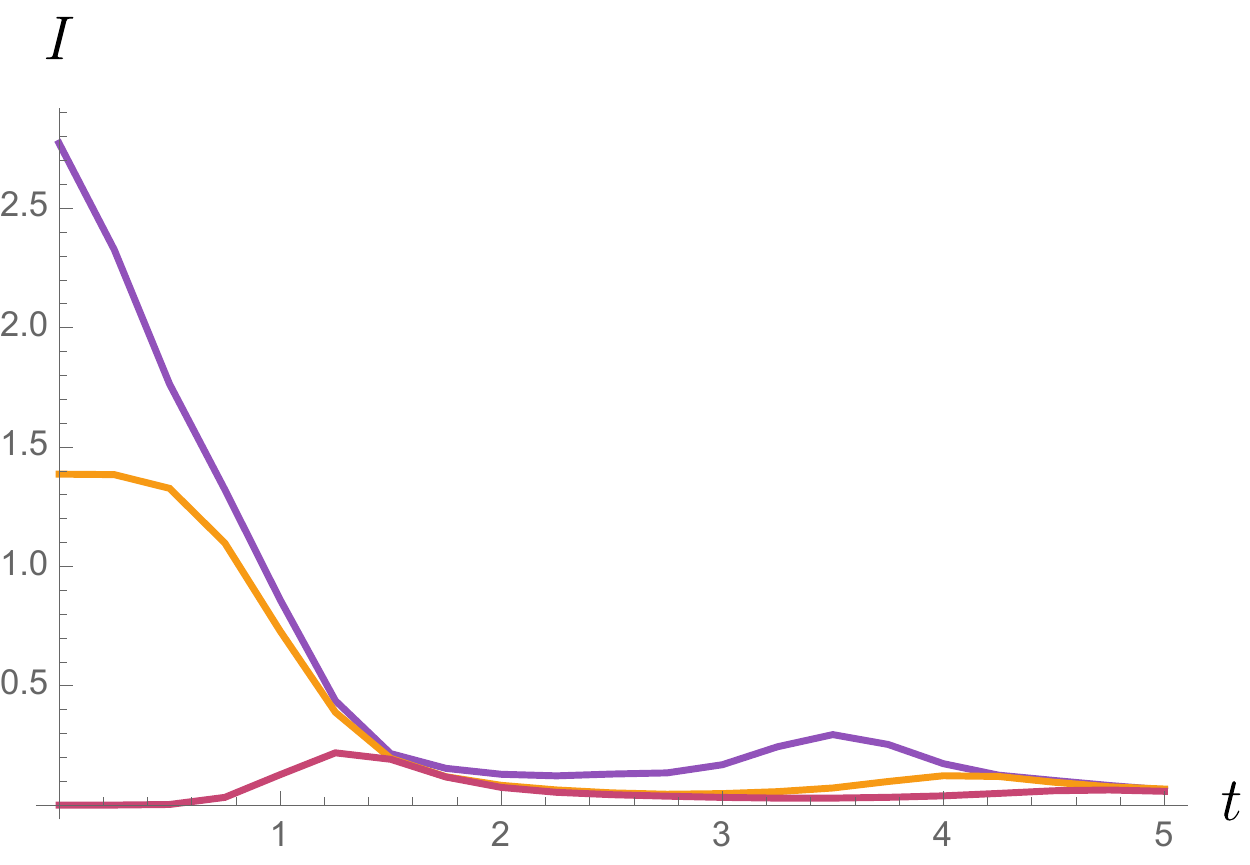}
    \includegraphics[height = 3.25cm]{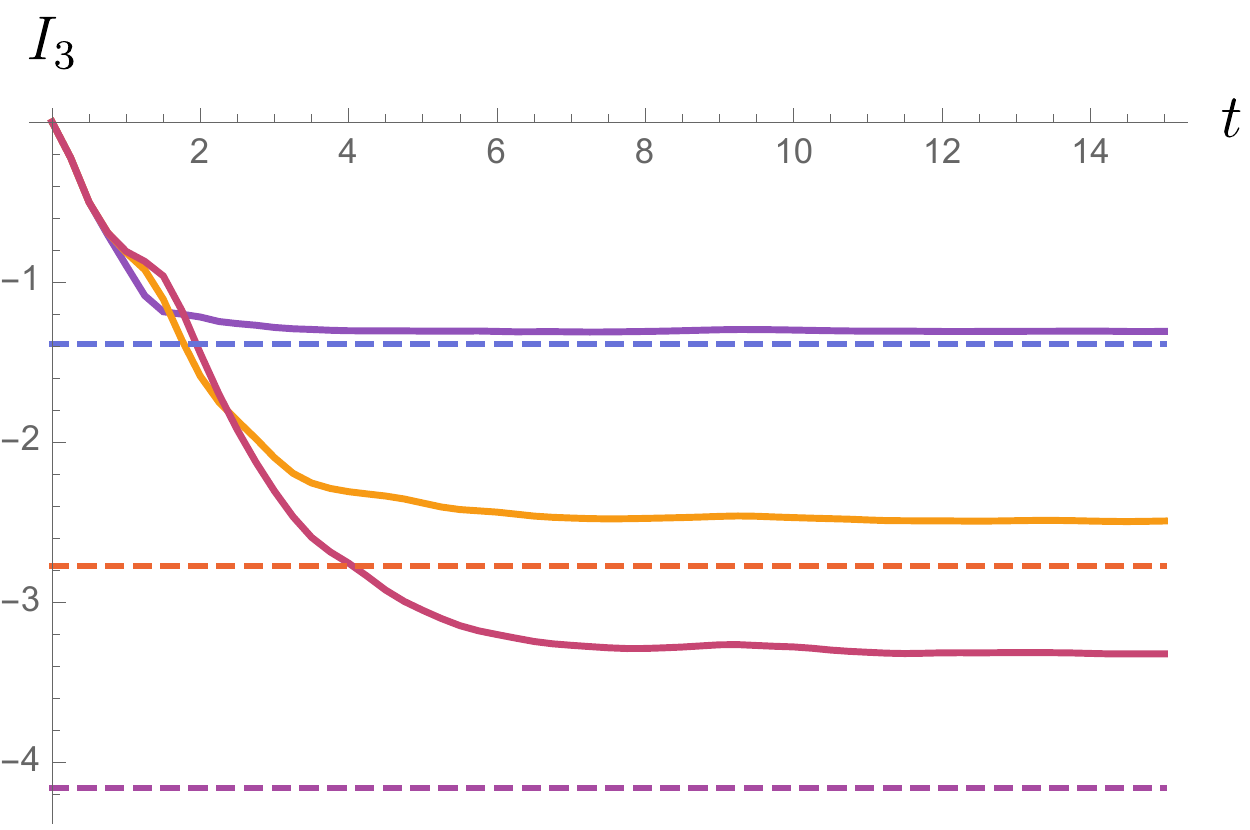}
    \caption{The operator entanglement is shown for the chaotic spin chain corresponding to Hamiltonian (\ref{sc_H}) with $g = -1.05$, $h=0.5$. (left) the BOLN and BOMI for overlapping intervals. The input interval is length 2 and the output varies as length 2 (purple), 3 (yellow), 4 (magenta) with the same starting position as the input interval. (center) the BOLN and BOMI for intervals of length 2 fully overlapped (purple) partially overlapped (yellow) and adjacent (magenta). (right) the TOLN and TOMI for input intervals of lengths 1 (purple), 2 (yellow), and 3 (magenta). The union of the other two intervals compose the entire output system. The dotted lines are the lower bounds for the TOLN and TOMI. We have seven input and output spins in all simulations.}
    \label{sc_erg}
\end{figure}

We note that the logarithmic negativity experiences non-analytic behavior commonly referred to as ``sudden death'' at finite temperatures \cite{2006OptCo.264..393Y}. This is an artifact of the non-analytic trace norm in the definition of negativity. In Fig.~\ref{sc_erg}, we find that the lower bound on the tripartite logarithmic negativity is saturated for the chaotic system at least for suitably small subsystem size relative to the total system, similar to the behavior of tripartite negativity in holography.

% Furthermore, we calculate the tripartite negativity in the spin chains in Fig.~\ref{sc_erg} and find it has .

\begin{figure}
    \centering
    \includegraphics[height = 3.25cm]{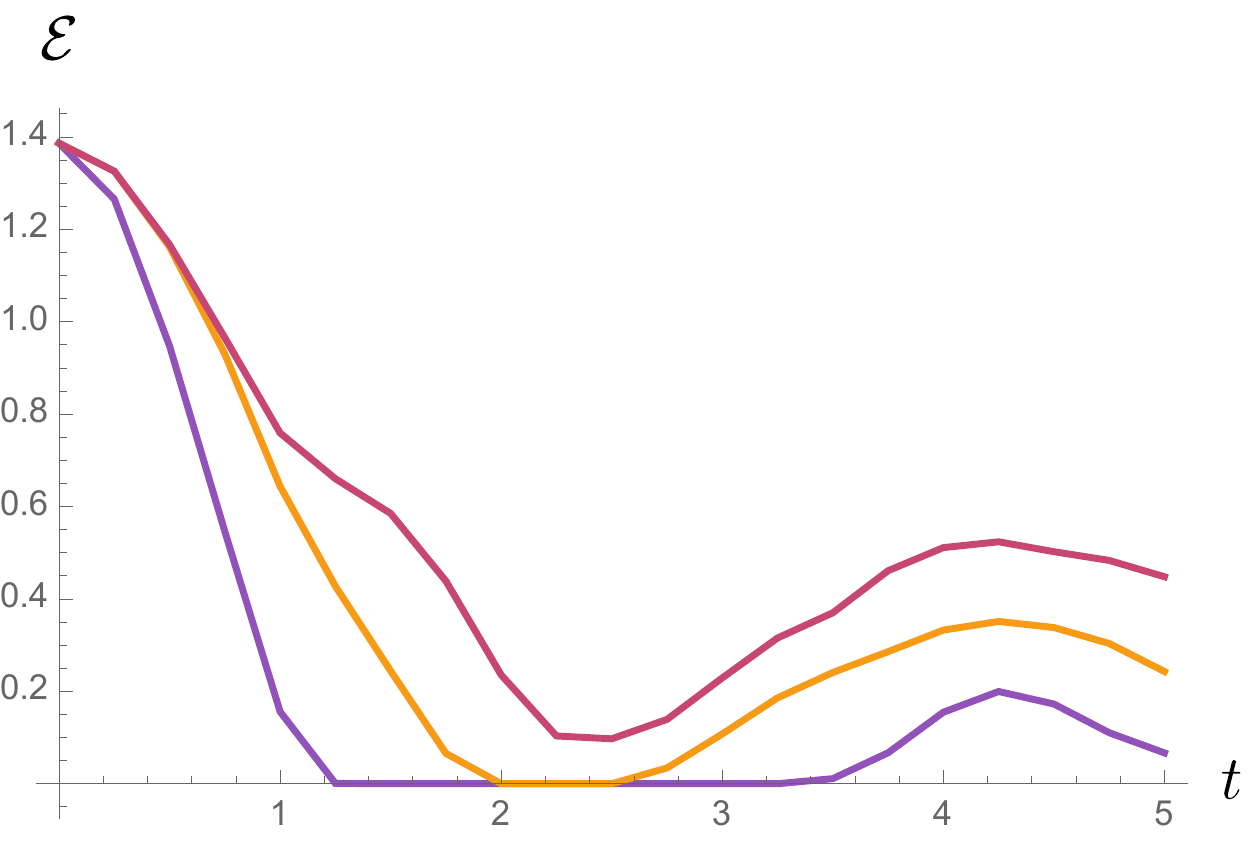}
    \includegraphics[height = 3.25cm]{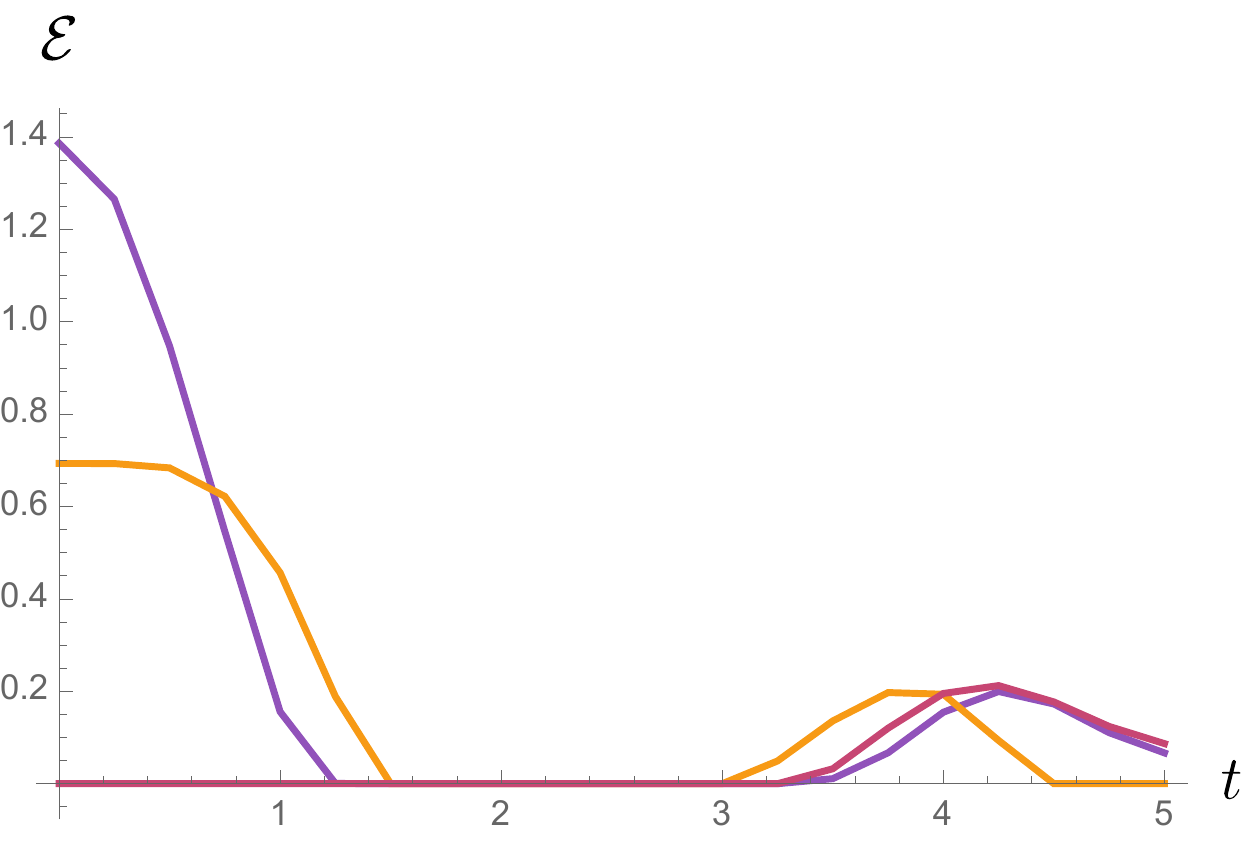}
    \includegraphics[height = 3.25cm]{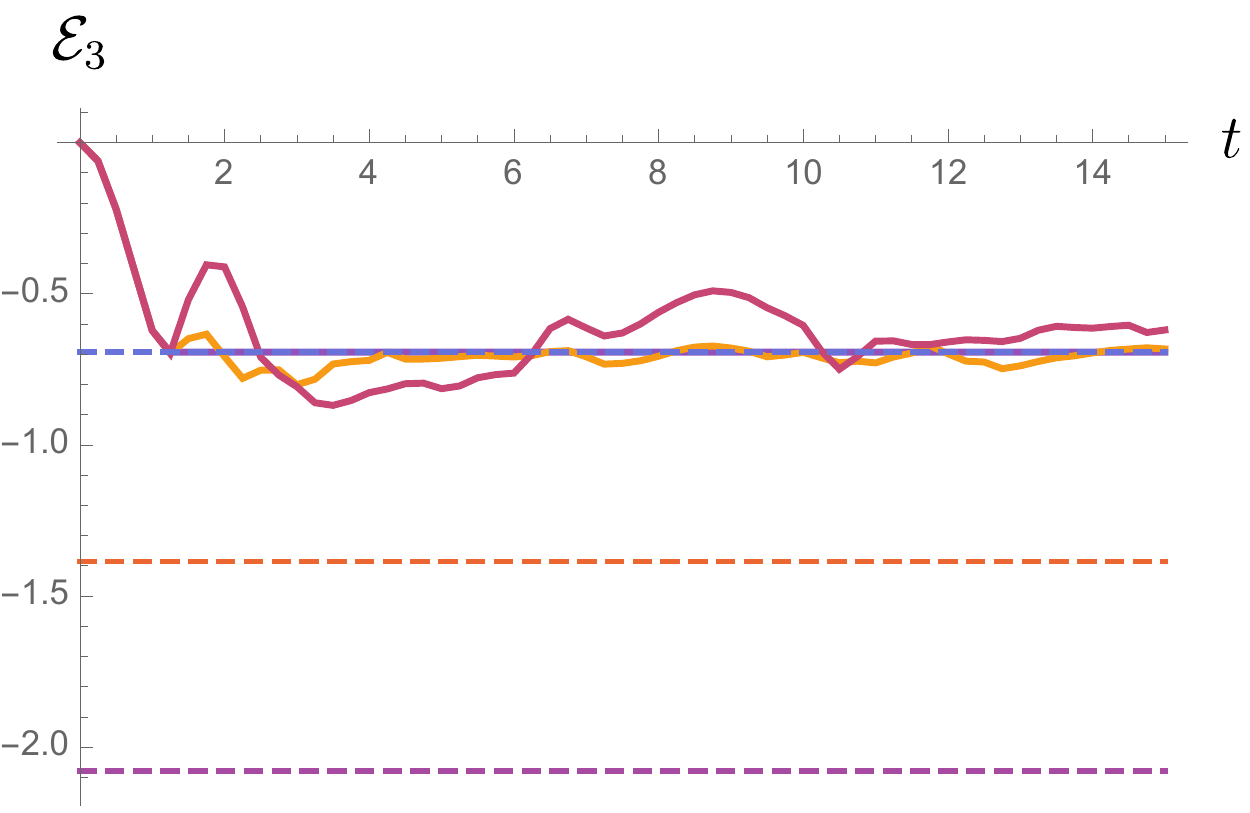}\\
    \includegraphics[height = 3.25cm]{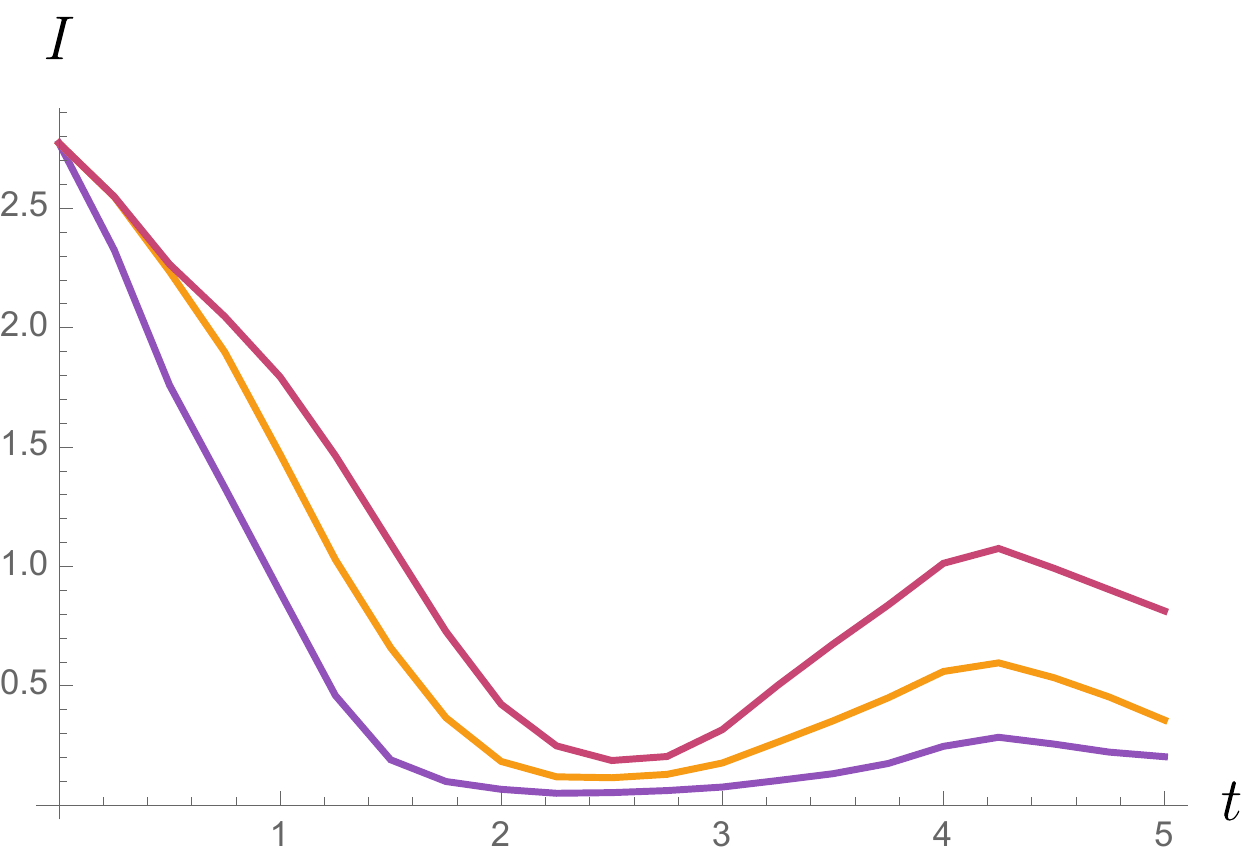}
    \includegraphics[height = 3.25cm]{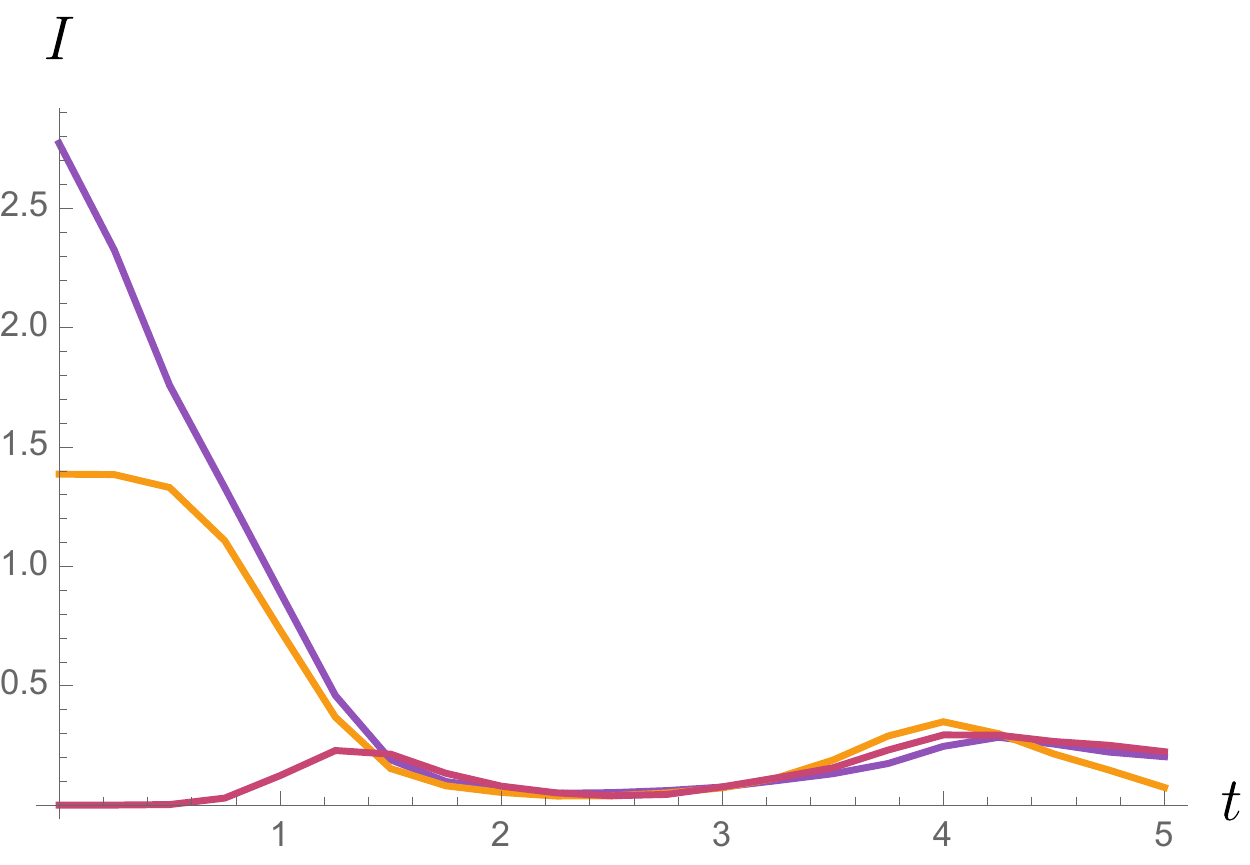}
    \includegraphics[height = 3.25cm]{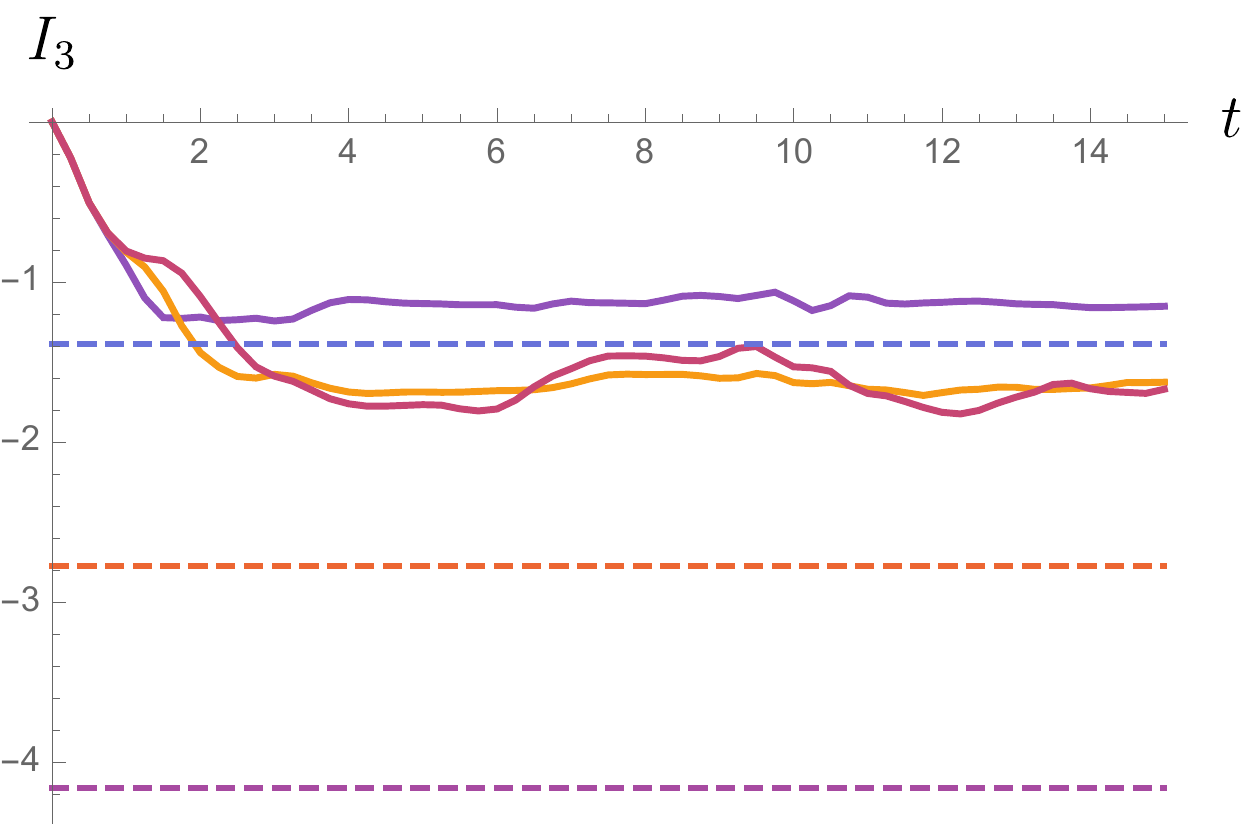} 
    \caption{The operator entanglement for the integrable spin chain (\ref{sc_H}) with $g = 1.0$, $h=0$. We use the same configurations as Fig.~\ref{sc_erg}.}
    \label{sc_int}
\end{figure}

We compare OMI and OLN and find signatures of ``spurious entanglement'' in the OMI at late times. In these situations, OLN appears to be a distinct quantity for characterizing the scrambling. This has practical implications for the ``simulatability'' of many-body systems because there is significantly less entanglement in the system than the mutual information seems to suggest. The classical correlations left over pose no threats to classical computation.

\section{Discussion}
\label{discussion_section}
In this work, we have initiated the study of operator logarithmic negativity to study the scrambling of quantum information in a variety of physical systems. This extends work focused on mutual information, which also captures classical correlations. We hope that this will lead to more studies comparing the negativity and mutual information to better understand the how classical and quantum information differ in dynamical settings. We conclude with some thoughts on future directions.
\subsection{Future directions}

\begin{figure}
\begin{center}
\begin{tabular}{c c c c} 
 & 
\includegraphics[width=1.6in]{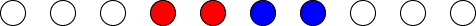} &
\includegraphics[width=1.6in]{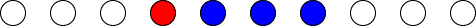} &
\includegraphics[width=1.6in]{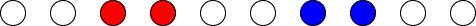}  \\ 
\rotatebox{90}{\hspace{0.2in} \textbf{Integrable}} &
\includegraphics[width=1.7in]{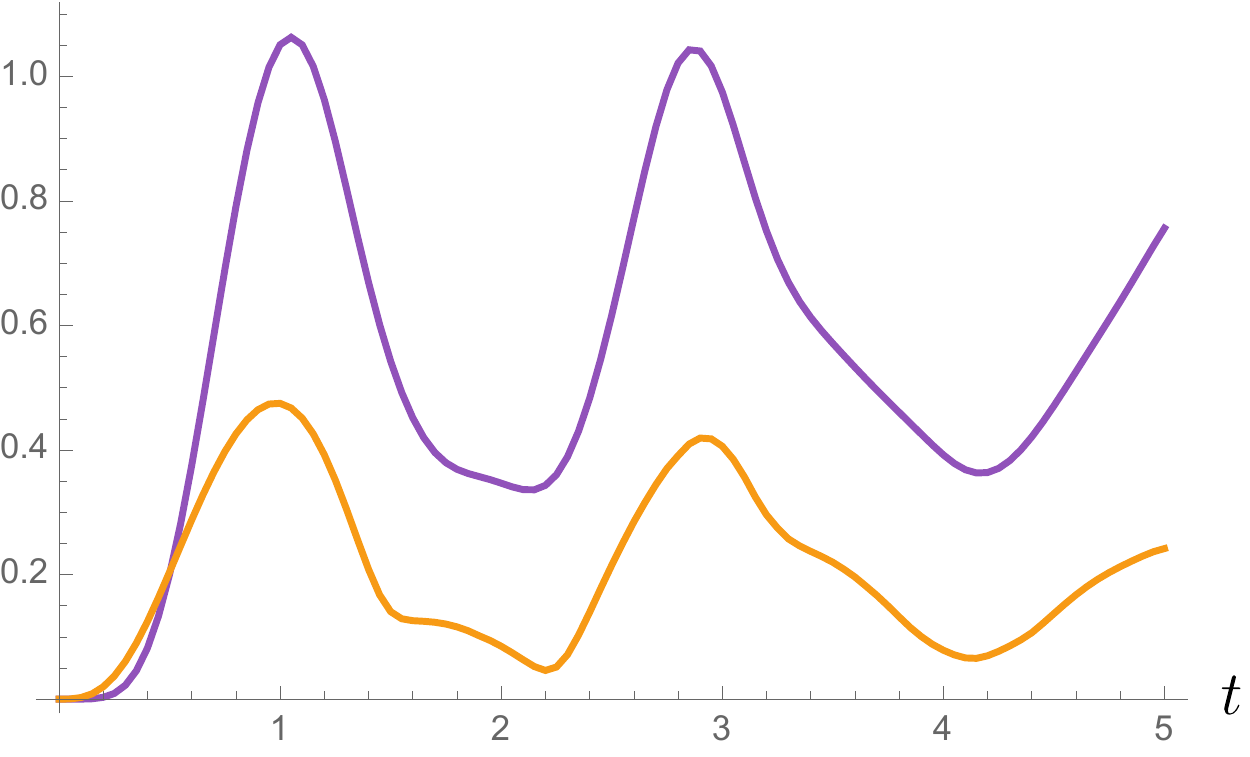} &
\includegraphics[width=1.7in]{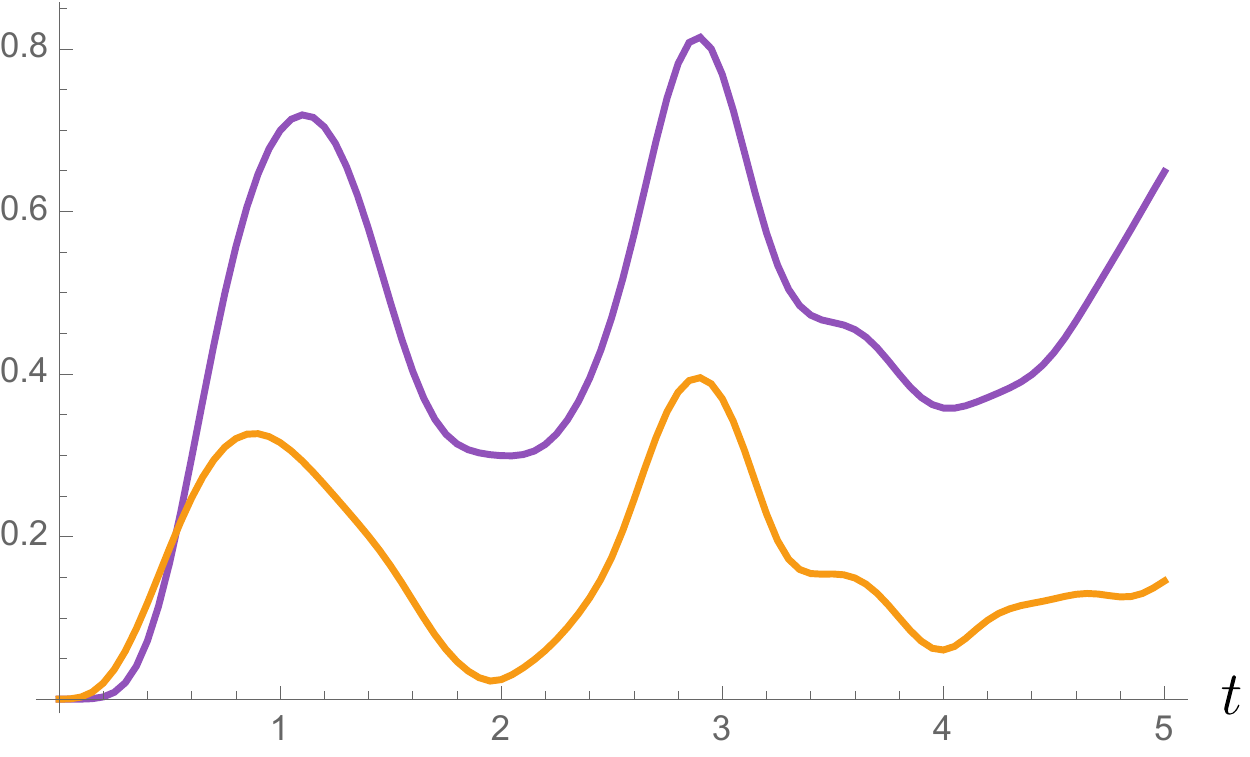} &
\includegraphics[width=1.7in]{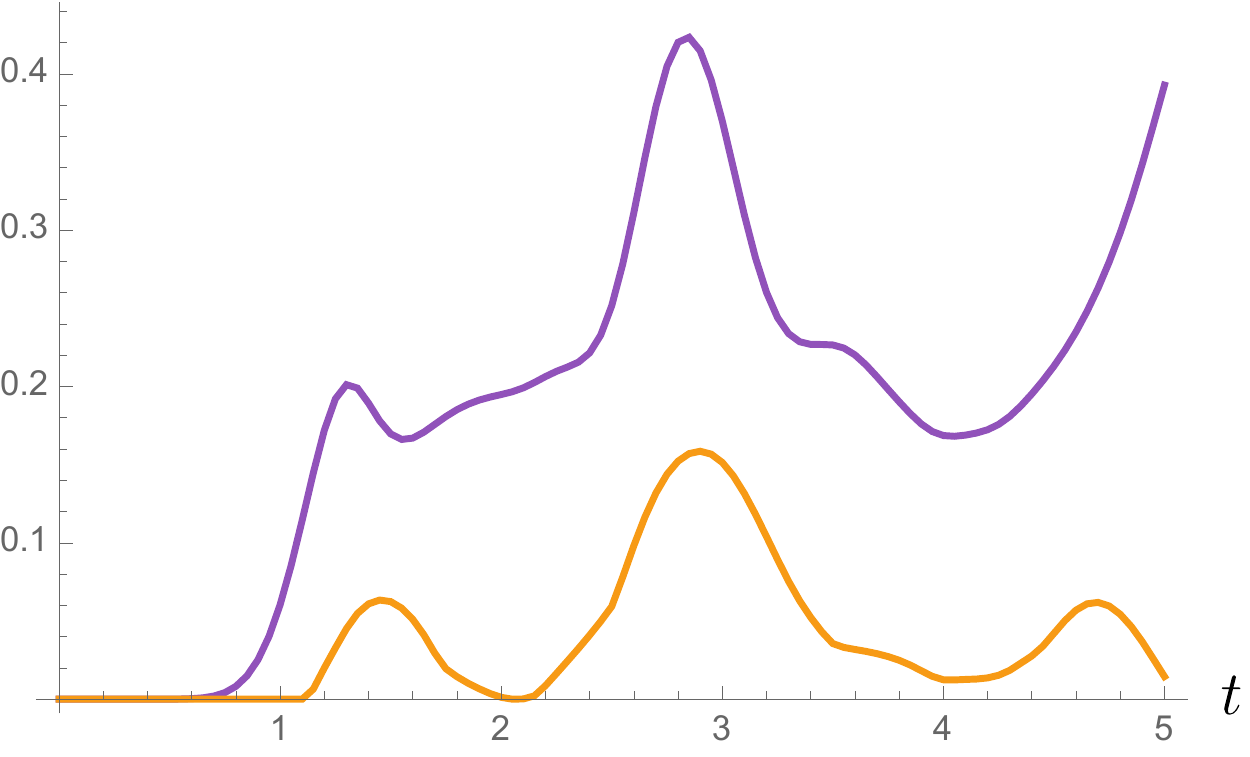} \\ 
\rotatebox{90}{\hspace{0.25in} \textbf{Chaotic}} &
\includegraphics[width=1.7in]{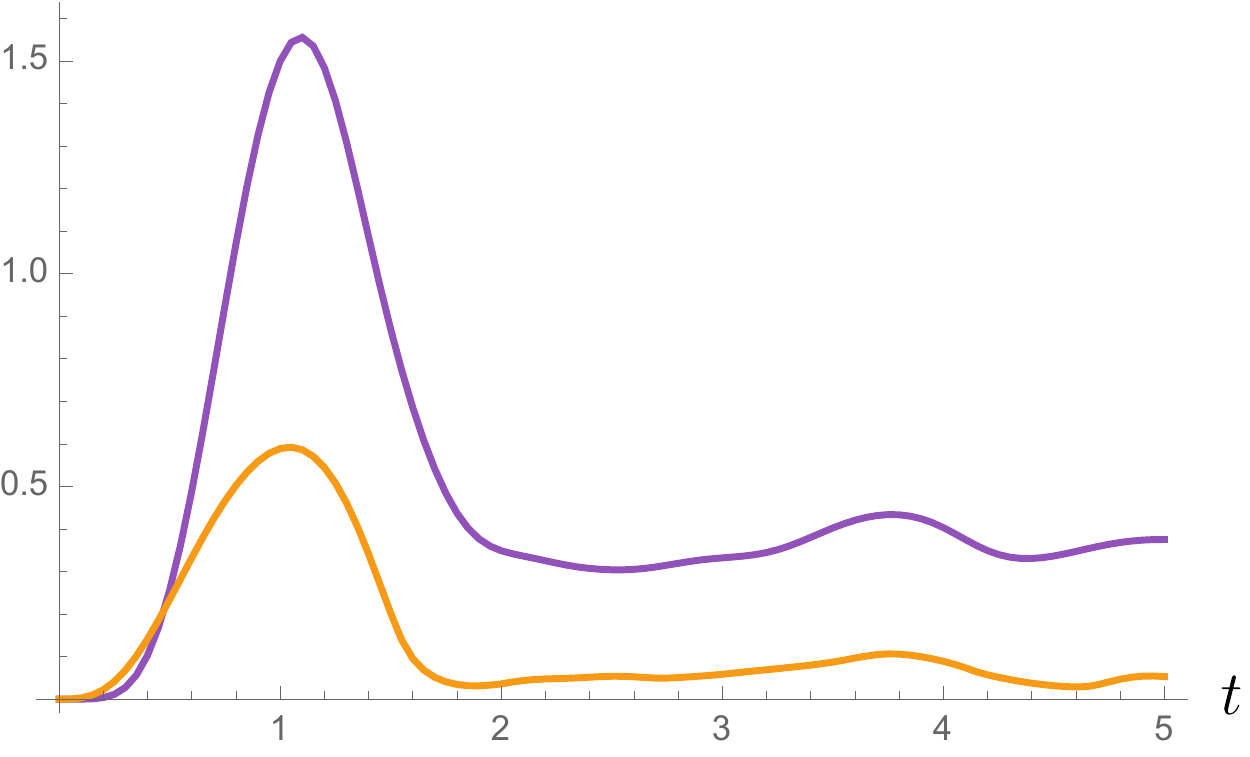} &
\includegraphics[width=1.7in]{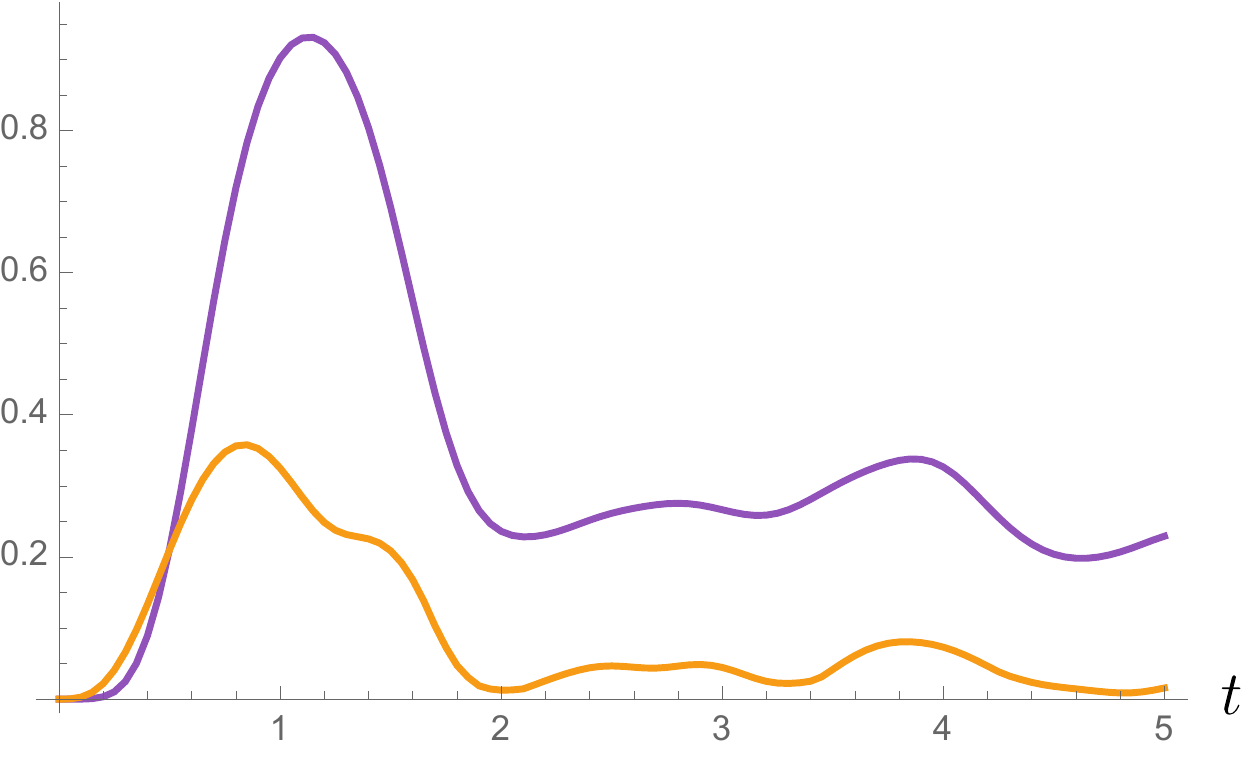} &
\includegraphics[width=1.7in]{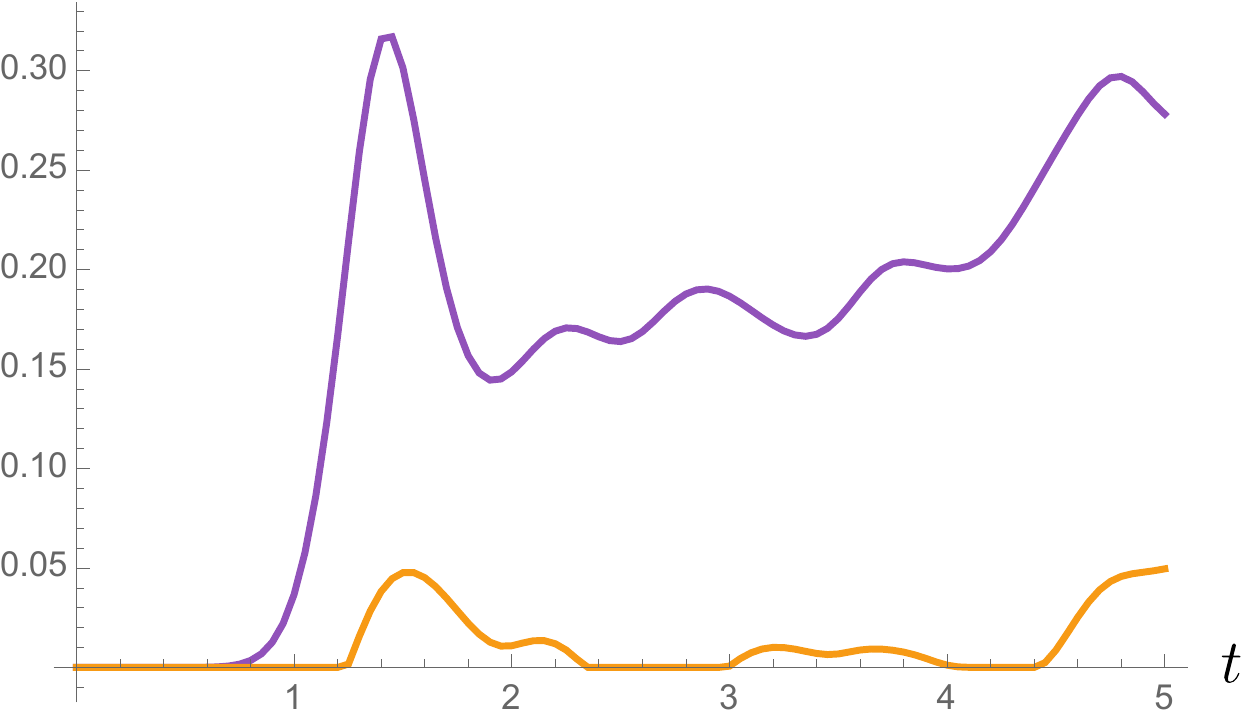} \\ 
\end{tabular}
    \caption{Numerical simulations of a global quench are shown for both integrable and chaotic quench Hamiltonians (\ref{sc_H}). We consider three configurations: adjacent symmetric, adjacent asymmetric and disjoint. These are labeled by the red and blue dots at the top of the columns. We use a 10 spin chain and begin will all spins up in the z-basis. The purple and yellow curves correspond to the mutual information and logarithmic negativity respectively.}
    \label{fig:quench}
\end{center}
\end{figure}

\paragraph{Quantum quench} We have exclusively been discussing the entanglement of operators, so we have never needed to chose a particular state. It would be interesting to study quantum quenches where one starts in an eigenstate of one Hamiltonian and suddenly changes the Hamiltonian, introducing nontrivial dynamics. Logarithmic negativity following quantum quenches has been studies for integrable systems \cite{2014JSMTE..12..017C,2015NuPhB.898...78H,2014NJPh...16l3020E,2015PhRvB..92g5109W,2018arXiv180909119A,2018arXiv181206258F}, though these tend to have identical dynamics as the (R\'enyi) mutual information, as formalized in Ref.~\cite{2018arXiv180909119A}. It would be particularly interesting to compare logarithmic negativity and mutual information for quantum quenches into non-integrable Hamiltonians. As a basic example, we have numerically computed such a quantum quench for the chaotic spin chains from the previous section which can be seen in Fig.~\ref{fig:quench}. Here, we can provide preliminary evidence of the prediction from \cite{2018arXiv180909119A} that the logarithmic negativity of disjoint intervals following a global quench in non-integrable theories is trivial in the space-time scaling limit. Additionally, we see further evidence of spurious entanglement and a violation of the proportionality between logarithmic negativity and (R\'enyi) mutual information that is seen in integrable systems. It would be fascinating to compute logarithmic negativity following quantum quenches in holographic conformal field theories.

\paragraph{Local operator entanglement} While the entanglement of the time evolution operator effectively characterizes the scrambling behavior of the Hamiltonian, it may also be illuminating to investigate the entanglement of Heisenberg evolved local operators. Local operator entanglement has a closer connection to OTOCs and local operators play a central role in the eigenstate thermalization hypothesis (ETH) \cite{1994PhRvE..50..888S}. Thus, one may presume that local operator entanglement may elucidate the connection between scrambling, chaos, and ETH.

\paragraph{Operator entanglement contour/ negativity contour}
While the saturation values of tripartite operator mutual information and negativity characterize the total scrambling ability of the Hamiltonian, they are agnostic to the fine-grained dynamics of the entanglement. The entanglement contour \cite{2014JSMTE..10..011C} is a quasi-local notion of entanglement that may be useful for understanding dynamics of quantum and classical information locally. For example, it has been studied following quantum quenches in both integrable and non-integrable conformal field theories \cite{2014JSMTE..10..011C,2019JPhA...52F5401K,2019arXiv190501144D}. It would thus be interesting to study (state-independent) scrambling with the entanglement contour (or negativity contour \cite{deNobili_thesis, 2019JPhA...52F5401K, 2019arXiv190807540K}). It is furthermore conceivable that the operator entanglement contour is a useful setting for studying the effective hydrodynamics of entanglement as it is a locally conserved ``fluid.''

\section*{Acknowledgements}

We thank Nicholas Hunter-Jones, Hassan Shapourian, Kotaro Tamaoka, and Xiao-Liang Qi for useful discussions.
SR thanks Tianci Zhou for a discussion on the line-tension picture
and its connection to the holographic calculations. 
SR is supported by a Simons Investigator Grant
from the Simons Foundation. MN is supported by JSPS Grant-in-Aid for Scientific Research (Wakate) No.19K14724, RIKEN iTHEMS Program, and the RIKEN Special Postdoctoral Researcher program. We thank the Yukawa Institute for Theoretical Physics (YITP) for hospitality during the completion of this work.

\appendix

\section{Logarithmic negativity at large central charge}
\label{app_neg_blocks}

In this appendix, we review the conjecture for the holographic dual of logarithmic negativity \cite{2019PhRvD..99j6014K} and derive new results using geodesic Witten diagrams. Unfortunately, the geometric dual for logarithmic negativity is more complicated that the dual to entanglement entropy due to gravitational backreaction in the bulk. This is a consequence of the logarithmic negativity reducing to the R\'enyi entropy at index $1/2$ when computed in pure states
\begin{align}
    \mathcal{E}\left(\rho_{AB}^{T_B}\right) = S_{1/2}\left( \rho_{A} \right) =S_{1/2}\left( \rho_{B} \right).
\end{align}
Replica numbers not equal to unity were shown to correspond to backreacted geometries in the bulk sourced by cosmic branes with tension \cite{2016NatCo...712472D}
\begin{align}
    T_n = \frac{n-1}{4 n G_N}.
    \label{tension}
\end{align}
Dong showed that a family of functions closely related to the R\'enyi entropies, called the \textit{modular entropies}, defined as
\begin{align}
   \tilde{S}_n \equiv  n^2 \partial_n \left( \frac{n-1}{n }S_n\right),
\end{align}
satisfy holographic area laws
\begin{align}
    \tilde{S}_n  = \frac{\mbox{Area(Cosmic Brane)}_n}{4 G_N}.
\end{align}
With this motivation, along with derivations of logarithmic negativity in quantum error correcting codes \cite{2019PhRvD..99j6014K}, the authors proposed that the holographic dual of \textit{modular logarithmic negativities} are branes, $E_{W,n}$, that have boundary conditions on the boundary of the entanglement wedge and tension (\ref{tension})
\begin{align}
    \tilde{\mathcal{E}}_{n}  = \frac{E_{W,n}}{4 G_N}.
\end{align}
% where the brane's tension is parametrized as
% \begin{align}
%     T_{n} = \frac{n-1}{4n G_N}.
% \end{align}
% Note the difference from the modular R\'enyi tensions\footnote{Further discussion of this and its origin in operator algebra quantum error correcting codes is the topic of ongoing work.}.
The modular negativities are related to the moments\footnote{Note that these are not quite the same moments described in earlier sections.} of the logarithmic negativity by
\begin{align}
   \tilde{\mathcal{E}}_{n} \equiv  n^2 \partial_{n} \left( \frac{n-1}{n }\mathcal{E}_{n}\right). 
\end{align}
The holographic negativity is then the limit
\begin{align}
    \mathcal{E} = \lim_{n \rightarrow 1/2} \mathcal{E}_{n}.
    \label{main_full}
\end{align}
There are many interesting limits where the backreaction can be accounted for accurately without having to explicitly solve Einstein's equations with a codimension-two source. In particular, these occur when the subregions are ball-shaped and the state is translationally invariant. Then, the holographic logarithmic negativity can be approximated by
\begin{align}
    \mathcal{E} = \mathcal{X}_d\frac{E_W}{4 G_N},
    \label{holo_neg_sphere}
\end{align}
where $\mathcal{X}_d$ is a constant that depends on the dimension of the entangling surface sphere
\begin{align}
  \mathcal{X}_d = \frac{1}{2}x_d^{d-2}\left(1 + x_d^2\right) - 1, 
\quad
   x_d = \frac{2}{d} \left(1 + \sqrt{1 - \frac{d}{2} + \frac{d^2}{4}}\right) 
\end{align}
for entangling surface $S^{d-1}$. For intervals in two dimensional CFTs, $\mathcal{X}_2 = 3/2$. The consistency of the holographic conjecture has been confirmed for single, adjacent, and disjoint intervals in the vacuum, single and adjacent intervals at finite temperature, and the thermofield double state. While the gravitational calculation for disjoint intervals is straightforward, the conformal field theory computation has some interesting subtleties and mysteries that were originally studied in Ref.~\cite{2014JHEP...09..010K}. We will look into these further after reviewing results in the simpler cases.

For a single interval at zero temperature, the negativity is computed via the following correlator of double twist fields
\begin{align}
    \mathcal{E} &= \lim_{n_e\rightarrow 1} \log \langle\sigma^2_{n_e}({x}_1) \bar{\sigma}^2_{n_e}({x}_2)\rangle. 
\end{align}
This is fixed by conformal invariance and is equivalent to the R\'enyi entropy at index 1/2
\begin{align}
    \mathcal{E} = \frac{c}{2}\log \left(\frac{|x_2 - x_1|}{\epsilon} \right).
\end{align}
When progressing to finite temperature, we cannot simply apply the conformal map from the complex plane to the cylinder, $w= e^{2\pi z/\beta}$. Rather, the negativity must be computed by the following four-point function \cite{2015JPhA...48a5006C}
\begin{align}
    \mathcal{E} = \lim_{L \rightarrow \infty} \lim_{n_e \rightarrow 1} \log \langle \sigma_n(-L) \bar{\sigma}^2_n(-l) \sigma^2_{n}(l) \bar{\sigma}_n(L)\rangle_{\beta},
    \label{single_correlator}
\end{align}
where the order of limits is important. This function is not universal and depends on the full operator content of the theory
\begin{align}
    \mathcal{E} = \frac{c}{2}\log \left(\frac{\beta}{\pi \epsilon}\sinh \frac{\pi l }{\beta} \right) - \frac{\pi c l}{2 \beta}+ g\left(e^{-2\pi l/\beta}\right),
\end{align}
where $g(x)$ is a theory dependent function. In Ref.~\cite{2019PhRvD..99j6014K}, the authors showed that in holographic CFTs, this reduces to 
\begin{align}
    \mathcal{E} = \frac{c}{2}\min \left[\log \left(\frac{\beta}{\pi \epsilon}\sinh \frac{\pi l }{\beta} \right), \log \left(\frac{\beta}{\pi \epsilon} \right) \right].
\end{align}
This was derived in the s-channel (large $l/\beta$) based on the assumption that we take the dominant conformal block as
\begin{align}
    \mathcal{F}_W = x^{c/8},
\end{align}
which is found by taking an equally weighted combination of the two solutions to the monodromy problem that we will review shortly \cite{2014JHEP...09..010K}. The entanglement wedge cross section is simply found to be
\begin{align}
        E_W = 2 \min \left[\log \left(\frac{\beta}{\pi \epsilon}\sinh \frac{\pi l }{\beta} \right), \log \left(\frac{\beta}{\pi \epsilon} \right) \right]
\end{align}
which precisely agrees with the CFT results after using (\ref{main}) and the Brown-Henneaux formula, $c = 3R/2G_N$ \cite{brown1986}.

The negativity of disjoint intervals in flat space is calculated from the four-point function
\begin{align}
    \mathcal{E} = \lim_{n_e\rightarrow 1}\log \langle \sigma_{n_e}(x_1)\bar{\sigma}_{n_e}(x_2)\bar{\sigma}_{n_e}(x_3)\sigma_{n_e}(x_4)\rangle.
    \label{disjoint_correlator}
\end{align}
This is universally solvable in the limit of adjacent intervals where it becomes
\begin{align}
        \mathcal{E} &= \lim_{n_e\rightarrow 1}\log \langle \sigma_{n_e}(x_1)\bar{\sigma}^2_{n_e}(x_2)\sigma_{n_e}(x_4)\rangle \\ \nonumber
        &= \frac{c}{4}\log \left(\frac{\left(x_4 - x_2 \right)\left(x_2 - x_1 \right)}{\epsilon \left( x_4 - x_1\right)} \right),
\end{align}
where we have introduced the regulator $\epsilon$. Similarly to the finite temperature case, we can solve the monodromy problem \cite{2016JHEP...01..146H}, use the series expansion, or explicitly calculate the geodesic Witten diagram to find
\begin{align}
\label{wedge_block}
  \mathcal{F}_W(x)
  &= \left(\frac{1}{4}\frac{1 + \sqrt{x}}{1-\sqrt{x}} \right)^{c/8},
                      \quad
    x = \frac{(x_4 - x_3)(x_2 - x_1)}{(x_3 - x_1)(x_4 - x_2)}.
\end{align}
This leads to a negativity of 
\begin{align}
    \mathcal{E} = \begin{cases} 0 & 0 < x < 1/2 \\ \frac{c}{4}\log  \left(\frac{1}{4}\frac{1 + \sqrt{x}}{1-\sqrt{x}} \right)
    \label{neg_disjoint} &1/2 <  x < 1 .
    \end{cases}
\end{align}

The minimal entanglement wedge cross-sectional area for disjoint intervals $(x_1, x_2)$ and $(x_3, x_4)$ is 
\begin{align}
  E_W &= \begin{cases}\log \left(1 + 2 y  + 2\sqrt{y (y+1)}\right)
  \label{two_ints},
 & y > 1, \\ 0 & y < 1,
    \end{cases} \\ \nonumber
      \quad
    y &= \frac{x}{1-x}.
\end{align}
This precisely matches the CFT results for both disjoint and adjacent intervals. 

We now study the origin of these conformal blocks that we have implicitly used for disjoint intervals and a single interval at finite temperature. As a fundamental building block of conformal field theories, conformal blocks have been the focus of intense study for many years. In particular, multiple techniques for computing conformal blocks at large central charge have been developed including Zamolodchikov's recursion relation \cite{zamolodchikov1984,1987TMP....73.1088Z}, the monodromy method \cite{1987TMP....73.1088Z,2005NuPhB.724..529H,2011JHEP...12..071H,1996NuPhB.477..577Z}, and geodesic Witten diagrams \cite{2016JHEP...01..146H, 2015JHEP...12..077H}. In this appendix, we review certain peculiarities arising in the computation of the conformal block relevant to logarithmic negativity for disjoint intervals \cite{2014JHEP...09..010K} and present some new peculiarities that are the focus of future work.

\paragraph{Series Expansion}We are interested in the conformal block that has external weights $h_i = \frac{c}{24}\left(n - \frac{1}{n} \right)$ and internal weight $h_p = \frac{c}{12}\left(\frac{n}{2} - \frac{2}{n} \right)$ at large central charge. The series expansion for the conformal block begins as \cite{Belavin:1984vu}
\begin{align}
    \label{expansion}
    \mathcal{F}(h_i, h_p, x) = x^{h_p- 2h_i}\left(1 + \frac{1}{2}h_p x + \frac{h_p(h_p+1)^2 }{4(2h_p+1)} x^2 + \frac{\left(h_p (1-h_p)-2 h_i(2h_p+1) \right)^2}{2(2h_p+1)(c(2h_p + 1)+2h_p(8h_p-5))}x^2 + \dots \right)
\end{align}
The series expansion takes into account all descendent operators of the primary $h_p$, hence the backreaction. In the replica limit, 
\begin{align}
    h_i \rightarrow 0, \quad h_p \rightarrow -\frac{c}{8}.
\end{align}
It ends up being important whether we take the limit of large c or the replica limit first\footnote{The same order of limits is also necessary for the holographic proposals for odd entropy \cite{2018arXiv180909109T} and reflected entropy \cite{2019arXiv190500577D}.}.  The limits do not commute. If we take c to be large for generic replica number and collect terms of order $(cx)^n$, then the series expansion precisely matches the conformal block computed from the entanglement wedge cross section (\ref{wedge_block})
\begin{align}
    \mathcal{F}\left(0, -\frac{c}{8},x\right) = x^{-c/8}\left( 1 - \frac{cx}{16}+ \frac{c^2x^2}{512} - \frac{c^3x^3}{24576}+ \dots\right) .
\end{align}
However, taking $h_p = -c/8$ first, the third and fourth terms of (\ref{expansion}) both become order $(c x)^2$ and divergent terms appear that are associated to the null state at level two in the limit of large c
\begin{align}
    L_{-2} \ket{\sigma_{n_e}^2} = 0, \quad n_e \rightarrow 1.
\end{align}
For any value of the central charge, the above state is of zero norm, though not necessarily a null vector. For generic values of $h_p$, the third term is of order $(cx)^2$ and the fourth is $cx^2$.

\paragraph{Monodromy Method}Alternative to the series expansion, the monodromy problem has been solved numerically for this conformal block at large c \cite{2014JHEP...09..010K}. The monodromy method is derived from Liouville theory at large c. In the semiclassical limit, $c\rightarrow \infty, h_i/c$ fixed, Virasoro blocks approximately exponentiate as
\begin{align}
    \mathcal{F}(h_i, h_p, x) \simeq \exp \left[-\frac{c}{6} f\left(\frac{h_p}{c}, \frac{h_i}{c}, x \right) \right] .   
\end{align}
$f(x)$ can then be computed by considering the differential equation
\begin{align}
    \label{mono_deq}
    \psi''(z) + T(z) \psi(z) = 0
\end{align}
with 
\begin{align}
    \label{mono_stress}
    T(z) = \sum_i \left(\frac{6h_i/c}{(z-x_i)^2} - \frac{c_i}{z-x_i}\right),
\end{align}
where the $x_i$'s are the positions of the external operators and the $c_i$'s are accessory parameters which are constrained by having $T(z)$ fall off as $x^{-4}$ at infinity, imposing
\begin{align}
    \sum_i c_i = 0, \quad \sum_{i} \left(c_ix_i -\frac{6h_i}{c} \right) = 0, \quad \sum_i \left(c_i x_i^2 -\frac{12h_i}{c}x_i \right)=0.
\end{align}
We can then write (\ref{mono_stress}) involving only a single independent accessory parameter
\begin{align}
    \frac{c}{6}T(z) = \frac{h_1}{z^2} + \frac{h_2}{(z-x)^2}+ \frac{h_3}{(1-z)^2} + \frac{h_1 + h_2+ h_3-h_4}{z(1-z)}-\frac{c}{6}\frac{c_2(x) x (1-x)}{z(z-x)(1-z)}
\end{align}
where $x$ is the cross ratio. The two solutions to (\ref{mono_deq}) must undergo a certain monodromy when encircling singular points
\begin{align}
    \begin{pmatrix}
    \psi_1 \\ \psi_2
    \end{pmatrix}
    \rightarrow
    M     \begin{pmatrix}
    \psi_1 \\ \psi_2
    \end{pmatrix}.
\end{align}
The trace of the monodromy matrix must be 
\begin{align}
    \Tr M = -2\cos \pi \Lambda_p, \quad h_p = \frac{c}{24}(1- \Lambda_p^2).
\end{align}
This fixes the remaining independent accessory parameter leading to 
\begin{align}
    \frac{\partial f}{\partial x} = c_2 (x).
\end{align}
Contrary to the usual computations, two solutions for the accessory parameter, $c_2(x)$, are found that satisfy the asymptotic condition
\begin{align}
    c_2(x) \simeq -\frac{3}{4}\frac{1}{1-x}, \quad x\rightarrow 1.
\end{align}
Interestingly, the solution corresponding to the entanglement wedge cross-section is precisely centered between the two solutions
\begin{align}
   x(1-x) c_{2}^{\pm}(1-x) =  -\frac{3}{4}\left( 1 - \left(\frac{1}{2}\pm \frac{1}{4}\right) (1-x) + \dots \right)
\end{align}
where the accessory parameter corresponding to conformal block ($\ref{wedge_block}$) is
\begin{align}
       x(1-x) c_{2}^{\pm}(1-x) =  -\frac{3}{4}\left( 1 - \frac{1}{2} (1-x) + \dots \right).
\end{align}
We have numerically solved the monodromy problem with the cross ratio off the real axis, which is important for operator entanglement, and have found the same two solutions to hold. 

\paragraph{Geodesic Witten diagrams}

We progress with new computations using the geodesic Witten diagram prescription for computing conformal partial waves \cite{2016JHEP...01..146H}. Much of this section is inspired by Refs.~\cite{2018PTEP.2018f3B03H, 2019arXiv190210161P}. We are concerned with the correlator
\begin{align}
    \langle\sigma_{n}(x_1)\bar{\sigma}_{n}(x_2)\bar{\sigma}_{n}(x_3)\sigma_{n}(x_4)\rangle.
\end{align}
There are two ways of taking the replica limit, taking $n$ to be even or odd. These correspond to calculating logarithmic negativity and odd entropy. We will do both computations simultaneously. We expand the correlation function in terms of partial waves
\begin{align}
    \langle\sigma_{n}(x_1)\bar{\sigma}_{n}(x_2)\bar{\sigma}_{n}(x_3)\sigma_{n}(x_4)\rangle = \sum_{\Delta, l} C_{\Delta,l}^{12}C_{\Delta, l}^{34}W_{\Delta, l}(x_i).
\end{align}
In the following,
%however, 
we only take the dominant channel. 

\begin{figure}
    \centering
    \includegraphics[height = 4cm]{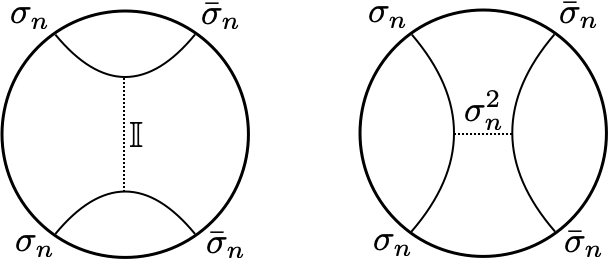}
    \caption{Depending on the cross-ratio, the dominant conformal block will either exchange the identity operator (left) or the double twist operator (right). This is a familiar story to the entanglement entropy, though there, the identity operator is exchanged in both channels. }
    \label{fig:gwd}
\end{figure}

For a scalar exchange
($l=0$), 
the partial wave is
\begin{align}
    W_{\Delta}(x_i) &= \frac{2 \Gamma(\Delta)}{\Gamma\left( \frac{\Delta+\Delta_{12}}{2}\right)\Gamma\left( \frac{\Delta-\Delta_{12}}{2}\right)} \frac{2 \Gamma(\Delta)}{\Gamma\left( \frac{\Delta+\Delta_{34}}{2}\right)\Gamma\left( \frac{\Delta-\Delta_{34}}{2}\right)} 
    \\ \nonumber &\times \int_{\gamma_{ij}}\int_{\gamma_{kl}}G_{b \partial}(y(\lambda),x_i) G_{b \partial}(y(\lambda),x_j)G_{bb}^{\Delta}(y(\lambda), y'(\lambda'))G_{b \partial}(y'(\lambda'),x_k)G_{b \partial}(y'(\lambda'),x_l),
\end{align}
where the integration is only over the geodesics connecting the boundary points 
(see Fig.\ \ref{fig:gwd}).
In Euclidean Poincare coordinates,
the bulk-bulk and bulk-boundary correlators are given by
\begin{align}
    G_{bb}^{\Delta}(y,y') &= e^{-\Delta\sigma(y,y')} {}{}_2F_1\left(\Delta, \frac{d}{2}; \Delta + 1-\frac{d}{2};e^{-2\sigma(y,y')}\right),
    \\ 
    G_{b\partial}(y(u,x),x_i) &= \left(\frac{u}{u^2+ |x-x_i|^2}\right)^{\Delta_i},
\end{align}
where $\sigma(y, y')$ is the geodesic distance between bulk points $y$ and $y'$ and $u,v$ are the cross ratios
\begin{align}
    u = \frac{x_{12}^2x_{34}^2}{x_{13}^2 x_{24}^2},\quad v = \frac{x_{14}^2x_{23}^2}{x_{13}^2 x_{24}^2}.
\end{align}
In two dimensions, 
\begin{align}
    G^{\Delta}_{bb} = \frac{e^{-\Delta\sigma(y,y')}}{1 - e^{-2\sigma(y,y')}} .
\end{align}
% When the geodesic distance is not small relative to the AdS radius, this is well approximated by
% \begin{align}
%     G^{\Delta_{n_e}^2}_{bb} \simeq {e^{-\Delta_{n_e}^2\sigma(y,y')}} .
% \end{align}
For the correlation function of twist-fields, the bulk to boundary propagators factor out of the integral as
\begin{align}
    G_{b \partial}(y(\lambda),x_1)G_{b \partial}(y(\lambda),x_2)G_{b \partial}(y'(\lambda'),x_3)G_{b \partial}(y'(\lambda'),x_4) = \left(|x_{12}||x_{34}| \right)^{-2 \Delta_{n}}.
\end{align}

When the intervals are well-separated, the dominant channel is the identity block and the partial wave is computed as
% \begin{align}
%     W_{\sigma_{n}^2} (x_i) =4 \frac{ \Gamma(\Delta_{\sigma_{n}^2})^2}{\Gamma\left(\frac{\Delta_{\sigma_{n}^2}}{2}\right)^4} \left(|x_{14}||x_{23}| \right)^{-2 \Delta_{n}}\int_{\gamma_{14}}\int_{\gamma_{23}} d\lambda d\lambda ' \frac{e^{-\Delta_{\sigma_{n}^2}\sigma(y, y')}}{1 - e^{-2 \sigma(y,y')}}.
% \end{align}
% For odd $n$ (relevant to the odd entropy), we have
\begin{align}
    W_{\Delta_{\mathbb{I}}}(x_i) =4 \frac{ \Gamma(\Delta_{\mathbb{I}})^2}{
    \Gamma\left(\Delta_{\mathbb{I}}/2\right)^4} \left(|x_{12}||x_{34}| \right)^{-2 \Delta_{n}}\int_{\gamma_{12}}\int_{\gamma_{34}} d\lambda d\lambda ' \frac{e^{-\Delta_{\mathbb{I}}\sigma(y, y')}}{1 - e^{-2 \sigma(y,y')}}
\end{align}
It was shown in Ref.~\cite{2019arXiv190210161P} that 
\begin{align}
    \lim_{\Delta_{\mathbb{I}} \rightarrow 0} 4 \frac{ \Gamma(\Delta_{\mathbb{I}})^2}
    {\Gamma\left(\Delta_{\mathbb{I}}/2\right)^4}\int_{\gamma_{12}}\int_{\gamma_{34}} d\lambda d\lambda ' \frac{e^{-\Delta_{\mathbb{I}}\sigma(y, y')}}{1 - e^{-2 \sigma(y,y')}} = 1.
    \label{vac_gwd_lim}
\end{align}
For the negativity ($n$ even), we find
\begin{align}
    \mathcal{E} = \lim_{n \rightarrow 1} \log \left(|x_{12}||x_{34}| \right)^{-2 \Delta_{n}} = 0,
\end{align}
which is consistent with the holographic conjecture
\begin{align}
    \mathcal{E} = \mathcal{X}_2\frac{E_W}{4G_N} = 0.
\end{align}
For odd entropy ($n$ odd), we have
\begin{align}
    S_o = \lim_{n_o \rightarrow 1}\frac{1}{1-n} \log \left(|x_{12}||x_{34}| \right)^{-2 \Delta_{n}} = 
    \frac{c}{3} \log  \left(|x_{12}||x_{34}| \right).
    \label{odd_disjoint}
\end{align}
In holographic theories, the following quantity, based on the odd entropy, was conjectured to calculate the entanglement wedge cross-section \cite{2018arXiv180909109T}
\begin{align}
    \mathcal{E}_W \equiv S_o - S = \frac{E_W}{4 G_N}.
\end{align}
Because (\ref{odd_disjoint}) is equivalent to the entanglement entropy for the two intervals, we confirm
\begin{align}
    \mathcal{E}_W = \frac{E_W}{4G_N} = 0. 
\end{align}

When the intervals are close, the dominant channel corresponds to the exchange of a double twist operator
\begin{align}
    \langle\sigma_{n}(x_1)\bar{\sigma}_{n}(x_2)\bar{\sigma}_{n}(x_3)\sigma_{n}(x_4)\rangle \propto  W_{\Delta_{\sigma_{n}^2}}(x_i).
\end{align}
The bulk to boundary propagators again factor out, this time as
\begin{align}
    G_{b \partial}(y(\lambda),x_1)G_{b \partial}(y(\lambda),x_2)G_{b \partial}(y'(\lambda'),x_3)G_{b \partial}(y'(\lambda'),x_4) = \left(|x_{14}||x_{23}| \right)^{-2 \Delta_{n}}.
\end{align}
The partial wave is then given by
\begin{align}
    W_{\sigma_{n}^2} (x_i) =4 \frac{ \Gamma(\Delta_{\sigma_{n}^2})^2}{
    \Gamma\left(\Delta_{\sigma_{n}^2}/2\right)^4} \left(|x_{14}||x_{23}| \right)^{-2 \Delta_{n}}\int_{\gamma_{14}}\int_{\gamma_{23}} d\lambda d\lambda ' \frac{e^{-\Delta_{\sigma_{n}^2}\sigma(y, y')}}{1 - e^{-2 \sigma(y,y')}}.
\end{align}

For $n$ odd,
% \begin{align}
%     W_{\sigma_{n}^2} (x_i) = 4 \frac{ \Gamma(\Delta_{\sigma_{n}^2})^2}
%     {\Gamma\left(\Delta_{\sigma_{n}^2}/2\right)^4} \left(|x_{14}||x_{23}| \right)^{-2 \Delta_{n}} \int_{\gamma_{14}}\int_{\gamma_{23}} d\lambda d\lambda' \frac{ e^{-\Delta_{\sigma_{n}^2}\sigma(y,y')}}{1- e^{-2\sigma(y,y')}}.
% \end{align}
we first take $c$ large and then the replica limit so that we can take the saddle point
\begin{align}
    S_o = \lim_{n\rightarrow 1}\frac{1}{1-n }\log \left[ 4 {\it const.}\frac{ \Gamma(\Delta_{\sigma_{n}^2})^2}
    {\Gamma\left(\Delta_{\sigma_{n}^2}/2\right)^4} \left(|x_{14}||x_{23}| \right)^{-2 \Delta_{n}}  \frac{ e^{-\Delta_{\sigma_{n}^2}E_W}}{1- e^{-2E_W}}\right].
\end{align}
Fixing the unknown constant and dropping the subleading term in the denominator, the odd entropy is
\begin{align}
    S_o &=  \lim_{n\rightarrow 1}\frac{1}{1-n} \left[\log \left(|x_{14}||x_{23}| \right)^{-2 \Delta_{n}} - \Delta_{\sigma_n^2} E_W \right] \\
    \nonumber
    &= \frac{c}{3} \log \left(|x_{14}||x_{23}| \right) + \frac{c}{6}  E_W ,
\end{align}
leading to 
\begin{align}
    \mathcal{E}_W  = \frac{E_W}{4G_N}.
    \label{oddentropy_holo}
\end{align}
% as conjectured in Ref.~\cite{2018arXiv180909109T}.

On the other hand,
for $n$ even, the exchanged operator is no longer light so there may be contributions from descendent states and not only the global conformal block. We proceed anyway to look at the global conformal block and discuss the corrections at the end of the section. The conformal partial wave is
\begin{align}
    W_{\sigma_{n_e}^2} (x_i) = 4 \frac{ \Gamma(\Delta_{\sigma_{n_e}^2})^2}{
    \Gamma\left(\Delta_{\sigma_{n_e}^2}/2\right)^4} \left(|x_{12}||x_{34}| \right)^{-2 \Delta_{n_e}} \int_{\gamma_{12}}\int_{\gamma_{34}} d\lambda d\lambda' \frac{e^{-\Delta_{\sigma_{n_e}^2}\sigma(y,y')}}{1- e^{-2 \sigma(y,y')}},
\end{align}
% \textcolor{red}{
% ($W_{\sigma^2_n}$
% should be 
% $W_{\sigma^2_{n_e}}$?
% $\Gamma(\Delta_{\sigma^2_n})$
% should be
% $\Gamma(\Delta_{\sigma^2_{n_e})}$?)}
% \textcolor{red}{(To go from (113)
% to (117), do we need to assume 
% $\Delta_{\sigma^2_n}$ small?)}
which is clearly divergent in the replica limit because $\Delta_{\sigma_{n_e}^2} \rightarrow -\frac{c}{4}$. In order to obtain a sensible answer, we first keep $\Delta = c a$ with $a>0$, and take large c, which allows us to take a saddle point approximation
\begin{align}
    \mathcal{E} = \lim_{n_e \rightarrow 1} \log \left[4 {\it const.}\frac{ \Gamma(\Delta_{\sigma_{n_e}^2})^2}
    {\Gamma\left(\Delta_{\sigma_{n_e}^2}/2\right)^4} \left(|x_{14}||x_{23}| \right)^{-2 \Delta_{n_e}}  \frac{ e^{ - \Delta_{\sigma_{n_e}^2} E_W }}{1- e^{- 2 E_W}}\right].
\end{align}
As for the odd entropy, we fix the unknown constant and drop the subleading term to find
\begin{align}
    \mathcal{E} = \frac{c}{4} E_W .
    % + \log\left(\frac{\it const.}{\left(1- e^{- 2 E_W}\right)}\right).
\end{align}
It is presently unclear how to treat the subleading term. We note that this term was dropped in the saddle-point calculation for the odd entropy because this was needed for consistency with the HHLL computation from Ref.~\cite{2018arXiv180909109T}.
Using the Brown-Henneaux formula, we may confirm
\begin{align}
    \mathcal{E} = \frac{3E_W}{8G_N} = \mathcal{X}_2 \frac{E_W}{4G_N}.
    \label{LN_vacuum}
\end{align}
By taking limiting cases of this formula, we can simply derive the formulas for adjacent and single intervals. It is important to re-emphasize that we have naively taken the global conformal block, disregarding the Virasoro descendents. We argue that including the Virasoro descendents will account for the full contribution to backreaction in the holographic conjecture (\ref{main_full}).

\begin{figure}
    \centering
    \includegraphics[height = 4cm]{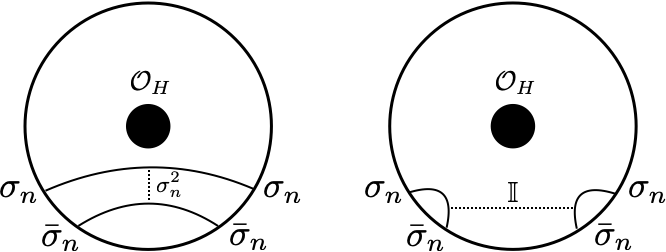} \quad
    \includegraphics[height = 4cm]{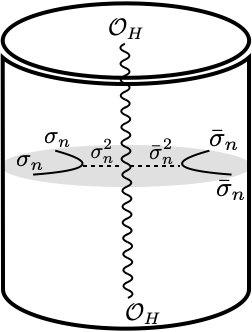}
    \caption{There are several dominant channels for the 6-point correlation function describing the logarithmic negativity and odd entropy in a high-energy eigenstate.}
    \label{fig:microstate_disjoint_gwd}
\end{figure}

We would additionally like to look at thermal states described by black holes in the bulk. We create black hole microstates by inserting asymptotic heavy operators ($\Delta \sim c$) into the correlation function
\begin{align}
    \langle\mathcal{O}_H(\infty)\sigma_{n_e}(x_1)\bar{\sigma}_{n_e}(x_2)\bar{\sigma}_{n_e}(x_3)\sigma_{n_e}(x_4)\mathcal{O}_H(-\infty)\rangle. 
\end{align}
There are a few channels to consider. Most simply, we can have the channel where the fusion of the two heavy operators is dominant. In this case, the correlation function factorizes
\begin{align}
&
    \langle\mathcal{O}_H(\infty)\sigma_{n_e}(x_1)\bar{\sigma}_{n_e}(x_2)\bar{\sigma}_{n_e}(x_3)\sigma_{n_e}(x_4)\mathcal{O}_H(-\infty)\rangle 
    \nonumber \\
    &\quad
    \sim \langle\sigma_{n_e}(x_1)\bar{\sigma}_{n_e}(x_2)\bar{\sigma}_{n_e}(x_3)\sigma_{n_e}(x_4)\rangle \langle \mathcal{O}_H(\infty)\mathcal{O}_H(-\infty)\rangle .
\end{align}
The fusion of the heavy operators creates a stationary black hole in the center of the bulk so that the correlation function of the light operators is now computed in the BTZ metric 
(see Fig.\ \ref{fig:microstate_disjoint_gwd}). The two point function is normalized, so again we find (\ref{oddentropy_holo}) and (\ref{LN_vacuum}), only now the minimal cross section is computed in the black hole background. Again, we have only considered the global conformal block.

In an alternate channel (see Fig.\ \ref{fig:microstate_disjoint_gwd}), the correlation function factors into two four-point HHLL correlators \cite{2016arXiv160106794B}
\begin{align}
&
    \langle\mathcal{O}_H(\infty)\sigma_{n_e}(x_1)\bar{\sigma}_{n_e}(x_2)\bar{\sigma}_{n_e}(x_3)\sigma_{n_e}(x_4)\mathcal{O}_H(-\infty)\rangle  \label{discon_corr}
    \nonumber \\
    &\quad \sim
    \langle\sigma_{n_e}(x_1)\sigma_{n_e}(x_4)\mathcal{O}_H(\infty)\mathcal{O}_H(-\infty)\rangle \langle \bar{\sigma}_{n_e}(x_2)\bar{\sigma}_{n_e}(x_3)\mathcal{O}_H(\infty)\mathcal{O}_H(-\infty)\rangle .
\end{align}
While this can be anticipated for the entanglement entropies\footnote{This can be seen by inserting the identity operator in the correlator. Then, at leading order in the OPE, the twist and anti-twist fields fuse to the identity operator and this fixes the projection operator to be the heavy state by the orthonormality of the two-point functions.}, it is a nontrivial result for the correlator that computes negativity.
% If the exchanged operator in the dominant conformal block were light, each of these four-point functions could be straightforwardly computed by the saddle point using the world-line approach to semiclassical conformal blocks \cite{2015JHEP...07..131H} with $\sigma_{min}$ now connecting the geodesics to the horizon of the black hole, hence representing the minimal entanglement wedge cross section. The exchanged operator turns out to be heavy (i.e.~a black hole) so it is not presently known precisely how the backreaction will effect the calculation \footnote{Discussion with Liam Fitzpatrick: there is no good gravity picture of this. Interestingly, these heavy blocks are also important at late Lorentzian times.}. It seems plausible that because the boundary to bulk geodesics are tensionless (because $\sigma_1 = 0$), the backreaction of the exchange operator. 
Let's focus on one correlator at a time and compute with geodesic Witten diagrams. As was formalized in Ref.~\cite{2015JHEP...12..077H}, the geodesic Witten diagram for HHLL conformal blocks is modified such that the bulk to bulk propagator and bulk to boundary propagator for the light operators is evaluated in the metric generated by the insertion of the heavy operator
\begin{align}
    ds^2 = \frac{\alpha^2}{\cos^2 \rho}\left(\frac{d \rho^2}{\alpha^2} + d\tau^2 + \sin^2 \rho d\phi^2 \right),
    \label{BTZ_metric}
\end{align}
where 
\begin{align}
    \alpha = \sqrt{1 - \frac{24h_H}{c}}.
\end{align}
For $\alpha^2 < 0$, this is a BTZ black hole geometry with horizon at $\rho=0$. 
This is motivated by the coordinate transformation of HHLL operators that takes care of the backreaction shown in Ref.~\cite{2015JHEP...11..200F}.
Like the vacuum state correlator, the bulk to boundary propagators factor out of the integral 
\begin{align}
    G_{b \partial}(y(\lambda),x_1)G_{b \partial}(y(\lambda),x_2)\tilde{G}_{b \partial}(y'(\lambda'),x_3)\tilde{G}_{b \partial}(y'(\lambda'),x_4) = \left( \frac{\beta}{\pi \epsilon} \sinh\frac{\pi |x_2 -x_1|}{\beta} \right)^{-2\Delta_{n}} \left(|x_3 - x_4|  \right)^{-2 \Delta_{\mathcal{O}}}
\end{align}
where the tilde means that it is evaluated in the BTZ metric (\ref{BTZ_metric}). For the odd entropy, we have the same evaluation as before except with the new metric. Including the other correlator from (\ref{discon_corr}), we find (fixing the overall constant)
\begin{align}
    S_o = \frac{c}{3}\log \left(\frac{\beta}{\pi \epsilon} \sinh \frac{\pi|x_2-x_1|}{\beta} + E_W\right).
\end{align}
Therefore, we have confirmed
\begin{align}
    \mathcal{E}_W = \frac{E_W}{4G_N}.
\end{align}

For $n$ even, 
we have
\begin{align}
    \langle\sigma_{n_e}(x_1){\sigma}_{n_e}(x_2)\mathcal{O}_{H}(x_3)\mathcal{O}_{H}(x_4)\rangle \propto \left(|x_3 - x_4|^2  \right)^{-\Delta_{\mathcal{O}}}  \int_{\gamma_{12}}\int_{\gamma_{34}} d\lambda d\lambda' G^{\Delta_{\sigma_{n_e}^2}}_{b b}(y(\lambda),y'(\lambda')),
\end{align}
giving 
\begin{align}
    \langle\sigma_{n_e}(x_1){\sigma}_{n_e}(x_2)\mathcal{O}_{H}(x_3)\mathcal{O}_{H}(x_4)\rangle \sim  e^{-\Delta_{\sigma_{n_e}^2}E_W}.
\end{align}
Including the other correlator from (\ref{discon_corr}), we find the negativity to be 
\begin{align}
    \mathcal{E} = \mathcal{X}_2 \frac{E_W}{4 G_N},
\end{align}
where the entanglement wedge cross section is disconnected and includes two radial geodesics that terminate on the horizon of the black hole.

% \subsection{Multiple intervals}

So far, we have considered each subsystem to be connected. In these set-ups, we have approximated the logarithmic negativity using only the global conformal block without all of its descendents. We know that this only should give us a first order approximation, but how off are we really by ignoring the ``graviton exchange''? This is effectively ignoring backreaction in the bulk. We consider (\ref{main_full}) in a perturbative expansion around $n=1$. Ultimately, we are interested in $n\rightarrow 1/2$. At leading order in $(n-1)$,
\begin{align}
    \langle \mathcal{O}_1 \dots \rangle = \mbox{gWd}^{(0)}_{AdS}\left[ \Delta = 
    \Delta_{\mathbb{I}}\right] + \mbox{gWd}_{AdS}^{(1)}\left[ \Delta = 
    \Delta_{\mathscr{O}(n-1)}\right] + 
    \mbox{gWd}^{(1)}_{{backreacted}_{\mathscr{O}(n-1)}}\left[ \Delta =
    \Delta_{\mathbb{I}}\right]  + \mathscr{O}\left((n-1)^2\right).
    \label{gwd_expansion}
\end{align}
The first term is the vacuum block computed in the non-backreacted geometry and is $\mathscr{O}(1)$ by (\ref{vac_gwd_lim}). The second term is the conformal block with the non-vacuum exchange in the non-backreacted geometry. This $\mathscr{O}(e^c)$ term is the term that we have been computing in this section and has reproduced the approximate holographic negativity formula (\ref{main}). The third term in (\ref{gwd_expansion}) corresponds to the vacuum diagram computed in the backreacted geometry. For the same reason as for the first term, this term provides a subleading $\mathscr{O}(1)$ contribution to the correlation function. In order to find the first nontrivial corrections to (\ref{main}), one must compute the non-vacuum exchange in the backreacted geometry. The identical argument applies to Renyi entropies expanded in $(n-1)$ with precisely the same terms contributing to the correlation function.

\begin{figure}
    \centering
    \includegraphics[height = 5cm]{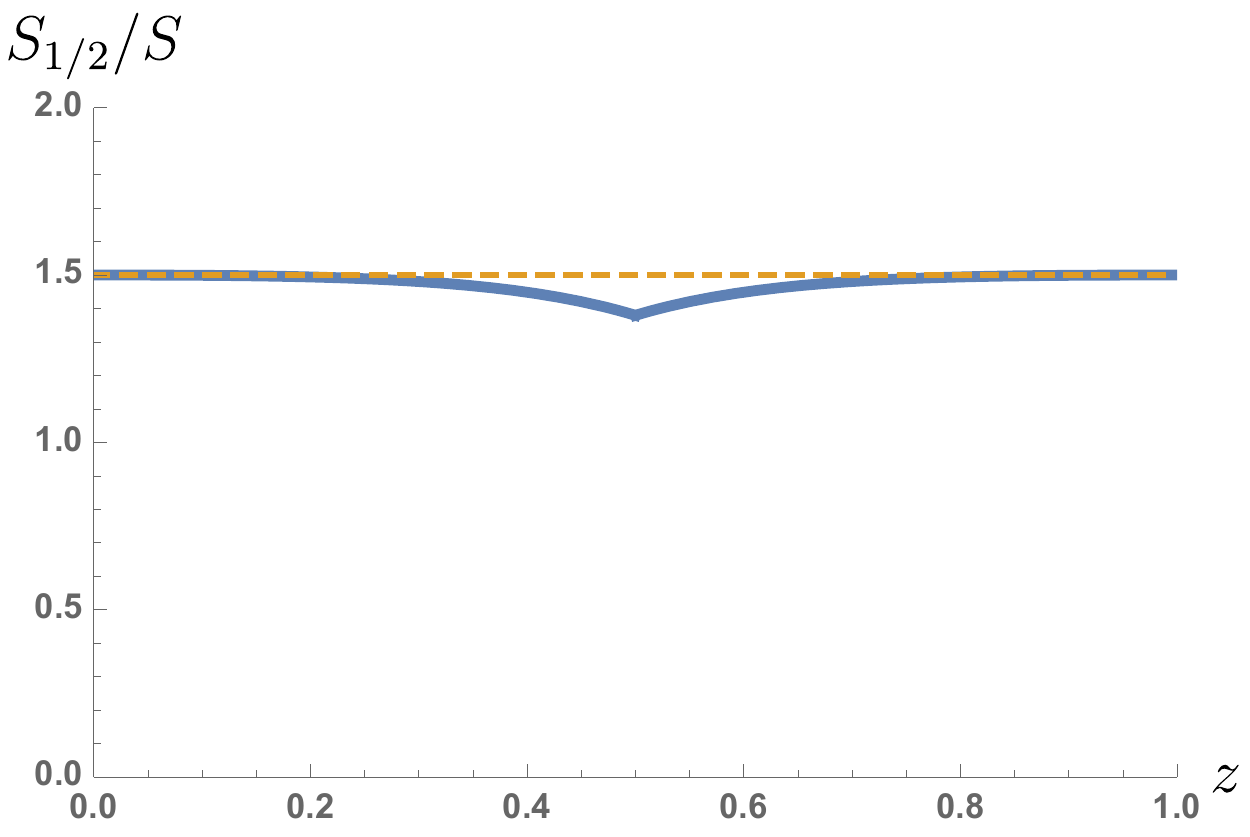}
    \caption{Displayed is the ratio between the R\'enyi entropy at index 1/2 and the von Neumann entropy for disjoint intervals at all values of the cross-ratio. The dotted orange line represents what the ratio would be if the cosmic branes do not backreact (global conformal block only) on each other while the blue curve is the full result (graviton exchange). Intuitively, as the intervals become very far apart or very close together, the backreaction is negligible. The cusp at $z = 1/2$ corresponds to the transition from connected to disconnected cosmic branes. We use the Zamolodchikov recursion formula to numerically compute the blue curve.}
    \label{fig:backreaction_ratio}
\end{figure}

Though, we are unable to directly check how much of an effect this backreaction has for the generic configurations, we are able to quantify this for the case that the subsystem of interest is the union of two disjoint intervals in the vacuum state
\begin{align}
    A = [x_1, x_2] \cup [x_3, x_4], \quad B = \bar{A}.
\end{align}
Because we are in the vacuum, we know the logarithmic negativity to equal the R\'enyi entropy at index 1/2. Using Zamolodchikov's recursion relation \cite{1987TMP....73.1088Z, 1996NuPhB.477..577Z}, we can accurately compute the Virasoro block for $S_{1/2}$. The ratio between $S_{1/2}$ and the von Neumann entropy characterizes the full backreaction from the cosmic branes in the bulk. Without backreaction, the ratio is $\mathcal{X}_2 = 3/2$. We find this to be quite accurate when the intervals are either very close ($z \rightarrow 1$) or very far away ($z \rightarrow 0$). At intermediate regimes, the backreaction has a small but noticeable effect on the ratio. The maximum change from $\mathcal{X}_2$ comes at $z = 1/2$ when the ratio is $\sim 1.38$ (see Fig.~\ref{fig:backreaction_ratio}). This gives us some idea of how much the descendent states may effect the simple formula (\ref{main}). In general, it seems that the backreaction must be accounted for using (\ref{main_full}). 

We proceed with the Witten diagram derivation of this result. The correlation function to compute the moments of the partial transpose of this configuration is
\begin{align}
    \Tr \left(\rho_{AB}^{T_B}\right)^n &= \lim_{x_1', x_2', x_3', x_4' \rightarrow x_1, x_2, x_3, x_4 }\braket{\sigma_n(x_1) \sigma_n(x_1')\bar{\sigma}_n(x_2) \bar{\sigma}_n(x_2')\sigma_n(x_3) \sigma_n(x_3')\bar{\sigma}_n(x_4) \bar{\sigma}_n(x_4')}
    \nonumber \\
    &= \braket{\sigma^2_{n}(x_1) \bar{\sigma}_n^2(x_2) \sigma^2_n(x_3) \sigma^2_n(x_4)}.
\end{align}
In both t and s channels, the dominant Virasoro block is the vacuum. The difference between the odd $n$ and even $n$ correlators is that for odd $n$, all external operators are light, while for even $n$, all external operators are heavy. With the exchange of the identity operator and all other operators light, the eight-point function factorizes into two four-point functions for n odd. Each four-point function has already been computed in the single interval case so the holographic formula is confirmed. This factorizing process can be iterated generally for any configuration of intervals for the odd entropy.

Now looking at the global conformal block for $n$ even, we have
\begin{align}
    \braket{\sigma^2_{n}(x_1) \bar{\sigma}_n^2(x_2) \sigma^2_n(x_3) \sigma^2_n(x_4)} \sim 4 \frac{ \Gamma(\Delta_{\mathbb{I}})^2}{\Gamma\left(\frac{\Delta_{\mathbb{I}}}{2}\right)^4} \left(|x_{12}||x_{34}| \right)^{-2 \Delta_{\sigma_n^2}}\int_{\gamma_{12}}\int_{\gamma_{34}} d\lambda d\lambda ' \frac{e^{-\Delta_{\mathbb{I}}\sigma(y, y')}}{1 - e^{-2 \sigma(y,y')}}.
\end{align}
Fixing the appropriate constant, we find the first order solution to the logarithmic negativity
\begin{align}
    \mathcal{E} = \frac{c}{2} \left(\log |x_{12}| + \log |x_{34}| \right).
\end{align}
In the other channel, we have
\begin{align}
    \mathcal{E} = \frac{c}{2} \left(\log |x_{14}| + \log |x_{23}| \right).
\end{align}
These correspond to (\ref{main}). We have captured the $z\rightarrow0,1$ limits of the logarithmic negativity using the global conformal block. As expected, the subleading graviton corrections that push the ratio $S_{1/2}/S$ down to $\sim 1.38$ are absent and we have the ratio precisely at $3/2$ for all values of the cross ratio. 
\section{Clifford circuits}
\label{clif_appendix}
In this appendix, we review the efficient numerical algorithm for computing entanglement entropy in Clifford circuits. We closely follow Ref.~\cite{2017PhRvX...7c1016N}. Under Clifford evolution, Pauli strings are mapped to Pauli strings. This highly constrains the Hilbert space that Clifford evolution samples. We are then able to specify the state of our system by the operators that stabilize the state
\begin{align}
    \mathcal{O}_i\ket{\psi} = \ket{\psi}.
\end{align}
The number of such operators is equal to the number of spins in our system. Furthermore, each stabilizer can be written as a Pauli string
\begin{align}
    \mathcal{O}_i \propto ( X_1^{v_{1x}} Z_1^{v_{1z}}) \otimes \dots \otimes ( X_L^{v_{Lx}} Z_L^{v_{Lz}}).
\end{align}
We can then fully specify the state by a $2L \times L$ matrix where each column is a vector specifying a particular stabilizer
\begin{align}
    \bar{v} = (v_{1x} , v_{1z} \dots, v_{Lx}, v_{Lz} ).
\end{align}
Each entry is either a $0$ or $1$. Clearly, this is a dramatic improvement to the standard $2^{L} \times 2^L$ matrix needed to express the entire Hilbert space. We now recall the update rules on each stabilizer operator under Clifford evolution. For a Hadamard operation on the Bloch sphere of site $i$
\begin{align}
    H = \frac{1}{\sqrt{2}}(X+Z),
\end{align}
we simply exchange $v_{ix}$ with $v_{iz}$ for all stabilizers. For a phase rotation about the $Z$ axis
\begin{align}
    P = \sqrt{Z},
\end{align}
we have $v_{iz} \rightarrow v_{iz} +v_{ix}$ where all operations are mod 2. Finally, for the CNOT operation, which can act either to the right or left,
\begin{align}
    \mbox{CNOT}^{(L)} &= \frac{1}{2}((1+ Z_1)+(1-Z_1)X_2), \\
     \mbox{CNOT}^{(R)} &= \frac{1}{2}((1+ Z_2)+(1-Z_2)X_1),
\end{align}
updates the strings as $v_{2x} \rightarrow v_{2x} + v_{1x}$ and $v_{1z} \rightarrow v_{1z} + v_{2z}$ for $\mbox{CNOT}^{(L)}$ and analogously for $\mbox{CNOT}^{(R)}$. 

To compute the entanglement entropy of a subregion $A$, we need to find the number of stabilizers that are independent once restricted to $A$, $I_A$. The entanglement entropy is then
\begin{align}
    S_A = (I_A - |A|) \log 2
\end{align}
where $A$ is the number of sites in $A$. To find the number of independent stabilizers, we are computing the Galois field ($\mathbb{Z}_2$) rank of the $2| A| \times L$ restriction of the $2L \times L$ matrix to the subsystem. This can be straightforwardly implemented in any programming language. 

In order to compute operator entanglement, our initial state must be composed of a tensor product of Bell pairs. This can be implemented by starting will all spins in the $Z$ direction and subsequently applying single Hadamard and CNOT gates to create the Bell pairs.

\begin{figure}
    \centering
    \includegraphics[width = 7cm]{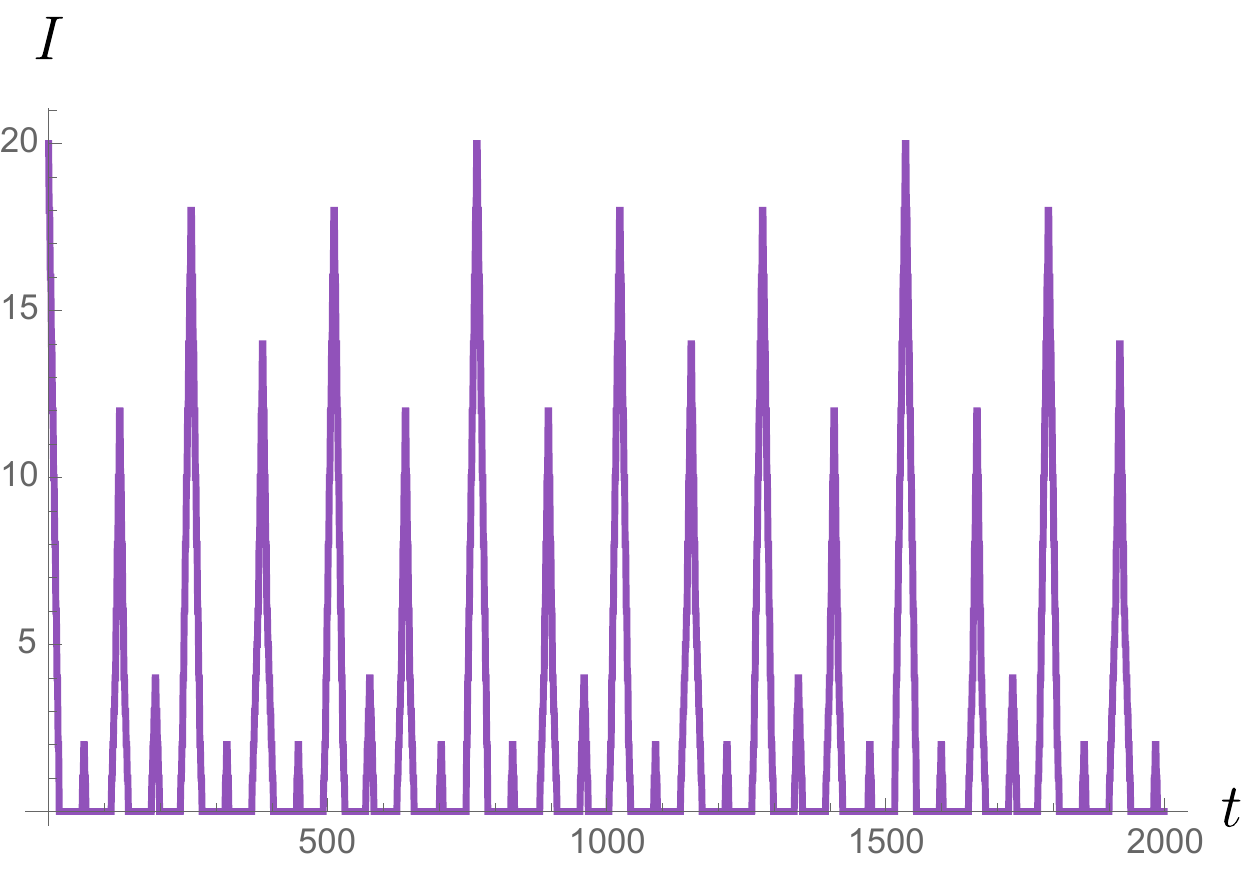}
    \caption{The $BOMI$ for symmetric intervals of length $10$ is shown for a Clifford circuit made exclusively of CNOT gates. We see recurrence at $t_{rec} \sim 800 \sim \mathcal{O}(L = 200)$. }
    \label{fig:recurrence}
\end{figure}

\begin{figure}
    \centering
    \includegraphics[height = 3.25cm]{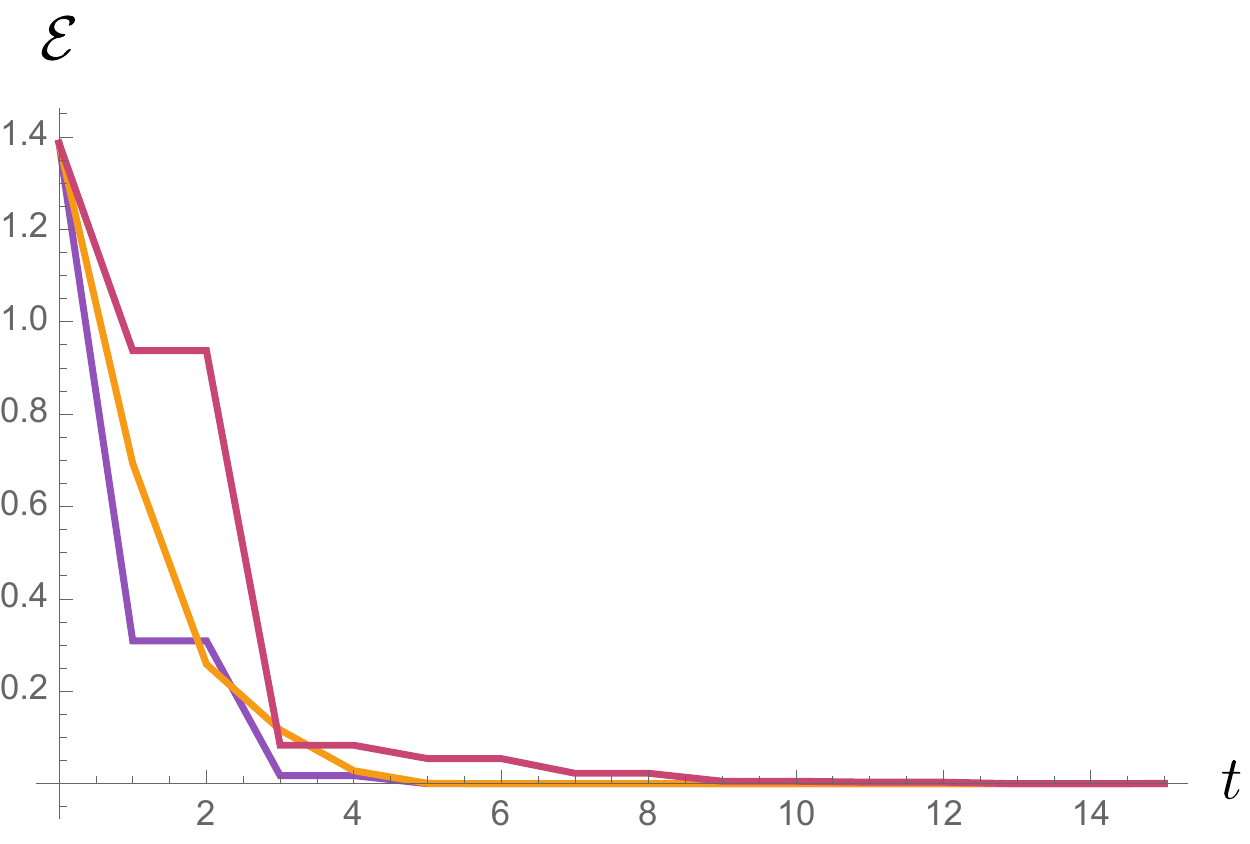}
    \includegraphics[height = 3.25cm]{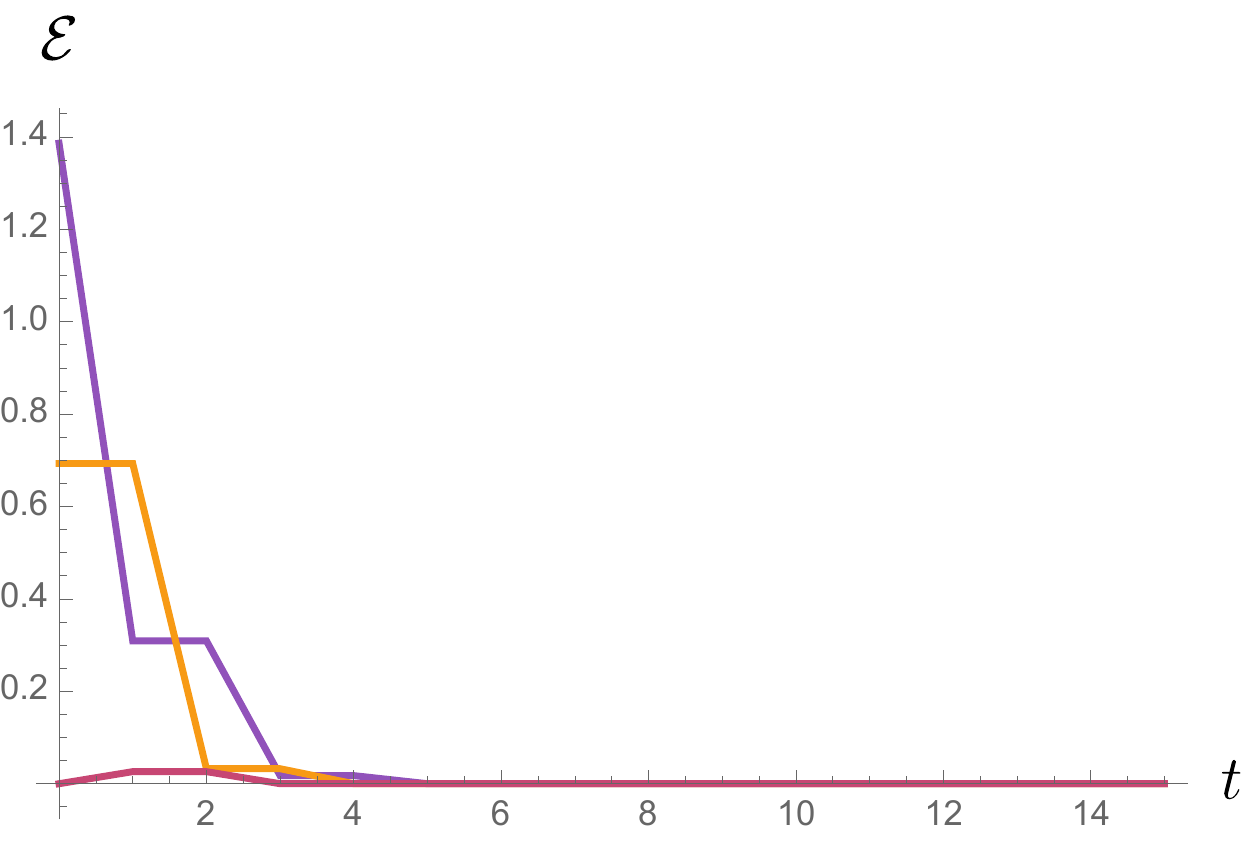}
    \includegraphics[height = 3.25cm]{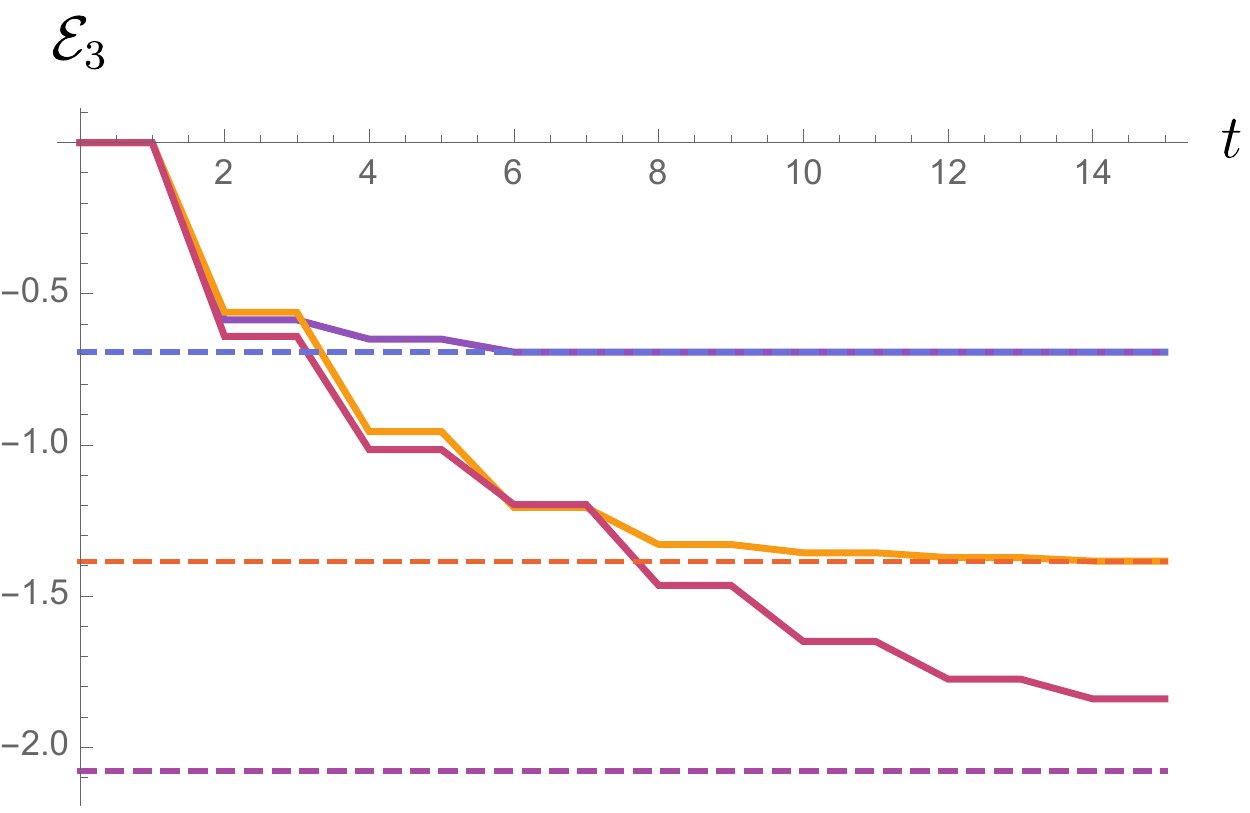}\\
    \includegraphics[height = 3.25cm]{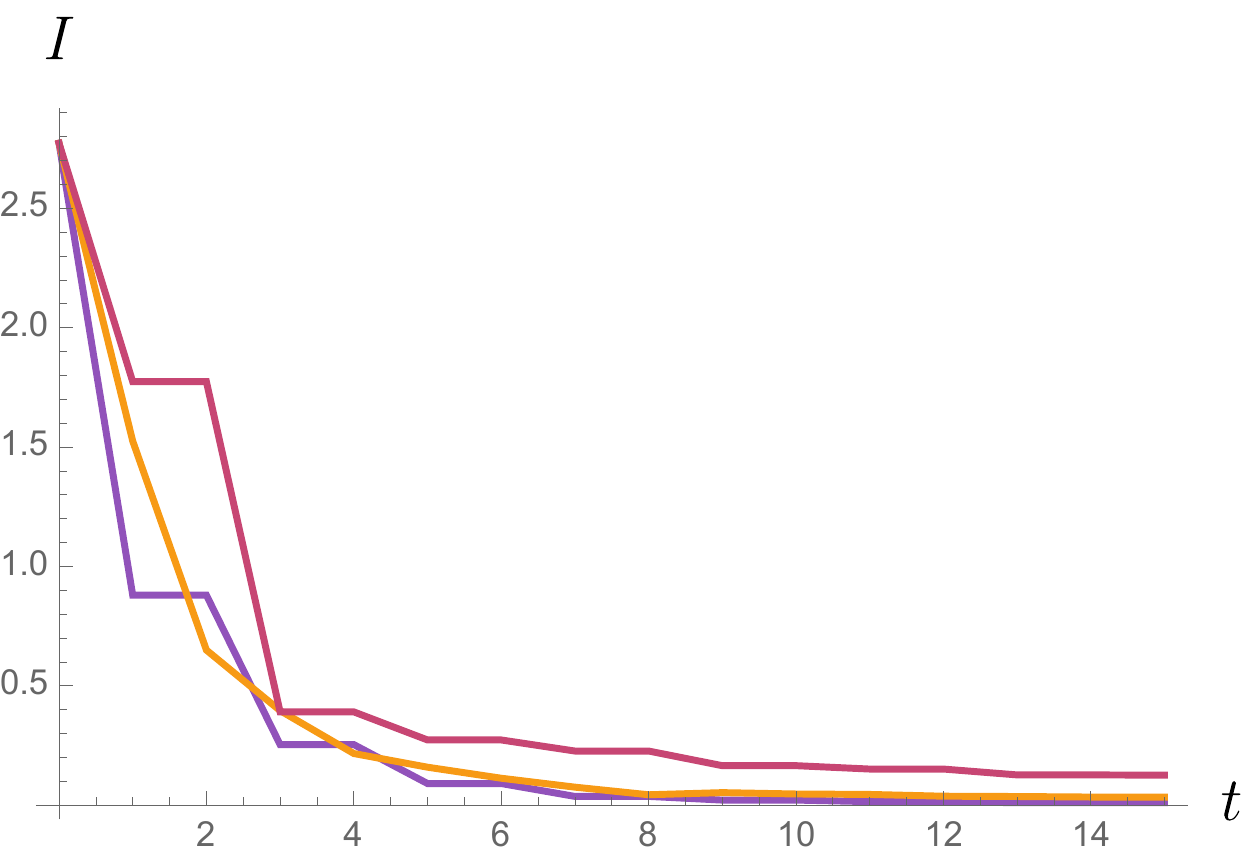}
    \includegraphics[height = 3.25cm]{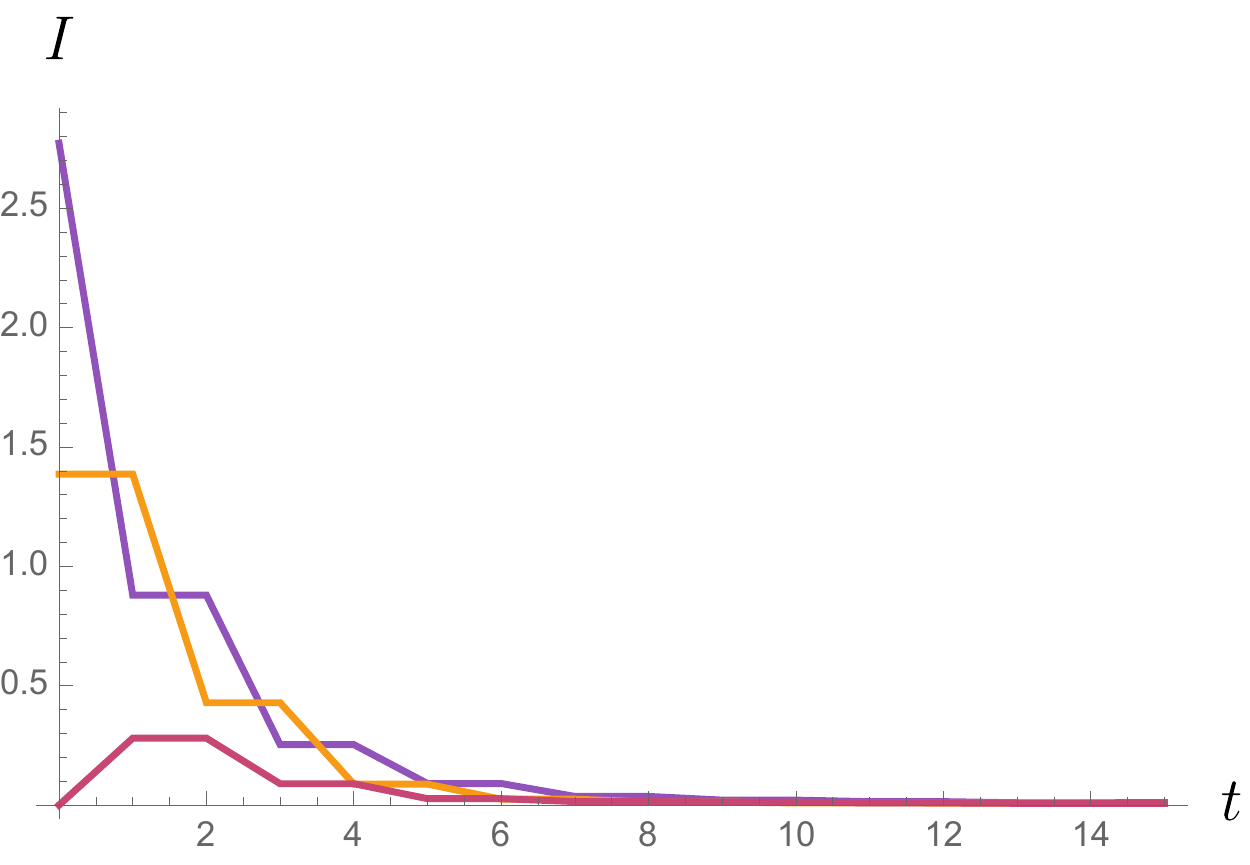}
    \includegraphics[height = 3.25cm]{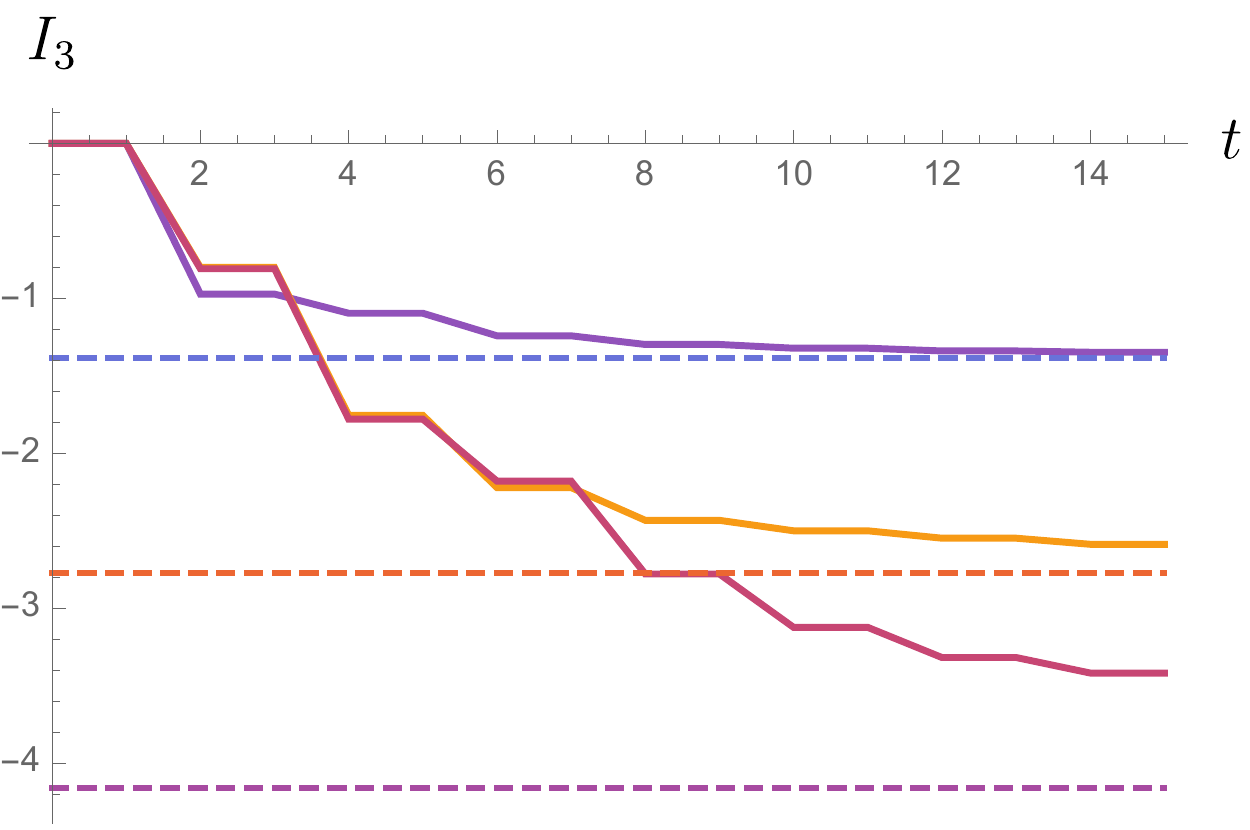}
    \caption{The operator entanglement for a random unitary circuit with seven input and output spins. The divergence from the line-tension picture from (\ref{linetension_eq}) can be attributed to finite size and finite $q$ ($q=2$) and size effects. We use the same configurations as Fig.~\ref{sc_erg}.}
    \label{randU}
\end{figure}

We demonstrate the importance of the randomly applied Hadamard and CNOT gates in Fig.\ 
\ref{fig:recurrence} where $\mathcal{O}(L)$ recurrence times are shown when the Clifford circuit is purely built from CNOT gates. Furthermore, we provide numerical results for operator entanglement of \textit{fully random} unitaries in Fig.\ \ref{randU}, though we are only able to do so for small system sizes.

\section{Details of calculation of OLN in CFT}
\label{CB_FF_details_App}
In this appendix, we detail the calculations of the R\'enyi operator logarithmic negativities for the free fermion and compact boson that we wrote without proof in the main text.  
\subsection{Free fermion}
As in Ref.~\cite{2016JSMTE..07.3102H}, the $2^{nd}$ R\'enyi OLN for the free fermion in the path-integral formalism is given by
\be
\begin{split}
\mathcal{E}^{(2)}= \log{\left[\int_{C_{A,B}}
D\Psi_1D\bar{\Psi}_1D\Psi_2D\bar{\Psi}_2\,  e^{-S_1-S_2}\right]} 
\end{split}
\ee
where the actions are given by
\be
S_i= \int d\tau d x\bar{\Psi}_i\gamma^{\mu}\partial_{\mu} \Psi_i.
\ee
We define a vector field ${\bf \Psi}$ by
\be
\begin{split}
{\bf \Psi}= \begin{pmatrix}
\Psi_1\\
\Psi_2 \\
\end{pmatrix}.
\end{split}
\ee
The boundary condition $C_{A,B}$ means that ${\bf \Psi}$ transforms when it rotates counterclockwise around boundaries of $A$ and B as follows,
\be
\begin{split}
 &T(X_2){\bf \Psi}=\begin{pmatrix}
 0&1\\
 -1&0 \\
 \end{pmatrix}{\bf \Psi}, ~~
 T(X_1){\bf \Psi}=\begin{pmatrix}
 0&-1\\
 1&0 \\
 \end{pmatrix}{\bf \Psi}, \\
 &T(Y_2){\bf \Psi}=\begin{pmatrix}
 0&-1\\
 1&0 \\
 \end{pmatrix}{\bf \Psi}, ~~
 T(Y_1){\bf \Psi}=\begin{pmatrix}
 0&1\\
 -1&0 \\
 \end{pmatrix}{\bf \Psi}.
\end{split}
\ee 

The eigenvalues $\lambda_{\pm}$ of $T(X_1), T(X_2), T(Y_1)$ and $T(Y_2)$ are 
\be
\lambda_{\pm}=\pm i,
\ee
and the eigenvector is ${\bf \Phi}$.
By redefining dynamical fields as
\be
\Phi_k(x)=e^{i\int^{x}_{x_0} dx'^{\mu}A^k_{\mu}(x')} \phi_k(x),
\ee
the action is given by
\be
\sum_{i=1}^2 S_i = \sum_{k=\pm1} \int d\tau dx \bar{\phi}_k(x)\gamma^{\mu}\left(\partial_{\mu}+iA^k_{\mu}\right)\phi_k(x).
\ee
Therefore, $2^{nd}$ OLN is given by
\be
\mathcal{E}^{(2)}=\log\left[\f{\prod_{k=\pm1}Z[k]}{Z^2}\right]=\log{ \left[\prod_{k=\pm1} \left\langle :e^{-i \int d\tau dx \bar{\Phi}_k\gamma^{\mu}A^k_{\mu}\Phi_k}:\right \rangle_{\beta}\right]},
\ee
where $A^k_{\mu}$ obeys the following conditions,
\be
\begin{split}
&
\oint_{C_{X_2}}dx^{\mu}A^1_{\mu}=
-\oint_{C_{X_2}}dx^{\mu}A^2_{\mu}=
-\oint_{C_{X_1}}dx^{\mu}A^1_{\mu}=
\oint_{C_{X_1}}dx^{\mu}A^2_{\mu}
=\f{\pi}{2}, 
\\
&
-\oint_{C_{Y_2+i\tau_1}}dx^{\mu}A^1_{\mu}=
\oint_{C_{Y_2+i\tau_1}}dx^{\mu}A^2_{\mu}=
\oint_{C_{Y_1+i\tau_1}}dx^{\mu}A^1_{\mu}=
-\oint_{C_{Y_1+i \tau_1}}dx^{\mu}A^2_{\mu}
=\f{\pi}{2}. \\
\end{split}
\ee
By performing bosonization, the current $j^{\mu}=\bar{\Phi}\gamma^{\mu}\Phi$ is written in terms of $\varphi$ by
\be
j^{\mu}_k=\bar{\Phi}_k\gamma^{\mu}\Phi_k=\epsilon^{\mu \nu}\partial_{\nu}\varphi_k.
\ee
Thus, the $2^{nd}$ OLN is given by
\be \label{2ndOLN}
\begin{split}
\prod_{\pm 1} \left\langle :e^{i \int d\tau dx \epsilon^{\nu \mu}\partial_{\mu}A^k_{\nu}\varphi_k}:\right \rangle_{\beta}&=\left\langle e^{\f{-i\pi}{4} \int d\tau dx\left[\delta (x-X_2)\delta(\tau)-\delta(x-X_1)\delta(\tau)-\delta(x-Y_2)\delta(\tau-\tau_1)+\delta(x-Y_1)\delta(\tau-\tau_1)\right]\varphi(x, \tau)} \right \rangle_{\beta} \\
&\times\left\langle e^{\f{-i\pi}{4} \int d\tau dx\left[-\delta (x-X_2)\delta(\tau)+\delta(x-X_1)\delta(\tau)+\delta(x-Y_2)\delta(\tau-\tau_1)-\delta(x-Y_1)\delta(\tau-\tau_1)\right]\varphi(x, \tau)} \right \rangle_{\beta} \\
&=\left\langle :e^{i\f{\varphi_1(w_1, \bar{w}_1)}{4}}: :e^{-i\f{\varphi_1(w_2, \bar{w}_2)}{4}}: :e^{-i\f{\varphi_1(w_3, \bar{w}_3)}{4}}: :e^{i\f{\varphi_1(w_4, \bar{w}_4)}{4}}:\right\rangle_{\beta} \\
&\times \left\langle :e^{-i\f{\varphi_2(w_1, \bar{w}_1)}{4}}: :e^{i\f{\varphi_2(w_2, \bar{w}_2)}{4}}: :e^{i\f{\varphi_2(w_3, \bar{w}_3)}{4}}: :e^{-i\f{\varphi_2(w_4, \bar{w}_4)}{4}}:\right\rangle_{\beta}, \\
\end{split}
\ee
where $w_1=X_2$, $w_2=X_1$, $w_3=Y_2+i\tau_1$ and $w_4=Y_1+i\tau_1$.
The chiral (anti-chiral) conformal dimension $h({\alpha})$ ($\bar{h}(\alpha)$) of $:e^{i \alpha \varphi} :$ is 
\be
h(\alpha) =\bar{h}(\alpha)=\f{\alpha^2}{8}.
\ee 
Therefore, $h$ and $\bar{h}$ of $:e^{\pm i\f{\varphi}{4}}:$ are given by
\be
h=\bar{h}=\f{1}{128}.
\ee
By performing the map $z=e^{\f{2\pi w}{\beta}}$, (\ref{2ndOLN}) is 
\be
\begin{split}
\prod_{\pm 1} \left\langle :e^{i \int d\tau dx \epsilon^{\nu \mu}\partial_{\mu}A^k_{\nu}\varphi_k}:\right \rangle_{\beta}&=\left(\f{2\pi}{\beta}\right)^{16h}e^{\f{8\pi h}{\beta}\left(X_1+X_2+Y_1+Y_2\right)}\\
&\times\left\langle :e^{i\f{\varphi_1(z_1, \bar{z}_1)}{4}}: :e^{-i\f{\varphi_1(z_2, \bar{z}_2)}{4}}: :e^{-i\f{\varphi_1(z_3, \bar{z}_3)}{4}}: :e^{i\f{\varphi_1(z_4, \bar{z}_4)}{4}}:\right\rangle \\
&\times \left\langle :e^{-i\f{\varphi_2(z_1, \bar{z}_1)}{4}}: :e^{i\f{\varphi_2(z_2, \bar{z}_2)}{4}}: :e^{i\f{\varphi_2(z_3, \bar{z}_3)}{4}}: :e^{-i\f{\varphi_2(z_4, \bar{z}_4)}{4}}:\right\rangle, \\
\end{split}
\ee
where $z_i=e^{\f{2\pi w_i}{\beta}}$.
Through Wick contractions, we find that the $2^{nd}$ operator logarithmic negativity is given by
\be
\begin{split}
&\mathcal{E}^{(2)}
=\log{\left[\Tr_{A\cup B}\left(\rho_{A\cup B}\right)^2\right]}
\\
&
=\f{1}{8}\log{\left[\f{|z_2-z_3|^2|z_1-z_4|^2}{|z_1-z_2|^2|z_1-z_3|^2|x_2-z_4|^2|z_3-z_4|^2}\right]}
+\f{1}{8}\log{\left(\f{\pi}{2\beta}\right)}
\\
&=\f{1}{8}\log{\left[\f{\pi}{2\beta}\right]}+\f{1}{16}\log{\left[\f{1}{\sinh^2{\left[\f{\pi}{\beta}(Y_1-Y_2)\right]}\sinh^2{\left[\f{\pi}{\beta}(X_1-X_2)\right]}}\right]}
\\
&+ \f{1}{16}\log{\left[\f{\cosh{\left[\f{\pi}{\beta}(X_1-Y_2+t)\right]}\cosh{\left[\f{\pi}{\beta}(X_1-Y_2-t)\right]}\cosh{\left[\f{\pi}{\beta}(X_2-Y_1+t)\right]}\cosh{\left[\f{\pi}{\beta}(X_2-Y_1-t)\right]}}{\cosh{\left[\f{\pi}{\beta}(X_1-Y_1+t)\right]}\cosh{\left[\f{\pi}{\beta}(X_1-Y_1-t)\right]}\cosh{\left[\f{\pi}{\beta}(X_2-Y_2+t)\right]}\cosh{\left[\f{\pi}{\beta}(X_2-Y_2-t)\right]}}\right]}.
\end{split}
\ee
This matches (\ref{ff_2nd_OLN_eq}) from the main text.

\subsection{Compact boson}
We proceed to the four point function of twist fields for the compactified free boson
% is given by
\begin{align}
&\langle\sigma_{n_e}(z_1,\bar{z}_1)\bar{\sigma}_{n_e}(z_2,\bar{z}_2) \sigma_{n_e}(z_3,\bar{z}_3) \bar{\sigma}_{n_e}(z_4,\bar{z}_4)  \rangle 
\nonumber \\
 &
=
\frac{1}{(z_{12}z_{34})^{2h_n}}\frac{1}{(\bar{z}_{12}\bar{z}_{34})^{2h_n}}\frac{F_n(x,\bar{x})}{(1-x)^{2h_n}(1-\bar{x})^{2h_n}} 
\nonumber \\
&= 
\frac{F_n(x,\bar{x})}{\left[e^{\frac{\pi}{\beta}(X_1+X_2+Y_1+Y_2)}\left(2\sinh\frac{\pi(X_1-X_2)}{\beta} \right)\left(2\sinh\frac{\pi(Y_1-Y_2)}{\beta} \right)\right]^{4h_n}(1-x)^{2h_n}(1-\bar{x})^{2h_n}},
\end{align}
where $F_n$ is given by 
% (2.4) of \cite{2017PhRvD..96d6020C}
\begin{equation}\label{Fn}
F_n(x,\bar{x}) = \frac{\Theta(0|T)^2}{\prod_{k=1}^{n-1}f_{k/n}(x)f_{k/n}(\bar{x})}
\end{equation}
and $f_{k/n}(x) = {}_2F_1\left(\frac{k}{n},1-\frac{k}{n},1,x\right)$. The Siegel Theta function $\Theta(0|T)$ is defined as 
% by (2.5) in Ref.~\cite{2017PhRvD..96d6020C}
\begin{equation}
\Theta(0|T) = \sum_{\vec{m},\vec{n} \in \mathbb{Z}^{n-1}} e^{\frac{\pi i}{2}\left(\vec{n}\sqrt{\eta}+\frac{\vec{m}}{\sqrt{\eta}} \right)\cdot \tau \cdot \left(\vec{n}\sqrt{\eta}+\frac{\vec{m}}{\sqrt{\eta}} \right)} e^{-\frac{\pi i}{2}\left(\vec{n}\sqrt{\eta}-\frac{\vec{m}}{\sqrt{\eta}} \right)\cdot \tau \cdot \left(\vec{n}\sqrt{\eta}-\frac{\vec{m}}{\sqrt{\eta}} \right)}
\end{equation}
where $\tau,\bar{\tau}$ are $(n-1) \times (n-1)$ matrices. The exponent can be written as
\begin{equation}\label{SigelThetaExponent}
\pi i
\begin{pmatrix}
\vec{n}^t & \vec{m}^t
\end{pmatrix}\underbrace{
\begin{pmatrix}
\eta \frac{\tau -\bar{\tau}}{2} & \frac{\tau +\bar{\tau}}{2} \\ 
 \frac{\tau +\bar{\tau}}{2}  & \frac{\tau -\bar{\tau}}{2\eta}
\end{pmatrix}}_{= T}
\begin{pmatrix}
\vec{n} \\
\vec{m}
\end{pmatrix}
\end{equation}
where $T$ is the modular matrix. The negativity becomes
\begin{equation}
\mathcal{E} = \lim_{n_e \rightarrow 1} \mathcal{E}^{(n_e)}
\end{equation}
where we have defined
\begin{equation}
\mathcal{E}^{(n)} = \log \left[ \left(\frac{\pi}{\beta} \right)^{8h_n} \frac{F_n(x,\bar{x})}{\left[\sinh\frac{\pi(X_1-X_2)}{\beta}\sinh\frac{\pi(Y_1-Y_2)}{\beta}\right]^{4h_n}(1-x)^{2h_n}(1-\bar{x})^{2h_n}} \right]
\end{equation}
We define the bipartite logarithmic negativity without the conformal factors 
\begin{equation}\label{BipartiteLogNegativity}
\mathcal{E}^{(n)}(A,B) = \log \left[  \frac{F_n(x,\bar{x})}{(1-x)^{2h_n}(1-\bar{x})^{2h_n}} \right].
\end{equation}
With this definition, the bipartite logarithmic negativity begins at zero for disjoint intervals, thus respecting causality.
% We also introduced a factor of $\frac{1}{n-1}$ to make a better comparison with the bipartite operator mutual information.
Let us first specialize to the case of $n=2$.

Note that our cross ratios are negative. Care must be taken to analytically continue the hypergeometric functions outside the range of $(0,1)$.
\begin{equation}\label{Hypergeometric1/2}
f_{1/2}(x)=
\begin{cases}
\frac{2}{\pi}\frac{1}{\sqrt{1-x}}K\left(\frac{x}{x-1} \right)& x<0 \\
\frac{2}{\pi}\frac{1}{\sqrt{x}}\left[K\left(\frac{1}{x} \right) -iK(1-\frac{1}{x})\right]& x>1
\end{cases}
\end{equation}
The modular parameters are
\begin{align}
\tau &= i \frac{{}_2F_1(1/2,1/2,1,1-x)}{{}_2F_1(1/2,1/2,1,x)} = i\frac{K\left(\frac{1}{1-x}\right)}{K\left(\frac{x}{x-1}\right)}+1,
\\
\bar{\tau} &= -i \frac{{}_2F_1(1/2,1/2,1,1-\bar{x})}{{}_2F_1(1/2,1/2,1,\bar{x})} = i\left(-\frac{K\left(\frac{1}{1-\bar{x}}\right)}{K\left(\frac{\bar{x}}{\bar{x}-1}\right)}\right)+1.
\end{align}
%For $n=2$, the exponent (\ref{SigelThetaExponent}) becomes
%\begin{align}
%\pi i \begin{pmatrix}n&m \end{pmatrix} T \begin{pmatrix} n \\ m \end{pmatrix}
%&= \pi i \begin{pmatrix}n&m \end{pmatrix} \underbrace{
%\begin{pmatrix}
%\eta i \frac{\Im(\tau)-\Im(\bar{\tau})}{2} & i \frac{\Im(\tau)+\Im(\bar{\tau})}{2}\\
%i \frac{\Im(\tau)+\Im(\bar{\tau})}{2} & i \frac{\Im(\tau)-\Im(\bar{\tau})}{2\eta}
%\end{pmatrix}}_{\equiv \tilde{T}}
% \begin{pmatrix} n \\ m \end{pmatrix}
% +
%\pi i \begin{pmatrix}n&m \end{pmatrix} 
%\begin{pmatrix}
%\eta  & 0\\
%0& \frac{1}{\eta}
%\end{pmatrix}
 %\begin{pmatrix} n \\ m \end{pmatrix}\\ \nonumber
%&= \pi i \begin{pmatrix}n&m \end{pmatrix} \tilde{T} \begin{pmatrix} n \\ m \end{pmatrix}
%+i \pi \left(\eta n^2+\frac{m^2}{\eta} \right)
%\end{align}
%$\tilde{T}$ is purely imaginary, so the first term is purely real and gives real contributions to $\Theta(0|T)$, which is what happens for the bipartite operator mutual information. The other term introduces complex phases into the sum.

Next, we compute the $2^{\text{nd}}$ tripartite logarithmic negativity
\begin{equation}\label{2ndTripartiteLogNegativity}
\mathcal{E}^{(2)}_3(A,B_1,B_2) = \mathcal{E}^{(2)}(A,B_1) +\mathcal{E}^{(2)}(A,B_2)-\mathcal{E}^{(2)}(A,B)
\end{equation}
where $B = B_1 \cup B_2$. We choose $B_1$ and $B_2$ to be a biparition of the output system. In other words, $B_1 = [Y_2,Y_1]$ and $B_2 = [Y_3, Y_2]$ with $Y_1 \rightarrow \infty$ and $Y_3 \rightarrow - \infty$. Keeping $A = [X_2, X_1]$ as before, the cross ratios reduce to
\begin{align}
x_{AB_1} &= -e^{\frac{\pi}{\beta}(X_1-t-Y_2)}\frac{\sinh\frac{\pi(X_1-X_2)}{\beta}}{\cosh\frac{\pi(X_2-Y_2-t)}{\beta}},
\quad
\bar{x}_{AB_1} = -e^{\frac{\pi}{\beta}(X_1+t-Y_2)}\frac{\sinh\frac{\pi(X_1-X_2)}{\beta}}{\cosh\frac{\pi(X_2-Y_2+t)}{\beta}}, \\ \nonumber
x_{AB_2} &= -e^{\frac{\pi}{\beta}(Y_2-X_2+t)}\frac{\sinh\frac{\pi(X_1-X_2)}{\beta}}{\cosh\frac{\pi(X_1-Y_2-t)}{\beta}},
\quad
\bar{x}_{AB_2} = -e^{\frac{\pi}{\beta}(Y_2-X_2-t)}\frac{\sinh\frac{\pi(X_1-X_2)}{\beta}}{\cosh\frac{\pi(X_1-Y_2+t)}{\beta}}, \\ \nonumber
x_{AB} &= \bar{x}_{AB} = 1 - e^{\frac{2\pi}{\beta}(X_1-X_2)}.
\end{align}

As the second R\'{e}nyi entropy and second logarithmic negativity are essentially the same, to observe any interesting behaviour, we need to go to larger R\'{e}nyi index. The formulas are mostly the same for $n=3$ with a few minor changes. First, the analytic continuation for the hypergeometric functions are different from (\ref{Hypergeometric1/2}). For $k/n \neq 1/2$,
\begin{align}
f_{k/n} (x) 
&= {}_2F_1(k/n,1-k/n,1,x) \\ \nonumber
&=
\begin{cases}
\frac{\Gamma(1-2k/n)}{\Gamma(1-k/n)^2}e^{-\frac{i \pi k}{n}}\left(\frac{1}{x} \right)^{k/n}\, {}_2F_1(\frac{k}{n},\frac{k}{n},\frac{2k}{n},\frac{1}{x}) 
\\ \quad 
-\frac{\Gamma(2k/n-1)}{\Gamma(k/n)^2}e^{\frac{i \pi k}{n}}\left(\frac{1}{x} \right)^{1-k/n}\,{}_2F_1(1-\frac{k}{n},1-\frac{k}{n},2(1-\frac{k}{n}),\frac{1}{x})& x>1 \\ \\
\left(\frac{1}{1-x}\right)^{k/n}\,{}_2F_1(\frac{k}{n},\frac{k}{n},1,\frac{x}{x-1})&x<0
\end{cases}
\end{align}
The modular parameters become two dimensional modular matrices
\begin{align}
(\tau)_{ij} &= \frac{2}{3}\sum_{k=1}^{2} p_{k/3}(x) \sin\left(\frac{\pi k}{3} \right) \cos\left(\frac{2\pi k(i-j)}{3} \right), \\
(\bar{\tau})_{ij} &= \frac{2}{3}\sum_{k=1}^{2} q_{k/3}(\bar{x}) \sin\left(\frac{\pi k}{3} \right) \cos\left(\frac{2\pi k(i-j)}{3} \right), 
\end{align}
where the functions $p_{k/n}$ and $q_{k/n}$ are
\begin{align}
&p_{k/n}= \\ \nonumber
& i \frac{\frac{\Gamma(1-\frac{2k}{n})}{\Gamma(1-\frac{k}{n})^2}e^{-\frac{i \pi k}{n}}\left(\frac{1}{1-x} \right)^{k/n}\, {}_2F_1(\frac{k}{n},\frac{k}{n},\frac{2k}{n},\frac{1}{1-x}) -\frac{\Gamma(\frac{2k}{n}-1)}{\Gamma(\frac{k}{n})^2}e^{\frac{i \pi k}{n}}\left(\frac{1}{1-x} \right)^{1-\frac{k}{n}}\,{}_2F_1(1-\frac{k}{n},1-\frac{k}{n},2(1-\frac{k}{n}),\frac{1}{1-x})}{\left(\frac{1}{1-x}\right)^{k/n}\,{}_2F_1(\frac{k}{n},\frac{k}{n},1,\frac{x}{x-1})} \\ \nonumber
&q_{k/n} =\\ \nonumber
& -i \frac{\frac{\Gamma(1-\frac{2k}{n})}{\Gamma(1-\frac{k}{n})^2}e^{\frac{i \pi k}{n}}\left(\frac{1}{1-\bar{x}} \right)^{k/n}\, {}_2F_1(\frac{k}{n},\frac{k}{n},\frac{2k}{n},\frac{1}{1-\bar{x}}) -\frac{\Gamma(\frac{2k}{n}-1)}{\Gamma(\frac{k}{n})^2}e^{-\frac{i \pi k}{n}}\left(\frac{1}{1-\bar{x}} \right)^{1-k/n}\,{}_2F_1(1-\frac{k}{n},1-\frac{k}{n},2(1-\frac{k}{n}),\frac{1}{1-\bar{x}})}{\left(\frac{1}{1-\bar{x}}\right)^{k/n}\,{}_2F_1(\frac{k}{n},\frac{k}{n},1,\frac{\bar{x}}{\bar{x}-1})} 
\end{align}
Since the cross ratios are negative, we must also replace the denominator of (\ref{BipartiteLogNegativity})
\begin{equation}
\prod_{k=1}^{n-1}f_{k/n}(x)f_{k/n}(\bar{x}) \rightarrow \prod_{k=1}^{n-1}\left[ 
\left(\frac{1}{1-x}\right)^{k/n}\,{}_2F_1\left(\frac{k}{n},\frac{k}{n},1,\frac{x}{x-1}\right)
\left(\frac{1}{1-\bar{x}}\right)^{k/n}\,{}_2F_1\left(\frac{k}{n},\frac{k}{n},1,\frac{\bar{x}}{\bar{x}-1}\right)\right].
\end{equation}
Apart from these modifications, all the formulas are the same as in the case of $n=2$.

% \bibliography{main}
% \bibliographystyle{JHEP.bst}
\providecommand{\href}[2]{#2}\begingroup\raggedright\endgroup

\end{document}